\documentclass[10pt,twocolumn,showpacs,preprintnumbers,amsmath,amssymb,aps,prb,longbibliography,superscriptaddress,nofootinbib]{revtex4-2}
\usepackage{array,braket,mathtools,siunitx}

\usepackage[unicode=true,
 bookmarks=true,bookmarksnumbered=true,bookmarksopen=false,
 breaklinks=true,pdfborder={0 0 0},backref=false,colorlinks=true]{hyperref}
\hypersetup{citecolor={blue},urlcolor={magenta}}

\usepackage{cleveref}
\crefname{appendix}{App.}{Apps.}
\crefname{equation}{Eq.}{Eqs.}
\crefname{figure}{Fig.}{Figs.}
\crefname{table}{Tab.}{Tabs.}
\crefname{section}{Sec.}{Secs.}

\newcommand{\Z}{\mathbb{Z}}
\newcommand{\R}{\mathbb{R}}
\newcommand{\br}{\mathbf{r}}
\newcommand{\bk}{\mathbf{k}}
\newcommand{\ba}{\mathbf{a}}
\newcommand{\bb}{\mathbf{b}}
\newcommand{\btau}{\boldsymbol\tau}
\newcommand{\bK}{\mathbf{K}}
\newcommand{\bM}{\mathbf{M}}
\newcommand{\bG}{\mathbf{G}}
\newcommand{\bQ}{\mathbf{Q}}
\newcommand{\bp}{\mathbf{p}}
\newcommand{\bq}{\mathbf{q}}
\newcommand{\bsigma}{\boldsymbol\sigma}
\newcommand{\bGamma}{\boldsymbol\Gamma}
\newcommand{\bhatx}{\mathbf{\hat{x}}}
\newcommand{\bhaty}{\mathbf{\hat{y}}}
\newcommand{\bhatz}{\mathbf{\hat{z}}}
\newcommand{\bhatn}{\mathbf{\hat{n}}}
\newcommand{\bu}{\mathbf{u}}
\newcommand{\bv}{\mathbf{v}}
\newcommand{\bzero}{\mathbf{0}}
\newcommand{\namezero}{\epsilon_0 = 0 \text{, }\theta_0 = 0}
\newcommand{\namesqthree}{\epsilon_0 = \ln\sqrt{3} \text{, }\theta_0 = 30^{\circ}}
\newcommand{\latconst}{a}

\begin{document}
\title{Kagome and honeycomb flat bands in moir\'e graphene}
\author{Michael G. Scheer}
\author{Biao Lian}
\affiliation{Department of Physics, Princeton University, Princeton, New Jersey 08544, USA}
\date{\today}

\begin{abstract}
We propose a class of graphene-based moir\'e systems hosting flat bands on kagome and honeycomb moir\'e superlattices. These systems are formed by stacking a graphene layer on a 2D substrate with lattice constant approximately $\sqrt{3}$ times that of graphene. When the moir\'e potentials are induced by a 2D irreducible corepresentation in the substrate, the model shows a rich phase diagram of low energy bands including eigenvalue fragile phases as well as kagome and honeycomb flat bands. Spin-orbit coupling in the substrate can lift symmetry protected degeneracies and create spin Chern bands, and we observe spin Chern numbers up to three. We additionally propose a moir\'e system formed by stacking two graphene-like layers with similar lattice constants and Fermi energies but with Dirac Fermi velocities of opposite sign. This system exhibits multiple kagome and honeycomb flat bands simultaneously. Both models we propose resemble the hypermagic model of [Scheer \textit{et al.}, Phys. Rev. B \textbf{106}, 115418 (2022)] and may provide ideal platforms for the realization of strongly correlated topological phases.
\end{abstract}

\maketitle

\section{Introduction}
Since the discovery of the fractional quantum Hall effect, nontrivial flat bands in two dimensions have become a dominant paradigm for strongly correlated topological phases of matter. In recent years, moir\'e systems such as twisted bilayer graphene (TBG) \cite{Bistritzer2011} have been predicted to host nontrivial flat bands \cite{Po2019,Ahn2019,Song2018,Song2021} and experiments have revealed remarkable strongly interacting phenomena such as unconventional superconductivity and correlated insulation \cite{Cao2018,Cao2018a,Yankowitz2019,Sharpe2019,Xie2019,Lu2019,Serlin2020}. Another way to realize nontrivial flat bands is to construct tight-binding models that exhibit destructive wavefunction interference \cite{Lieb1989,Mielke1991,Calugaru2022}. For example, a nearest neighbor tight-binding model with one symmetric orbital per site on a kagome lattice has one exactly flat band \cite{Bergman2008,Xu2015}. Interacting systems based on this tight-binding model have long been investigated as candidates for strongly correlated phases such as Mott insulators or spin liquids \cite{Yan2011,Yin2022}. Similarly, a nearest neighbor tight-binding model with $p_x$ and $p_y$ orbitals on each site of a honeycomb lattice has two exactly flat bands when certain parameters are neglected \cite{Wu2007}. For completeness, we review these flat band tight-binding models in \cref{app:kagome-honeycomb-models,app:kagome-honeycomb-bcl}. Despite extensive searches for crystalline materials realizing such nontrivial flat bands, they remain quite rare. Additionally, it is often difficult to tune the Fermi level into a flat band \cite{Liu2021,Regnault2022,Jovanovic2022,Yin2022}.

\begin{figure}[h]
	\centering
	\includegraphics{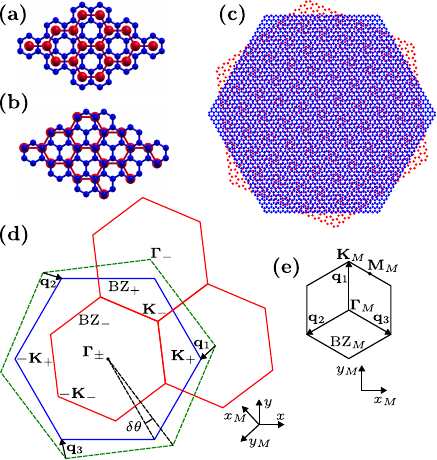}
	\caption{Illustration of the coupled-valley graphene model. \textbf{(a)}, \textbf{(b)} Real space commensurate stacking configurations described by \cref{eq:commensurate-sqrt3-30deg} for two honeycomb lattice materials (made with \cite{Kokalj1999}). \textbf{(c)} The moir\'e pattern of two honeycomb lattices described by \cref{eq:moire-condition,eq:commensurate-sqrt3-30deg} with $\delta\epsilon = -0.15$ and $\delta\theta = 7^{\circ}$. \textbf{(d)} A momentum space diagram illustrating the moir\'e vectors $\bq_j$, the graphene coordinate system $(x,y)$, and the moir\'e coordinate system $(x_M,y_M)$. The graphene and substrate BZs are shown in blue and red, respectively. The green dashed hexagon is formed from substrate $\bGamma_-$ points and coincides with the graphene BZ when $\delta\epsilon = \delta\theta = 0$. \textbf{(e)} The moir\'e BZ and high symmetry moir\'e quasimomenta in the moir\'e coordinate system $(x_M,y_M)$. This diagram pertains to both the coupled-valley graphene and opposite-velocity models.}
	\label{fig:sqrt3-diagram}
\end{figure}

A natural question is whether kagome or honeycomb flat band tight-binding models can be realized in moir\'e materials, which generally enjoy highly tunable Fermi levels. Recent studies on moir\'e models for twisted crystalline materials with triangular Bravais lattices and low energy physics near the $\bGamma$ point have found kagome and honeycomb flat bands. These so-called $\bGamma$-valley models have been derived for transition metal dichalcogenides (TMDs) \cite{Angeli2021,Xian2021}, interfaces between topological insulators and ferromagnetic insulators \cite{Liu2022}, and semiconductors \cite{Wang2022}. Kagome and honeycomb moir\'e superlattices have been observed with scanning tunneling microscopy in twisted bilayer $\text{WSe}_2$, although the orbitals from which they emerge remain obscure \cite{Pei2022}.

In this paper, we introduce a graphene-based moir\'e model that realizes kagome and honeycomb lattice flat bands. We consider a graphene layer stacked on a two-dimensional crystalline substrate with a triangular Bravais lattice and a lattice constant approximately $\sqrt{3}$ times that of graphene. At a twist angle near $30^{\circ}$, the two layers are nearly commensurate and produce a moir\'e pattern as shown in \cref{fig:sqrt3-diagram}\textbf{(c)}. The $\bK$ and $-\bK$ graphene valleys are both folded close to the $\bGamma$ point of the substrate, and are thus coupled by van der Waals interactions. Assuming the substrate states near the $\bGamma$ point are gapped at the graphene Fermi energy, the low energy physics in the graphene layer can be described by a continuum model consisting of two Dirac cones with intervalley and intravalley moir\'e potentials. We refer to this model as the \emph{coupled-valley graphene model}. When the substrate has maximal symmetry and the interlayer twist angle is exactly $30^{\circ}$, the model respects the magnetic space group $P6mm1'$ (No. 183.186 in the BNS setting \cite{Gallego2012}). A particularly interesting limit is that in which the moir\'e potentials are produced entirely by substrate states occupying a two-dimensional (2D) spinless irreducible corepresentation (coirrep) \cite{Newmarch1982}. Without spin-orbit coupling (SOC), we find a rich phase diagram of low energy bands including eigenvalue fragile phases \cite{Song2020} and magic parameters with kagome or honeycomb flat bands near charge neutrality. With SOC in the substrate, the $z$ component of spin is approximately conserved. We find moir\'e bands with spin Chern numbers up to three, allowing realization of the quantum spin Hall effect \cite{Kane2005,Bernevig2006}.

We note that commensurate bilayers with exactly $\sqrt{3}$ lattice constant ratio and perfect $30^{\circ}$ alignment have been studied using density functional theory \cite{Cai2013,Hamid2021}. Moir\'e materials near this configuration have also been studied theoretically and experimentally \cite{Wallbank2013,Dunbrack2021,Lu2022,Wu2022}. However, these studies did not consider the case in which the moir\'e potentials are produced by a 2D spinless coirrep in the layer with a larger lattice constant, nor did they report kagome or honeycomb moir\'e flat bands.

We additionally propose a moir\'e model consisting of two stacked graphene-like materials with nearly equal lattice constants, both of which have Dirac cones at $\bK$ and $-\bK$. We show that if the two layers have nearly opposite Dirac Fermi velocities and nearly equal Fermi energies, the system may have multiple kagome and honeycomb moir\'e flat bands simultaneously near charge neutrality. We refer to this model as the \emph{opposite-velocity model}. Both models we propose bear strong mathematical resemblance to the hypermagic model of Ref. \cite{Scheer2022}.

\begin{table*}
	\centering
	\begin{tabular}{c|c|c|c|c}
	\hline\textbf{(a)} $T_{s,\bq}$ and $S_{s,\eta,\bq}$ & $\sigma_0$ & $\bsigma \cdot \bhatn_{\zeta_j + \phi_1 - \pi/2}$ & $\bsigma \cdot \bhatn_{\zeta_j + \phi_1}$ & $\sigma_z$\\
	\hline $T_{s,m\bq_j}$ & $is\tilde{w}_{|m|,0}$ & $0$ & $w_{|m|,y}$ & $iw_{|m|,z}$\\
	$S_{s,\eta,\bzero}$ & $w_{0,0}$ & $0$ & $0$ & $s\tilde{w}_{0,z}$\\
	$S_{s,\eta,\gamma(\bq_{1+j} - \bq_{2+j})}$ & $w_{\sqrt{3},0} + is\eta\gamma\tilde{w}_{\sqrt{3},0}$ & $\eta w_{\sqrt{3},x} + is \gamma\tilde{w}_{\sqrt{3},x}$ & $0$ & $s\tilde{w}_{\sqrt{3},z} + i\eta\gamma w_{\sqrt{3},z}$\\
	& & & & \\
	\hline\textbf{(b)} 1D coirrep & $0$ & $x$ & $y$ & $z$\\
	\hline $E_0 w_{0,\mu}$ & $-3 |v|^2$ & $0$ & $0$ & $0$\\
	$E_0 w_{1,\mu}$ & $0$ & $0$ & $-2v^2$ & $M_{y,-} v^2$\\
	$E_0 w_{2,\mu}$ & $0$ & $0$ & $-v^2$ & $-M_{y,-} v^2$\\
	$E_0 w_{\sqrt{3},\mu}$ & $\frac{1}{2}|v|^2$ & $-M_{y,-}|v|^2$ & $0$ & $-\frac{\sqrt{3}}{2}|v|^2$\\
	& & & & \\
	\hline\textbf{(c)} 2D coirrep & $0$ & $x$ & $y$ & $z$\\
	\hline $E_0 w_{0,\mu}$ & $-3(|v_0|^2 + |v_x|^2)$ & $0$ & $0$ & $0$\\
	$E_0 w_{1,\mu}$ & $0$ & $0$ & $2v_0 v_x$ & $v_0^2 - 2v_x^2$\\
	$E_0 w_{2,\mu}$ & $0$ & $0$ & $-2v_0 v_x$ & $-(v_0^2 + v_x^2)$\\
	$E_0 w_{\sqrt{3},\mu}$ & $\frac{1}{2}(|v_0|^2 - 2|v_x|^2)$ & $\frac{1}{2}(v_0 v_x^* + v_0^* v_x)$ & $0$ & $\frac{\sqrt{3}}{2}|v_0|^2$\\
	& & & & \\
	\hline\textbf{(d)} 2D coirrep (SOC) & $0$ & $x$ & $y$ & $z$\\
	\hline $E_0 \tilde{w}_{0,\mu}$ & $0$ & $0$ & $0$ & $3\lambda(-|v_0|^2 + |v_x|^2)$\\
	$E_0 \tilde{w}_{1,\mu}$ & $\lambda(v_0^2 + 2v_x^2)$ & $0$ & $0$ & $0$\\
	$E_0 \tilde{w}_{2,\mu}$ & $\lambda(-v_0^2 + v_x^2)$ & $0$ & $0$ & $0$\\
	$E_0 \tilde{w}_{\sqrt{3},\mu}$ & $\frac{\lambda\sqrt{3}}{2}|v_0|^2$ & $-\frac{\lambda\sqrt{3}}{2}(v_0 v_x^* + v_0^* v_x)$ & $0$ &  $\frac{\lambda}{2}(|v_0|^2 + 2|v_x|^2)$
	\end{tabular}
	\caption{\textbf{(a)} Coefficients for expansions of $T_{s,\bq}$ and $S_{s,\eta,\bq}$ in \cref{eq:coupled-valley-T-S-expansion} with respect to the matrices shown in the column titles. The parameters $w_{m,\mu}$ and $\tilde{w}_{m,\mu}$ are real and have energy units, and $\tilde{w}_{m,\mu}=0$ in the absence of SOC. The angle $\phi_1$ is defined in \cref{eq:coupled-valley-phi_1}, $\zeta_j = \frac{2\pi}{3}(j-1)$, $\bhatn_\phi = R_\phi \bhatx$, and $\bsigma = \sigma_x \bhatx + \sigma_y \bhaty$. The index values are $j \in \{1, 2, 3\}$, $m \in \{1, -2\}$, and $\gamma \in \{1, -1\}$. \textbf{(b)} Expressions for $w_{m,\mu}$ in terms of $E_0$ and $v$ for a 1D spinless coirrep. \textbf{(c)} Expressions for $w_{m,\mu}$ in terms of $E_0$, $v_0$, and $v_x$ for a 2D spinless coirrep. \textbf{(d)} Expressions for $\tilde{w}_{m,\mu}$ in terms of $E_0$, $\lambda$, $v_0$, and $v_x$ for a 2D spinless coirrep with SOC.}
	\label{tbl:moire-model-form}
\end{table*}

\section{Coupled-valley graphene model}
We consider a system consisting of a graphene layer stacked on top of a 2D crystalline substrate with a triangular Bravais lattice. The lattice constants of graphene and the substrate are $\latconst \approx \SI{0.246}{\nano\meter}$ and $e^\epsilon \latconst$, respectively, for some real number $\epsilon$. Additionally, the substrate Bravais lattice has a counterclockwise rotation of angle $\theta$ relative to that of the graphene layer. In order to create a moir\'e pattern, we take
\begin{equation}\label{eq:moire-condition}
\epsilon = \epsilon_0 + \delta\epsilon,\quad \theta = \theta_0 + \delta\theta, \quad |\delta\epsilon|, |\delta\theta| \ll 1,
\end{equation}
where $\epsilon_0$ and $\theta_0$ describe some commensurate configuration\footnote{These commensurate configurations are fully classified in \cref{app:commensurate-configurations}, and \cref{apptbl:commensurate-configurations} lists the first $28$ of them. Additionally, \cref{app:general-moire-models} discusses the form for moir\'e models near arbitrary commensurate configurations.}. For the coupled-valley graphene model, we take
\begin{equation}\label{eq:commensurate-sqrt3-30deg}
\epsilon_0 = \ln\sqrt{3},\quad \theta_0 = 30^{\circ}.
\end{equation}

Hereafter, we use $+$ and $-$ subscripts to denote momenta in the top (graphene) and bottom (substrate) layers, respectively. As illustrated in \cref{fig:sqrt3-diagram}\textbf{(d)}, when $\delta\epsilon = \delta\theta = 0$ the momentum $\bK_+$ is an element of the commensurate reciprocal lattice. As a result, when $\delta\epsilon$ and $\delta\theta$ are not both zero, graphene states near $\bK_+$, $-\bK_+$, and $\bGamma_+$ are van der Waals coupled to substrate states near $\bGamma_-$, yielding a moir\'e Brillouin zone (BZ) as shown in \cref{fig:sqrt3-diagram}\textbf{(e)}. The graphene states near $\bGamma_+$ are far from the graphene Fermi energy and can be ignored. Assuming the substrate states around $\bGamma_-$ are also highly detuned, we can use Schrieffer-Wolff perturbation theory to project out the substrate states and derive a moir\'e model within the graphene layer involving only states near $\bK_+$ and $-\bK_+$. The result is the \emph{coupled-valley graphene model} which is derived in detail in \cref{app:coupled-valley-spinless,app:coupled-valley-spinful}, and which we now present.

We assume that SOC is present only in the substrate and not in the interlayer hoppings. Due to the high symmetry of the $\bGamma_-$ point, the $z$ component of electron spin is conserved. As a result, we can describe the system with a moir\'e Hamiltonian $H_s$ for electrons of each spin $s \in \{\uparrow, \downarrow\}$. In a convenient real space basis, this Hamiltonian takes the form
\begin{equation}\label{eq:coupled-valley-hamiltonian}
H_s = \begin{pmatrix}
S_{s,+}(\br) - i\hbar v_F \bsigma_M \cdot \nabla & T_s(\br)\\
T^\dagger_s(\br) & S_{s,-}(\br) -i\hbar v_F \bsigma_M \cdot \nabla
\end{pmatrix}
\end{equation}
where $T_s(\br)$ and $S_{s,\eta}(\br)$ are intervalley and intravalley moir\'e potentials, $\eta \in \{+, -\}$ stands for graphene valley, $v_F \approx 10^6 \si{\meter\per\second}$ is the graphene Fermi velocity, $\bsigma_M = \sigma_x \bhatx_M + \sigma_y \bhaty_M$ is a vector of Pauli matrices, and $\bhatx_M, \bhaty_M$ are axis unit vectors for the moir\'e coordinate system in \cref{fig:sqrt3-diagram}\textbf{(d)}, \textbf{(e)}. We choose the zero energy point to be the graphene Fermi energy. As explained in \cref{app:coupled-valley-spinless,app:coupled-valley-spinful}, we have chosen a basis in which the two Dirac cones in \cref{eq:coupled-valley-hamiltonian} have the same form even though they originate from opposite graphene valleys.

To leading order, $T_s(\br)$ and $S_{s,\eta}(\br)$ originate from second order hopping processes among the valleys, namely $\eta\bK_+\rightarrow \bGamma_-\rightarrow -\eta\bK_+$ and $\eta\bK_+\rightarrow \bGamma_-\rightarrow \eta\bK_+$, respectively. They can be expanded as
\begin{equation}\label{eq:coupled-valley-T-S-expansion}
\begin{split}
T_s(\br) &= \sum_{j=1}^3\sum_{m=1,-2} T_{s,m\bq_j} e^{im\bq_j\cdot \br}\\
S_{s,\eta}(\br) &= S_{s,\eta,\bzero} + \sum_{j=1}^3 \sum_{\gamma=\pm} S_{s,\eta,\gamma(\bq_{1+j} - \bq_{2+j})} e^{i \gamma(\bq_{1+j} - \bq_{2+j})\cdot \br}
\end{split}
\end{equation}
for $2\times 2$ complex matrices $T_{s,\bq}$ and $S_{s,\eta,\bq}$. The $\bq_j$ vectors are defined by
\begin{equation}\label{eq:define-q_j}
\bq_j = R_{\zeta_j}\bK_M=|\bK_M| R_{\zeta_j} \bhaty_M
\end{equation}
where
\begin{equation}\label{eq:coupled-valley-K_M}
\bK_M=(1-e^{-\delta\epsilon}R_{\delta\theta})\bK_+,\quad |\bK_M|\approx \sqrt{\delta\epsilon^2+\delta\theta^2}|\bK_+|.
\end{equation}
Here, $R_\phi$ denotes rotation by angle $\phi$ about $\bhatz$ and $\zeta_j = \frac{2\pi}{3}(j-1)$. The $\bq_j$ vectors, moir\'e coordinate system, moir\'e BZ, and high symmetry moir\'e quasimomenta are illustrated in \cref{fig:sqrt3-diagram}\textbf{(d)}, \textbf{(e)}.

The graphene electronic states carry a corepresentation (corep) of the magnetic space group $P6mm1'$, which is generated by Bravais lattice translations, $C_{6z}$ (rotation by $\pi/3$ about $\bhatz$), $M_y$ (reflection through the $xz$ plane), and $\mathcal{T}$ (spinful time-reversal). We assume that the substrate states also carry a corep of $P6mm1'$. When $\delta\theta = 0$, the moir\'e model in \cref{eq:coupled-valley-hamiltonian} then also carries a corep of $P6mm1'$. In this case, the symmetries constrain the moir\'e potentials to the form in \cref{tbl:moire-model-form}\textbf{(a)}, where the parameters $w_{m,\mu}$ and $\tilde{w}_{m,\mu}$ are real and have energy units, the $\tilde{w}_{m,\mu}$ parameters vanish in the absence of SOC, and
\begin{equation}\label{eq:coupled-valley-phi_1}
\phi_1 = \arg\left(e^{-\delta\epsilon - i\delta\theta}-1\right)\approx \arg\left(-\delta\epsilon - i\delta\theta\right).
\end{equation}
When $\delta\theta$ is nonzero but small, the $M_y$ symmetry is weakly broken. The moir\'e potentials may then gain small symmetry-breaking perturbations on the order of $\delta\theta$, but these are typically negligible. See Sec. III of Ref. \cite{Scheer2022} for a discussion of a similar symmetry-breaking term in the model for TBG near a commensurate twist angle.

We first examine the case without SOC. The moir\'e potentials are then spin-independent and are determined by the spinless interlayer coupling and substrate Hamiltonian $\mathcal{H}_-$ at $\bGamma_-$. $\mathcal{H}_-$ carries a spinless corep $\rho_-$ of the magnetic point group $6mm1'$ (see \cref{app:point-groups}). By Schur's lemma, one can diagonalize $\mathcal{H}_-$ while also decomposing $\rho_-$ into coirreps. As a result, we can expand the moir\'e potentials in \cref{eq:coupled-valley-T-S-expansion} as a sum of contributions of coirreps in $\rho_-$.

The spinless coirreps of $6mm1'$ all have dimension $1$ or $2$ (see \cref{apptbl:characters}). The moir\'e potential contribution from a 1D spinless coirrep depends on its energy $E_0$ in $\mathcal{H}_-$ (measured relative to the graphene Fermi energy) and an interlayer hopping parameter $v$, while the contribution from a 2D spinless coirrep depends on its energy $E_0$ and two interlayer hopping parameters, $v_0$ and $v_x$. The parameter $v$ is either real or imaginary, depending on the coirrep. Likewise, $v_0$ and $v_x$ are either both real or both imaginary. The $w_{m,\mu}$ parameters are given in terms of $E_0$, $v$, $v_0$, and $v_x$ in \cref{tbl:moire-model-form}\textbf{(b)}, \textbf{(c)}.

\begin{figure}[h]
	\centering
	\includegraphics{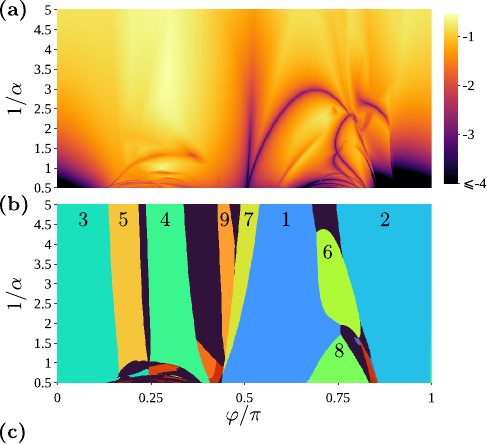}
	\begin{tabular}{c|cc|c}
		phase & min & max & $P6mm1'$ EBR decomps\\
		\hline $1$ & $-2$ & $2$ & $(E)_{2b}$\\
		$2$ & $-2$ & $1$ & $(E_2)_{1a} \oplus (B_2)_{1a}$\\
		$3$ & $-1$ & $1$ & $(A_2)_{2b}$\\
		$4$ & $-1$ & $1$ & $(A_2)_{3c} \oplus (E_1)_{1a} \oplus (A_1)_{1a} \boxminus (E)_{2b}$\\
		$5$ & $-1$ & $3$ & $(A_2)_{2b} \oplus (E_1)_{1a}$\\
		$6$ & $-2$ & $2$ & $(A_1)_{3c} \oplus (E_2)_{1a} \oplus (B_1)_{1a} \boxminus (A_1)_{2b}$\\
		$7$ & $-2$ & $1$ & $(A_1)_{3c}$\\
		$8$ & $-2$ & $1$ & $(A_2)_{3c}$\\
		$9$ & $-3$ & $1$ & $(A_1)_{3c} \oplus (A_1)_{2b} \boxminus (A_1)_{1a}$
	\end{tabular}
	\caption{Low energy bandwidths and phase diagram of low energy bands for the coupled-valley graphene model without SOC and with moir\'e potentials arising from a 2D spinless coirrep at $\bGamma_-$. We take $\delta\theta = 0$, $\delta\epsilon < 0$, $\phi_1 = 0$, $v_0, v_x \in \R$, $\lambda = 0$, and $E_0 < 0$. \textbf{(a)} The base $10$ logarithm of the narrowest bandwidth (in units of $\hbar v_F |\bK_M|$) among the first three valence bands and first three conduction bands at charge neutrality. The bandwidth is computed with moir\'e quasimomenta $\bGamma_M$, $\bK_M$, $\bM_M$, $\bK_M/2$, and $\bM_M/2$. \textbf{(b)} Phase diagram of low energy bands. All parameters for which the low energy band structure has more than four bands are shown in black. \textbf{(c)} For each of the nine largest phases in \textbf{(b)}, we show the band indices and a linear combination of EBRs of $P6mm1'$ for the low energy band structure. The $n$th conduction (valence) band has index $n$ ($-n$). The symbols $\oplus$ and $\boxminus$ indicate sum and difference of EBRs, respectively. The full list of EBRs for $P6mm1'$ can be found on the Bilbao Crystallographic Server \cite{Elcoro2021,Xu2020}.}
	\label{fig:sqrt3-phases}
\end{figure}

We now suppose that the substrate has SOC, and note that any effect of SOC on the interlayer couplings can be neglected at leading order. Since all spinful coirreps of $6mm1'$ have dimension $2$ (see \cref{apptbl:characters}), the inclusion of spin maps 1D spinless coirreps to 2D spinful coirreps and 2D spinless coirreps to 4D spinful reducible coreps. As a result, SOC can only modify contributions arising from 2D spinless coirreps, and in this case it simply splits the 4D spinful corep into two spinful coirreps with energies $E_1$ and $E_2$. We define
\begin{equation}\label{eq:E0-lambda-SOC}
E_0 = \frac{2}{E_1^{-1} + E_2^{-1}},\quad \lambda = \frac{E_1^{-1} - E_2^{-1}}{E_1^{-1} + E_2^{-1}}
\end{equation}
where $\lambda$ characterizes the SOC strength. The parameters $w_{m,\mu}$ and $\tilde{w}_{m,\mu}$ for a 2D spinless coirrep with SOC are given in \cref{tbl:moire-model-form}\textbf{(c)}, \textbf{(d)} in terms of $E_0$, $\lambda$, $v_0$, and $v_x$.

Since $w_{m,\mu}$ and $\tilde{w}_{m,\mu}$ for a single coirrep vary inversely with $E_0$, the moir\'e potentials will typically be dominated by coirreps near the graphene Fermi energy. We consider now the intriguing limit in which the moir\'e potentials arise entirely from a single 2D spinless coirrep. Without loss of generality, we assume $v_0, v_x \in \R$ (see \cref{app:phase-diagram-parameters}). In this case, the model can be parameterized by the dimensionless quantities
\begin{equation}\label{eq:define-alpha-varphi}
\alpha = \frac{|v_0|^2 + |v_x|^2}{|E_0| \hbar v_F |\bK_M|},\quad \varphi = \arg(v_0 + i v_x),
\end{equation}
$\lambda$, $\phi_1$, and the sign of $E_0$. The parameter $\alpha$ resembles the parameter of the same name in the BM model \cite{Bistritzer2011} in that larger $\alpha$ corresponds to stronger interlayer coupling and larger moir\'e lattice constant.

\begin{figure}[h]
	\centering
	\includegraphics{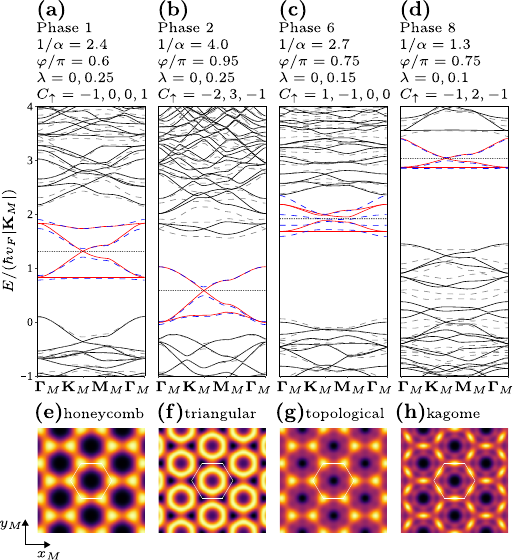}
	\caption{\textbf{(a)}-\textbf{(d)} Moir\'e band structures of the coupled-valley model in \cref{eq:coupled-valley-hamiltonian} for phases $1$, $2$, $6$, and $8$ of \cref{fig:sqrt3-phases}\textbf{(b)}, with parameters written above the panels. Band structures without (with) SOC are shown with solid (dashed) lines. The black dotted lines indicate charge neutrality. The low (high) energy bands are shown without SOC in red (black) and with SOC in blue (gray). The spin Chern numbers are written in increasing order of energy above the panels. \textbf{(e)}-\textbf{(h)} Real space charge density distributions from fully filling only the low energy spinless bands shown in red in \textbf{(a)}-\textbf{(d)}. Lighter colors indicate higher charge densities. A moir\'e unit cell is shown as a white hexagon. A description of the EBR decomposition given in \cref{fig:sqrt3-phases}\textbf{(c)} is written above each panel. The low energy spinless bands illustrated in \textbf{(c)} and \textbf{(g)} have at least a fragile topology and therefore do not admit exponentially localized symmetric Wannier functions \cite{Po2018,Bouhon2019,Song2020a,Song2020}. The highest two spinless valence bands in \textbf{(c)} are compatible with the EBR decomposition $(A_1)_{3c} \oplus (B_1)_{1a} \boxminus (A_1)_{2b}$ while the lowest two spinless conduction bands are compatible with $(E_2)_{1a}$. \cref{appfig:sqrt3-bands} contains similar plots for the other phases tabulated in \cref{fig:sqrt3-phases}\textbf{(c)}.}
	\label{fig:sqrt3-bands}
\end{figure}

We focus on the case with $\delta\theta = 0$ and $\delta\epsilon\neq 0$. Without loss of generality, we choose $\delta\epsilon < 0$, $\phi_1 = 0$, and $E_0 < 0$ (see \cref{app:phase-diagram-parameters}). We then explore the moir\'e band structure without SOC as a function of $\alpha \in (0, \infty)$ and $\varphi \in [0, \pi)$. For each set of parameters, we identify a low energy band structure, by which we mean a minimal set of bands containing the first valence and conduction bands such that the symmetry coirreps at high symmetry points (i.e., $\bGamma_M$, $\bK_M$, and $\bM_M$) are well defined and satisfy the momentum space compatibility relations for $P6mm1'$ from magnetic topological quantum chemistry\footnote{If more than one minimal set of bands exists, we choose the set containing the fewest number of conduction bands.} \cite{Bradlyn2017,Kruthoff2017,Elcoro2021}. We then define the phases by identifying an integer linear combination of elementary band representations (EBRs) with a minimal sum for the negative coefficients compatible with each low energy band structure. \cref{fig:sqrt3-phases}\textbf{(b)} shows all phases for which the low energy band structure has at most four bands and \cref{fig:sqrt3-phases}\textbf{(c)} tabulates the EBR decompositions for the nine largest such phases. The full list of EBRs for each magnetic space group can be found on the Bilbao Crystallographic Server \cite{Elcoro2021,Xu2020}. Example band structures are shown with solid lines in \cref{fig:sqrt3-bands}\textbf{(a)}-\textbf{(d)} and \cref{appfig:sqrt3-bands}\textbf{(a)}-\textbf{(e)}. The low (high) energy bands are shown in red (black).

Remarkably, phase $1$ is compatible with the EBR for the honeycomb lattice flat band model \cite{Wu2007} (see \cref{app:honeycomb-two-orbital}), while phases $7$ and $8$ are compatible with EBRs for the kagome lattice flat band model \cite{Bergman2008,Xu2015} (see \cref{app:kagome-one-orbital}). Moreover, the linear combinations of EBRs for phases $4$, $6$, and $9$ include a subtraction, implying that these phases have at least a fragile topology when the low energy bands are isolated\footnote{The low energy band structure is neither guaranteed to be connected nor isolated. For example, in \cref{fig:sqrt3-bands}\textbf{(c)} they are isolated but not connected, while in \cref{appfig:sqrt3-bands}\textbf{(e)} they are connected but not isolated.} \cite{Po2018,Bouhon2019,Song2020a,Song2020}. Real space charge density distributions corresponding to the low energy band structures are shown in \cref{fig:sqrt3-bands}\textbf{(e)}-\textbf{(h)} and \cref{appfig:sqrt3-bands}\textbf{(f)}-\textbf{(j)}. Triangular, honeycomb, and kagome lattice patterns are clearly visible for phases $2$, $1$, and $8$, respectively, in agreement with \cref{fig:sqrt3-phases}\textbf{(c)}.

In addition to identifying phases, it is also important to search for flat bands. The dark regions in \cref{fig:sqrt3-phases}\textbf{(a)} indicate parameters for which there is a flat band among the first three valence and first three conduction bands. We see that several phases admit flat bands, and the parameters for \cref{fig:sqrt3-bands}\textbf{(a)}, \textbf{(c)}, \textbf{(d)} were chosen to exhibit extremely flat bands near charge neutrality.

With SOC, the symmetry protected degeneracies are generically split and the bands gain spin Chern numbers $C_\uparrow$ and $C_\downarrow = -C_\uparrow$. The dashed lines in \cref{fig:sqrt3-bands}\textbf{(a)}-\textbf{(d)} and \cref{appfig:sqrt3-bands}\textbf{(a)}-\textbf{(e)} show moir\'e band structures with SOC. For example, \cref{fig:sqrt3-bands}\textbf{(d)} has a band with $C_\uparrow=3$. We note that the spin Chern numbers with fixed $\lambda$ are not necessarily constant within each spinless phase.

\section{Opposite-velocity model}
We now introduce another moir\'e construction yielding kagome and honeycomb flat bands (see \cref{app:opposite-velocity-model} for a detailed derivation). We consider a stack of two 2D materials with triangular Bravais lattices and $P6mm1'$ symmetry. We assume that both layers have Dirac cones at their $\bK$ and $-\bK$ points centered at their Fermi energies, carrying the same coreps of $P6mm1'$ as that of graphene. The bottom layer to top layer lattice constant ratio $e^\epsilon$ and counterclockwise twist angle $\theta$ satisfy \cref{eq:moire-condition} with
\begin{equation}\label{eq:commensurate-0-0}
\epsilon_0 = 0,\quad \theta_0 = 0
\end{equation}
so that
\begin{equation}
\epsilon = \delta\epsilon,\quad \theta = \delta\theta,\quad |\epsilon|, |\theta| \ll 1.
\end{equation}
We study only the case without SOC for simplicity. The system is illustrated in \cref{fig:hypermagic}\textbf{(a)}, \textbf{(b)}.

As in the BM model for TBG, the two valleys $\bK_\pm$ and $-\bK_{\pm}$ are nearly decoupled and are related by time-reversal. In a convenient real space basis, the Hamiltonian for the $\bK_\pm$ valley takes the form
\begin{equation}\label{eq:opposite-velocity-hamiltonian}
H = \begin{pmatrix}
S_+(\br) - i\hbar v_+ \bsigma_M \cdot \nabla & T(\br)\\
T^\dagger(\br) & S_-(\br) +i\hbar v_- \bsigma_M \cdot \nabla
\end{pmatrix}
\end{equation}
where $T(\br)$ and $S_l(\br)$ are interlayer and intralayer moir\'e potentials, $l \in \{+, -\}$ indicates layer, $v_l$ is the Dirac cone Fermi velocity for layer $l$, $\bsigma_M = \sigma_x \bhatx_M + \sigma_y \bhaty_M$ is a vector of Pauli matrices, and $\bhatx_M, \bhaty_M$ are axis unit vectors for the moir\'e coordinate system in \cref{fig:hypermagic}\textbf{(b)} and \cref{fig:sqrt3-diagram}\textbf{(e)}. As explained in \cref{app:opposite-velocity-model}, we have chosen a basis in which the sign for the bottom layer Dirac cone is negated.

To leading order, the moir\'e potentials can be expanded as
\begin{equation}
T(\br) = \sum_{j=1}^3 T_{\bq_j} e^{i\br \cdot \bq_j},\quad S_l(\br) = S_{l,\bzero}
\end{equation}
for $2\times 2$ complex matrices $T_\bq$ and $S_{l,\bq}$. The $\bq_j$ vectors are defined by \cref{eq:define-q_j} with
\begin{equation}\label{eq:opposite-velocity-K_M}
\bK_M=(e^{-\epsilon}R_{\theta}-1)\bK_+,\quad |\bK_M|\approx \sqrt{\epsilon^2+\theta^2}|\bK_+|.
\end{equation}
The $\bq_j$ vectors, moir\'e coordinate system, moir\'e BZ, and high symmetry moir\'e quasimomenta are illustrated in \cref{fig:hypermagic}\textbf{(b)} and \cref{fig:sqrt3-diagram}\textbf{(e)}.

When $\theta = 0$, the moir\'e model in \cref{eq:opposite-velocity-hamiltonian} inherits all valley preserving symmetries from the spinless coreps of $P6mm1'$ on the two layers. Since $C_{2z}$ (rotation by $\pi$ about $\bhatz$) and $\mathcal{T}$ reverse valley, the moir\'e model carries a spinless corep of the magnetic space group $P6'm'm$ (No. 183.187 in the BNS setting \cite{Gallego2012}) which is generated by Bravais lattice translations, $C_{3z}$ (rotation by $2\pi/3$ about $\bhatz$), $C_{2z}\mathcal{T}$, and $M_y$. In this case, the symmetries imply
\begin{equation}
\begin{split}
T_{\bq_j} &= w_1 \bsigma \cdot \bhatn_{\zeta_j + \phi_1} + iw_0 \sigma_z\\
S_{l,\bzero} &= \left(E_F + \frac{lE_\Delta}{2} \right) \sigma_0
\end{split}
\end{equation}
for real parameters $w_0$, $w_1$, $E_F$, and $E_\Delta$. Here, $\bhatn_\phi = R_\phi \bhatx$,  $\bsigma = \sigma_x \bhatx + \sigma_y \bhaty$, and
\begin{equation}
\phi_1 = \arg\left(1-e^{-\epsilon - i\theta}\right)\approx \arg\left(\epsilon + i\theta\right).
\end{equation}
Without loss of generality, we set $E_F = 0$. When $\theta$ is nonzero but small, the $M_y$ symmetry is weakly broken. The moir\'e potentials may then gain small perturbations on the order of $\theta$, but these are typically negligible. For simplicity, we assume $\theta = 0$ and $\epsilon \neq 0$ hereafter, in which case $\phi_1 \in \{0, \pi\}$.

\begin{figure}[h]
	\centering
	\includegraphics{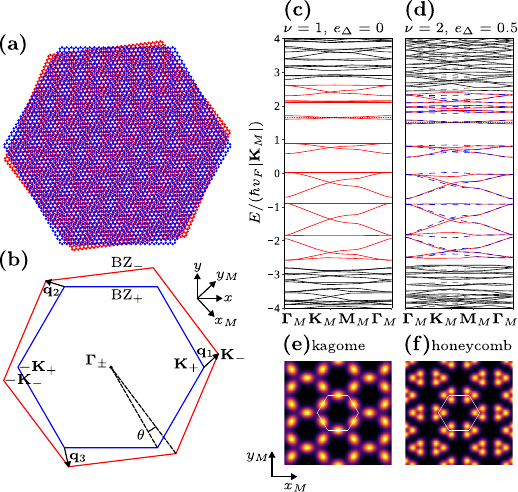}
	\caption{Illustration of the opposite-velocity model. \textbf{(a)} A moir\'e pattern for two honeycomb lattices with $\epsilon = -0.15$ and $\theta = 7^{\circ}$. \textbf{(b)} A momentum space diagram illustrating the $\bq_j$ moir\'e vectors, the top layer coordinate system $(x,y)$, and the moir\'e coordinate system $(x_M,y_M)$. The top (bottom) layer BZ is shown in blue (red). \textbf{(c)}, \textbf{(d)} Moir\'e band structures with $w_0/|w_1| = 0.8$, $\phi_1\in\{0,\pi\}$, $w_1\cos\phi_1 > 0$, and $1/\alpha = 0.5$, with $\nu$ and $e_\Delta$ given above each panel. The black dotted lines indicate charge neutrality. The parameters for $\textbf{(c)}$ realize the hypermagic model and have an emergent $C_{2z}$ symmetry. In \textbf{(d)}, the solid lines show bands along $\bGamma_M \to \bK_M \to \bM_M \to \bGamma_M$ while the dashed lines shown bands along $\bGamma_M \to -\bK_M \to -\bM_M \to \bGamma_M$. These bands differ because the emergent $C_{2z}$ symmetry is absent. \textbf{(e)} The real space charge density distribution from fully filling only bands $-4$ to $-2$ in \textbf{(c)}, which forms a kagome lattice pattern. \textbf{(f)} The real space charge density distribution from fully filling only bands $-1$ and $1$ in \textbf{(c)}, which forms a honeycomb lattice pattern. Here, the $n$th conduction (valence) band has index $n$ ($-n$). See \cref{appfig:hypermagic} for charge density distributions for the other red bands in \textbf{(c)} as well as EBR decompositions for the red bands in \textbf{(c)} and the red and blue bands in \textbf{(d)}.}
	\label{fig:hypermagic}
\end{figure}

We refer to this model in the regime $v_-v_+ < 0$ as the \emph{opposite-velocity model}. If both layers are graphene-like honeycomb lattices, this requires their inplane nearest neighbor hoppings to have opposite signs\footnote{Although the sign of the Dirac Fermi velocity is basis dependent, it is well defined in the presence of a fixed form for the symmetry operators.}. We parameterize the model by the dimensionless quantities
\begin{equation}
\nu = -\frac{v_+}{v_-}, \quad \alpha = \frac{|w_1|}{\hbar v_F |\bK_M|}, \quad e_\Delta = \frac{E_\Delta}{\hbar v_F |\bK_M|},
\end{equation}
$w_0/|w_1|$, and the sign of $w_1 \cos\phi_1$, where $v_F = \sqrt{|v_+ v_-|}$. In particular, when $\nu = 1$ and $e_\Delta = 0$ (i.e., the two layers have opposite Dirac Fermi velocities and equal Fermi energies), the Hamiltonian gains an emergent effective $C_{2z}$ symmetry that sends $\br \mapsto -\br$ and interchanges the layers, and the space group is enriched to $P6mm1'$. In this case, the Hamiltonian in \cref{eq:opposite-velocity-hamiltonian} has the same form as the hypermagic model given in Eqs. (71), (74), and (75) of Ref. \cite{Scheer2022} with $|\phi_0| = \pi/2$, and the $T_{\bq_j}$,  $S_{l,\bzero}$ matrices have the same form as the $T_{s,\bq_j}$, $S_{s,\eta,\bzero}$ matrices in \cref{tbl:moire-model-form}\textbf{(a)} with $\tilde{w}_{m,\mu}=0$. This happens because the coreps of $P6mm1'$ in the coupled-valley graphene model without SOC in \cref{eq:coupled-valley-hamiltonian} and the hypermagic model are the same.

\cref{fig:hypermagic}\textbf{(c)} shows an example band structure with $\nu = 1$ and $e_\Delta = 0$, which simultaneously exhibits at least eight flat bands. All the low energy bands in red are compatible with EBRs of $P6mm1'$ supported on honeycomb or kagome lattices. The EBR for each group of bands is shown in \cref{appfig:hypermagic}\textbf{(f)}. In particular, four (two) groups of connected bands correspond to kagome (honeycomb) lattice flat band models (see \cref{app:kagome-honeycomb-models}), which can be observed in the real space charge density distributions in \cref{fig:hypermagic}\textbf{(e)}, \textbf{(f)} and \cref{appfig:hypermagic}\textbf{(a)}-\textbf{(e)}.

\cref{fig:hypermagic}\textbf{(d)} shows an example band structure identical to that in \cref{fig:hypermagic}\textbf{(c)}, except that $\nu = 2$ and $e_\Delta = 0.5$ so the emergent $C_{2z}$ symmetry is absent. Interestingly, all of the flat bands in \cref{fig:hypermagic}\textbf{(a)} remain quite flat in \cref{fig:hypermagic}\textbf{(b)}. The main change is that the EBR of $P6mm1'$ corresponding to the honeycomb lattice flat band model becomes a composite band representation of $P6'm'm$, and one of the two groups of bands with this EBR splits into two disconnected groups. The groups of bands supported on kagome lattices remain connected. The EBR decomposition for each group of bands is shown in \cref{appfig:hypermagic}\textbf{(g)}.

\section{Discussion}
A considerable number of 2D materials (e.g., germanene and CdS) are known to have lattice constant approximately $\sqrt{3}$ times that of graphene \cite{Wallbank2013,Lu2022}. These materials can be considered candidate substrates for the coupled-valley graphene model in \cref{eq:coupled-valley-hamiltonian}. In addition to the appropriate lattice constant, a substrate material must have a 2D spinless coirrep at the $\bGamma$ point near the graphene Fermi energy in order to realize the phases in \cref{fig:sqrt3-phases}\textbf{(b)}. In order to find such materials, \textit{ab initio} studies are needed.

We note that for a given substrate, the moir\'e potentials may be tuned with pressure (which modulates interlayer hopping) and out-of-plane displacement field (which modulates the relative energies of states in the two layers). Additionally, if the substrate material is placed both above and below the graphene layer, the moir\'e potentials will be enhanced by a factor of two. This is similar to the case of symmetric twisted trilayer graphene \cite{Khalaf2019,Calugaru2021}.

In comparison to $\bGamma$-valley models which also have kagome or honeycomb moir\'e flat bands \cite{Angeli2021,Xian2021,Liu2022,Wang2022}, the coupled-valley graphene model has the advantage that the electrons are localized within graphene, which is a clean and theoretically well understood material. Additionally, our model shows a rich phase diagram of low energy bands including eigenvalue fragile phases in addition to flat bands.

It is less clear how to realize the opposite-velocity model in \cref{eq:opposite-velocity-hamiltonian}. However, this model can host many kagome or honeycomb flat bands simultaneously, so a realization could provide a variety of interacting phases within a single sample. It is worth noting additionally that both models we propose could potentially be realized with metamaterials \cite{Wu2018}.

\begin{acknowledgments}
We thank Jonah Herzog-Arbeitman, Yuanfeng Xu, Aaron Dunbrack, and Jennifer Cano for valuable discussions. This work is supported by the Alfred P. Sloan Foundation, the National Science Foundation through Princeton University’s Materials Research Science and Engineering Center DMR-2011750, and the National Science Foundation under award DMR-2141966. Additional support is provided by the Gordon and Betty Moore Foundation through Grant GBMF8685 towards the Princeton theory program.
\end{acknowledgments}

\bibliography{bibliography}
\clearpage
\appendix
\onecolumngrid

\begin{center}
    {\bf \large Appendices}    
\end{center}

\begin{figure}[h]
	\centering
	\includegraphics{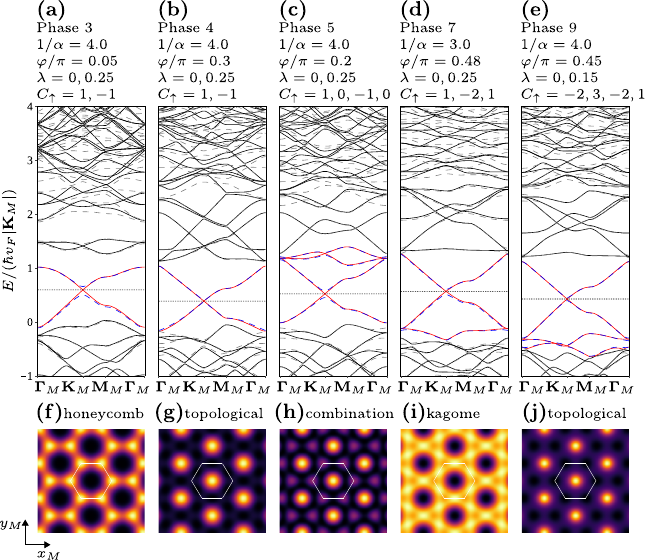}
	\caption{Continuation of \cref{fig:sqrt3-bands} for phases $3$, $4$, $5$, $7$, and $9$. See the caption of \cref{fig:sqrt3-bands} for more information.}
	\label{appfig:sqrt3-bands}
\end{figure}

\begin{figure}[h]
	\centering
	\includegraphics{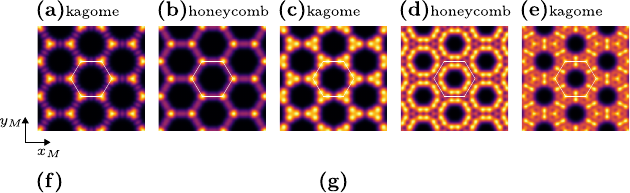}\\
	\hspace{1.2cm}
	\begin{tabular}{cc|c}
		min & max & $P6mm1'$ EBRs\\
		\hline $6$ & $8$ & $(B_1)_{3c}$\\
		$2$ & $5$ & $(E)_{2b}$\\
		$-1$ & $1$ & $(A_2)_{2b}$\\
		$-4$ & $-2$ & $(A_1)_{3c}$\\
		$-7$ & $-5$ & $(B_2)_{3c}$\\
		$-11$ & $-8$ & $(E)_{2b}$\\
		$-14$ & $-12$ & $(A_1)_{3c}$\\
		& &	
	\end{tabular}\hspace{0.9cm}
	\begin{tabular}{cc|c}
		min & max & $P6'm'm$ EBR decomps\\
		\hline $6$ & $8$ & $(A')_{3c}$\\
		$4$ & $5$ & $(^1 E)_{2b}$\\
		$2$ & $3$ & $(E)_{1a}$\\
		$-1$ & $1$ & $(A_1)_{2b}$\\
		$-4$ & $-2$ & $(A')_{3c}$\\
		$-7$ & $-5$ & $(A'')_{3c}$\\
		$-11$ & $-8$ & $(^1 E)_{2b} \oplus (E)_{1a}$\\
		$-14$ & $-12$ & $(A')_{3c}$
	\end{tabular}
	\caption{Continuation of \cref{fig:hypermagic}. \textbf{(a)}-\textbf{(e)} Real space charge density distributions for bands $-14$ to $-12$, $-11$ to $-8$, $-7$ to $-5$, $2$ to $5$, and $6$ to $8$, respectively in \cref{fig:hypermagic}\textbf{(c)}. \textbf{(f)} $P6mm1'$ EBRs for the bands shown in red in \cref{fig:hypermagic}\textbf{(c)}. \textbf{(g)} $P6'm'm$ EBR decompositions for the bands shown in red and blue in \cref{fig:hypermagic}\textbf{(d)}. The $n$th conduction (valence) band has index $n$ ($-n$). The symbol $\oplus$ indicates sum of EBRs. The full list of EBRs for $P6mm1'$ and $P6'm'm$ can be found on the Bilbao Crystallographic Server \cite{Elcoro2021,Xu2020}.}
	\label{appfig:hypermagic}
\end{figure}

\section{Notations}\label{app:notations}
Throughout this paper we consider bilayer structures formed from two crystalline materials with triangular Bravais lattices. We use $l = +$ and $l = -$ to denote the top and bottom layers, respectively. The top layer has lattice constant $\latconst$ and the bottom layer has lattice constant $e^\epsilon \latconst$ for some real number $\epsilon$. Additionally, the bottom layer is rotated counterclockwise by angle $\theta$ relative to the top layer. We denote the Bravais lattice, reciprocal lattice, primitive unit cell, and Brillouin zone of layer $l \in \{+, -\}$ by $L_l$, $P_l$, $\Omega_l$, and $\text{BZ}_l$, respectively. We use $\bhatx$, $\bhaty$, and $\bhatz$ for unit vectors in $\R^3$ and use primitive vectors
\begin{equation}\label{eq:define-a1-a2}
\ba_1 = \latconst\bhatx,\quad \ba_2 = R_{-\pi/3}\ba_1
\end{equation}
for $L_+$ and
\begin{equation}\label{eq:define-b1-b2}
\bb_1 = R_{2\pi/3}\bb_2,\quad \bb_2 = -4\pi\bhaty/(\latconst\sqrt{3})
\end{equation}
for $P_+$. We define high symmetry crystal momenta
\begin{equation}
\bGamma_+ = \bzero,\quad \bK_+ = \frac{2}{3}\bb_1 + \frac{1}{3}\bb_2,\quad \bM_+ = \frac{1}{2}\bb_1 + \frac{1}{2}\bb_2
\end{equation}
for the top layer and
\begin{equation}
\bGamma_- = \bzero,\quad \bK_- = e^{-\epsilon}R_\theta \bK_+,\quad \bM_- = e^{-\epsilon}R_\theta \bM_+
\end{equation}
for the bottom layer. We write $|S|$ for the area of a region $S \subset \R^2$ such as $\text{BZ}_l$ or $\Omega_l$. $R_\phi$ denotes rotation by angle $\phi$ about $\bhatz$ and $\mathcal{R}_\bhatn$ denotes reflection through the plane orthogonal to the vector $\bhatn \in \R^3$. If $V$, $V_1$, and $V_2$ are sets of vectors, $\bu$ is a vector, and $r$ is a real number, we define
\begin{equation}
V_1 + V_2 = \{\bv_1 + \bv_2 | \bv_1 \in V_1, \bv_2 \in V_2\},\quad \bu + V = \{\bu + \bv | \bv \in V\},\quad rV = \{r\bv | \bv \in V\}.
\end{equation}
Finally, we denote the Pauli matrices by $\sigma_x$, $\sigma_y$, and $\sigma_z$, we denote the $2\times 2$ identity matrix by $\sigma_0$, and we use the vectors of Pauli matrices $\bsigma = \sigma_x \bhatx + \sigma_y \bhaty$ and $\bsigma^* = \sigma_x \bhatx - \sigma_y \bhaty$.

\section{Commensurate configurations}\label{app:commensurate-configurations}
In the following subsections we enumerate all commensurate configurations, explain some of their properties, and classify them into four types.

\subsection{Conditions for commensuration}\label{app:commensuration-conditions}
We say that two Bravais lattices are commensurate if they share a nonzero element. If the triangular Bravais lattices $L_-$ and $L_+$ defined in \cref{app:notations} are commensurate then their intersection $L_c = L_- \cap L_+$ is another triangular Bravais lattice called the commensuration superlattice. In this section, we derive all values of $\epsilon$ and $\theta$ such that $L_-$ and $L_+$ are commensurate. Our approach is similar to that of App. D. in Ref. \cite{Scheer2022}, which covers the case in which $\epsilon = 0$.

Let $\tilde{a}$ and $\tilde{b}$ be matrices with columns $(\ba_1, \ba_2)$ and $(\bb_1, \bb_2)$, respectively. Then every element of $L_+$ takes the form $\tilde{a} \bu_+$ for some integer vector $\bu_+$ and every element of $L_-$ takes the form $e^\epsilon R_\theta \tilde{a}\bu_-$ for some integer vector $\bu_-$. It follows that $L_-$ and $L_+$ are commensurate if and only if
\begin{equation}
\bu_+ = e^\epsilon \tilde{a}^{-1} R_\theta \tilde{a} \bu_-
\end{equation}
is satisfied by some nonzero integer vectors $\bu_-$ and $\bu_+$, or equivalently if all matrix elements of $e^\epsilon \tilde{a}^{-1} R_\theta \tilde{a}$ are rational. One can compute
\begin{equation}\label{appeq:scaled-a_inv-R_theta-a}
e^\epsilon \tilde{a}^{-1} R_\theta \tilde{a} = \begin{pmatrix}
x_0 + y_0 & 2y_0\\
-2y_0 & x_0 - y_0
\end{pmatrix}\\
\end{equation}
where $e^{\epsilon + i\theta} = x_0 + y_0 i\sqrt{3}$. It follows that $L_-$ and $L_+$ are commensurate if and only if $x_0$ and $y_0$ are both rational.

Applying the same argument to the reciprocal lattices, we see that $P_-$ and $P_+$ are commensurate if and only if all matrix elements of $e^{-\epsilon} \tilde{b}^{-1} R_\theta \tilde{b}$ are rational. Since
\begin{equation}\label{appeq:scaled-b_inv-R_theta-b}
e^{-\epsilon} \tilde{b}^{-1} R_\theta \tilde{b} = e^{-2\epsilon}\begin{pmatrix}
x_0 - y_0 & 2y_0\\
-2y_0 & x_0 + y_0
\end{pmatrix}
\end{equation}
we see that $P_-$ and $P_+$ are commensurate if and only if $L_-$ and $L_+$ are commensurate.

For commensurate $L_-$ and $L_+$, we can write
\begin{equation}\label{appeq:define-mu-nu-rho}
e^{-l(\epsilon + i \theta)} = \frac{\mu_l + \nu_l i\sqrt{3}}{\rho_l}
\end{equation}
for $l \in \{+, -\}$ and integers $\mu_l$, $\nu_l$, $\rho_l$ with $\rho_l \geq 1$ and $\gcd(\mu_l, \nu_l, \rho_l) = 1$. Any commensurate configuration is equivalent up to an isometry to one with
\begin{equation}\label{appeq:epsilon-theta-constraints}
\epsilon \geq 0 \text{ and } 0 \leq \theta \leq \pi/6.
\end{equation}
\cref{apptbl:commensurate-configurations} shows the parameters for several commensurate configurations satisfying \cref{appeq:epsilon-theta-constraints}.

\begin{table}[h]
	\centering
	\begin{tabular}{c|c|c|c|c|c|c|c|c|c|c}
		$\#$ & Type & $\mu_+$ &  $\nu_+$ & $\rho_+$ & $\mu_-$ & $\nu_-$ & $\rho_-$ & $\theta$ & $N_+$ & $N_-$\\
		\hline 1 & I$+$ & $1$ & $0$ & $1$ & $1$ & $0$ & $1$ & $0.0^{\circ}$ & $1$ & $1$\\
		2 & II$+$ & $3$ & $-1$ & $6$ & $3$ & $1$ & $2$ & $30.0^{\circ}$ & $3$ & $1$\\
		3 & I$-$ & $1$ & $0$ & $2$ & $2$ & $0$ & $1$ & $0.0^{\circ}$ & $4$ & $1$\\
		4 & I$+$ & $5$ & $-1$ & $14$ & $5$ & $1$ & $2$ & $19.11^{\circ}$ & $7$ & $1$\\
		5 & II$+$ & $1$ & $0$ & $3$ & $3$ & $0$ & $1$ & $0.0^{\circ}$ & $9$ & $1$\\
		6 & II$-$ & $3$ & $-1$ & $4$ & $3$ & $1$ & $3$ & $30.0^{\circ}$ & $4$ & $3$\\
		7 & II$+$ & $3$ & $-1$ & $12$ & $3$ & $1$ & $1$ & $30.0^{\circ}$ & $12$ & $1$\\
		8 & I$-$ & $7$ & $-1$ & $26$ & $7$ & $1$ & $2$ & $13.9^{\circ}$ & $13$ & $1$\\
		9 & I$+$ & $1$ & $0$ & $4$ & $4$ & $0$ & $1$ & $0.0^{\circ}$ & $16$ & $1$\\
		10 & I$+$ & $4$ & $-1$ & $19$ & $4$ & $1$ & $1$ & $23.41^{\circ}$ & $19$ & $1$\\
		11 & II$-$ & $9$ & $-1$ & $14$ & $9$ & $1$ & $6$ & $10.89^{\circ}$ & $7$ & $3$\\
		12 & II$+$ & $9$ & $-1$ & $42$ & $9$ & $1$ & $2$ & $10.89^{\circ}$ & $21$ & $1$\\
		13 & I$-$ & $1$ & $0$ & $5$ & $5$ & $0$ & $1$ & $0.0^{\circ}$ & $25$ & $1$\\
		14 & II$+$ & $3$ & $-1$ & $18$ & $9$ & $3$ & $2$ & $30.0^{\circ}$ & $27$ & $1$\\
		15 & I$-$ & $5$ & $-1$ & $7$ & $5$ & $1$ & $4$ & $19.11^{\circ}$ & $7$ & $4$\\
		16 & I$-$ & $5$ & $-1$ & $28$ & $5$ & $1$ & $1$ & $19.11^{\circ}$ & $28$ & $1$\\
		17 & I$+$ & $11$ & $-1$ & $62$ & $11$ & $1$ & $2$ & $8.948^{\circ}$ & $31$ & $1$\\
		18 & II$+$ & $2$ & $0$ & $3$ & $3$ & $0$ & $2$ & $0.0^{\circ}$ & $9$ & $4$\\
		19 & II$+$ & $1$ & $0$ & $6$ & $6$ & $0$ & $1$ & $0.0^{\circ}$ & $36$ & $1$\\
		20 & I$+$ & $11$ & $-3$ & $74$ & $11$ & $3$ & $2$ & $25.28^{\circ}$ & $37$ & $1$\\
		21 & II$-$ & $6$ & $-1$ & $13$ & $6$ & $1$ & $3$ & $16.1^{\circ}$ & $13$ & $3$\\
		22 & II$+$ & $6$ & $-1$ & $39$ & $6$ & $1$ & $1$ & $16.1^{\circ}$ & $39$ & $1$\\
		23 & I$-$ & $13$ & $-1$ & $86$ & $13$ & $1$ & $2$ & $7.589^{\circ}$ & $43$ & $1$\\
		24 & II$-$ & $3$ & $-1$ & $8$ & $6$ & $2$ & $3$ & $30.0^{\circ}$ & $16$ & $3$\\
		25 & II$+$ & $3$ & $-1$ & $24$ & $6$ & $2$ & $1$ & $30.0^{\circ}$ & $48$ & $1$\\
		26 & I$+$ & $1$ & $0$ & $7$ & $7$ & $0$ & $1$ & $0.0^{\circ}$ & $49$ & $1$\\
		27 & I$-$ & $13$ & $-3$ & $14$ & $13$ & $3$ & $14$ & $21.79^{\circ}$ & $7$ & $7$\\
		28 & I$-$ & $13$ & $-3$ & $98$ & $13$ & $3$ & $2$ & $21.79^{\circ}$ & $49$ & $1$
	\end{tabular}
	\caption{Parameters for all commensurate configurations satisfying \cref{appeq:epsilon-theta-constraints} and $N_-N_+ \leq 49$ in increasing order of $N_-N_+$. The parameters $\mu_+$, $\nu_+$, $\rho_+$, $\mu_-$, $\nu_-$, $\rho_-$, and $\theta$ are defined in \cref{app:commensuration-conditions} while $N_+$ and $N_-$ are defined by \cref{appeq:define-N_l}. By \cref{appeq:ratio-Nplus-Nminus}, $e^\epsilon = \sqrt{N_+/N_-}$. The four types of commensurate configuration (I$+$, I$-$, II$+$, and II$-$) are described in \cref{app:equivalence-class-K_l}. The values shown for $\theta$ are rounded to four significant figures.}
	\label{apptbl:commensurate-configurations}
\end{table}

\subsection{Primitive vectors of $L_c$}\label{app:primitive-vectors-L_c}
We will now derive the primitive vectors of $L_c$ when $L_-$ and $L_+$ are commensurate. To do so, we first compute the primitive vectors $\bu_l^1$ and $\bu_l^2$ of the Bravais lattice consisting of all integer vectors $\bu_l$ such that
\begin{equation}\label{appeq:commensurate-lattice-condition}
e^{-l\epsilon}\tilde{a}^{-1}R_{-l\theta}\tilde{a}\bu_l \in \Z^2.
\end{equation}
The primitive vectors of $L_c$ will then be $\tilde{a} \bu_+^1$ and $\tilde{a} \bu_+^2$, or alternatively $e^\epsilon R_\theta \tilde{a} \bu_-^1$ and $e^\epsilon R_\theta \tilde{a} \bu_-^2$.

By \cref{appeq:scaled-a_inv-R_theta-a,appeq:define-mu-nu-rho}, we have
\begin{equation}
e^{-l\epsilon}\tilde{a}^{-1} R_{-l\theta} \tilde{a} = \frac{1}{\rho_l}\begin{pmatrix}
\mu_l+\nu_l & 2\nu_l\\
-2\nu_l & \mu_l-\nu_l
\end{pmatrix}.
\end{equation}
Taking $\bu_l = x\bhatx + y\bhaty$, \cref{appeq:commensurate-lattice-condition} becomes a congruence
\begin{equation}\label{appeq:commensurate-congruence-1}
\begin{pmatrix}
\mu_l+\nu_l & 2\nu_l\\
-2\nu_l & \mu_l-\nu_l
\end{pmatrix}
\begin{pmatrix}
x \\ y
\end{pmatrix} \equiv \begin{pmatrix}
0 \\ 0
\end{pmatrix}
\pmod{\rho_l}.
\end{equation}
Multiplying by the adjugate gives $\rho_l|(\mu_l^2 + 3\nu_l^2)x$ and $\rho_l|(\mu_l^2+3\nu_l^2)y$. Defining $d_l = \gcd(\rho_l, \mu_l^2 + 3\nu_l^2)$, we then have $(\rho_l/d_l) | x$ and $(\rho_l/d_l) | y$. \cref{appeq:commensurate-congruence-1} is then equivalent to
\begin{equation}\label{appeq:commensurate-congruence-2}
\begin{pmatrix}
\mu_l+\nu_l & 2\nu_l\\
-2\nu_l & \mu_l-\nu_l
\end{pmatrix}
\begin{pmatrix}
x' \\ y'
\end{pmatrix} \equiv \begin{pmatrix}
0 \\ 0
\end{pmatrix}
\pmod{d_l}
\end{equation}
where $x = (\rho_l/d_l)x'$ and $y = (\rho_l/d_l)y'$. Since $d_l | (\mu_l^2 + 3\nu_l^2)$ and $d_l | \rho_l$, any common prime factor of $d_l$ and $\nu_l$ also divides $\mu_l$ and $\rho_l$, which is not possible. This implies $\gcd(d_l, \nu_l) = 1$, and we choose $\nu_l^{-1}$ to be an inverse of $\nu_l$ modulo $d_l$. We now consider several cases.
\begin{enumerate}
\item $d_l \equiv 1 \pmod{2}$. In this case, either $\rho_l$ is odd or $\rho_l$ is even and $\mu_l + \nu_l$ is odd. Since
\begin{equation}
(\mu_l-\nu_l)(\mu_l+\nu_l) + 4\nu_l^2 \equiv 0 \pmod{d_l},
\end{equation}
we have $\gcd(d_l, \mu_l-\nu_l) = \gcd(d_l, \mu_l+\nu_l) = \gcd(d_l, 2\nu_l) = 1$. It follows that the two congruences in \cref{appeq:commensurate-congruence-2} are redundant, so we only need to solve
\begin{equation}
(\mu_l+\nu_l)x' + 2\nu_l y' \equiv 0 \pmod{d_l}.
\end{equation}
We solve this as $y' \equiv -2^{-1}\nu_l^{-1}(\mu_l+\nu_l)x' \pmod{d_l}$ where $2^{-1}$ is an inverse of $2$ modulo $d_l$. The two primitive vectors for \cref{appeq:commensurate-lattice-condition} are then
\begin{equation}
\begin{split}
\bu_l^1 &= (\rho_l/d_l)(\bhatx - 2^{-1}\nu_l^{-1}(\mu_l+\nu_l)\bhaty)\\
\bu_l^2 &= \rho_l\bhaty.
\end{split}
\end{equation}

\item $d_l \equiv 2 \pmod{4}$. In this case, $\rho_l \equiv 2 \pmod{4}$ and $\mu_l$ and $\nu_l$ are both odd. \cref{appeq:commensurate-congruence-2} simplifies to
\begin{equation}\label{appeq:commensurate-congruence-3}
\begin{pmatrix}
(\mu_l+\nu_l)/2 & \nu_l\\
-\nu_l & (\mu_l-\nu_l)/2
\end{pmatrix}
\begin{pmatrix}
x' \\ y'
\end{pmatrix} \equiv \begin{pmatrix}
0 \\ 0
\end{pmatrix}
\pmod{d_l/2}.
\end{equation}
Since
\begin{equation}
(\mu_l-\nu_l)(\mu_l+\nu_l)/2 + 2\nu_l^2 \equiv 0\pmod{d_l/2},
\end{equation}
we have $\gcd(d_l/2, (\mu_l-\nu_l)/2) = \gcd(d_l/2, (\mu_l+\nu_l)/2) = \gcd(d_l/2, \nu_l) = 1$.
It follows that the two congruences in \cref{appeq:commensurate-congruence-3} are redundant, so we only need to solve
\begin{equation}
(\mu_l + \nu_l)x'/2 + \nu_l y' \equiv 0 \pmod{d_l/2}.
\end{equation}
We solve this as $y' = -\nu_l^{-1}(\mu_l + \nu_l)x'/2 \pmod{d_l/2}$. The two primitive vectors for \cref{appeq:commensurate-lattice-condition} are then
\begin{equation}
\begin{split}
\bu_l^1 &= (\rho_l/d_l)(\bhatx - \nu_l^{-1}(\mu_l+\nu_l)\bhaty/2)\\
\bu_l^2 &= (\rho_l/2)\bhaty.
\end{split}
\end{equation}

\item $d_l \equiv 4 \pmod{8}$. In this case, $\rho_l \equiv 0 \pmod{4}$ and $\mu_l$ and $\nu_l$ are both odd. \cref{appeq:commensurate-congruence-3} holds in this case as well. Since $d_l/2$ is even and $(\mu_l + \nu_l)/2 + (\mu_l - \nu_l)/2 \equiv 1 \pmod{2}$, \cref{appeq:commensurate-congruence-3} implies that $x'$ and $y'$ are both even. \cref{appeq:commensurate-congruence-3} then further simplifies to
\begin{equation}\label{appeq:commensurate-congruence-4}
\begin{pmatrix}
(\mu_l+\nu_l)/2 & \nu_l\\
-\nu_l & (\mu_l-\nu_l)/2
\end{pmatrix}
\begin{pmatrix}
x'' \\ y''
\end{pmatrix} \equiv \begin{pmatrix}
0 \\ 0
\end{pmatrix}
\pmod{d_l/4}
\end{equation}
where $x' = 2x''$ and $y' = 2y''$. Since
\begin{equation}
(\mu_l-\nu_l)(\mu_l+\nu_l)/4 + \nu_l^2 \equiv 0 \pmod{d_l/4}
\end{equation}
we have $\gcd(d_l/4, (\mu_l -\nu_l)/2) = \gcd(d_l/4, (\mu_l + \nu_l)/2) = \gcd(d_l/4, \nu_l) = 1$. It follows that the two congruences in \cref{appeq:commensurate-congruence-4} are redundant and we only need to solve
\begin{equation}
(\mu_l + \nu_l)x''/2 + \nu_l y'' \equiv 0 \pmod{d_l/4}.
\end{equation}
We solve this as $y'' = -\nu_l^{-1}(\mu_l + \nu_l)x''/2 \pmod{d_l/4}$. The two primitive vectors are then
\begin{equation}
\begin{split}
\bu_l^1 &= 2(\rho_l/d_l)(\bhatx - \nu_l^{-1}(\mu_l + \nu_l)\bhaty/2)\\
\bu_l^2 &= (\rho_l/2)\bhaty.
\end{split}
\end{equation}
\end{enumerate}
Note that $\mu_l^2 + 3\nu_l^2 \not\equiv 0 \pmod{8}$ if at least one of $\mu_l$ and $\nu_l$ is odd, so these are all the possible cases.

To summarize, $\bu_l^1$ and $\bu_l^2$ can be written in the form
\begin{equation}
\begin{split}
\bu_l^1 &= f_l(\rho_l/d_l)(\bhatx - h_l\bhaty)\\
\bu_l^2 &= g_l(\rho_l/2)\bhaty
\end{split}
\end{equation}
where
\begin{equation}
\begin{split}
f_l &= \begin{cases}
1 & \text{when } 4 \nmid d_l\\
2 & \text{when } 4 | d_l
\end{cases}\\
g_l &= \begin{cases}
1 & \text{when } 2 | d_l\\
2 & \text{when } 2 \nmid d_l
\end{cases}
\end{split}
\end{equation}
and where $h_l$ satisfies
\begin{equation}\label{appeq:h_l-congruence}
2\nu_l h_l \equiv \mu_l + \nu_l \pmod{d_l}.
\end{equation}

Since $\tilde{a} \bu_+^1$ and $\tilde{a} \bu_+^2$ are primitive vectors for $L_c$, the area of the primitive unit cell $\Omega_c$ of $L_c$ is
\begin{equation}
|\Omega_c| = \frac{f_+ g_+ \rho_+^2}{2d_+} |\Omega_+|.
\end{equation}
Similarly, since $e^\epsilon R_\theta \tilde{a} \bu_-^1$ and $e^\epsilon R_\theta \tilde{a} \bu_-^2$ are primitive vectors for $L_c$, we have
\begin{equation}
|\Omega_c| = \frac{f_- g_- \rho_-^2}{2d_-} |\Omega_-|.
\end{equation}
It follows that $\Omega_c$ contains $N_l$ elements of $L_l$ where
\begin{equation}\label{appeq:define-N_l}
N_l = \frac{|\Omega_c|}{|\Omega_l|} = \frac{f_l g_l \rho_l^2}{2d_l}
\end{equation}
is a positive integer. Additionally,
\begin{equation}\label{appeq:ratio-Nplus-Nminus}
\frac{N_+}{N_-} = \frac{|\Omega_-|}{|\Omega_+|} = \frac{|\text{BZ}_+|}{|\text{BZ}_-|} = e^{2\epsilon}.
\end{equation}

\subsection{Primitive vectors of $P_- \cap P_+$}\label{app:primitive-vectors-Pminus-Pplus}
We will now derive the primitive vectors of $P_- \cap P_+$ when $P_-$ and $P_+$ are commensurate. As in \cref{app:primitive-vectors-L_c} we first compute the primitive vectors $\bv_l^1$ and $\bv_l^2$ of the Bravais lattice consisting of all integer vectors $\bv_l$ such that
\begin{equation}
e^{l\epsilon}\tilde{b}^{-1}R_{-l\theta}\tilde{b}\bv_l \in \Z^2.
\end{equation}
The primitive vectors of $P_- \cap P_+$ will then be $\tilde{b} \bv_+^1$ and $\tilde{b} \bv_+^2$, or alternatively $e^{-\epsilon} R_\theta \tilde{b} \bv_-^1$ and $e^{-\epsilon} R_\theta \tilde{b} \bv_-^2$.

Since
\begin{equation}
e^{l(\epsilon -i\theta)} = \frac{\mu_{-l} - \nu_{-l}i\sqrt{3}}{\rho_{-l}},
\end{equation}
\cref{appeq:scaled-b_inv-R_theta-b} implies
\begin{equation}\label{appeq:scaled-b_inv-R_theta-b-with-l}
e^{l\epsilon} \tilde{b}^{-1} R_{-l\theta} \tilde{b} = \frac{1}{\rho_{-l}}\begin{pmatrix}
\mu_{-l}+\nu_{-l} & -2\nu_{-l}\\
2\nu_{-l} & \mu_{-l}-\nu_{-l}
\end{pmatrix}.
\end{equation}
By an argument similar to that in \cref{app:primitive-vectors-L_c}, we find
\begin{equation}
\begin{split}
\bv_l^1 &= \mathcal{R}_{\bhaty} \bu_{-l}^1\\
\bv_l^2 &= \bu_{-l}^2.
\end{split}
\end{equation}
Furthermore, each primitive unit cell of $P_- \cap P_+$ contains $N_{-l}$ elements of $P_l$.

\subsection{Commensuration reciprocal lattice}
Suppose that $L_-$ and $L_+$ are commensurate and let $P_c$ be the reciprocal lattice of $L_c$. Since $L_c$ is a triangular Bravais lattice, $P_c$ is as well. Clearly, $P_-$ and $P_+$ are both subsets of $P_c$. It follows that $P_c$ also contains the Bravais lattice $P_- + P_+$. However, since the reciprocal lattice of $P_- + P_+$ is a subset of $L_c$, it follows that $P_c = P_- + P_+$. We call $P_c$ the commensuration reciprocal lattice and we denote the Brillouin zone of $P_c$ by $\text{BZ}_c$.

We can write the primitive vectors for $P_c$ corresponding to $\tilde{a}\bu_+^1$ and $\tilde{a} \bu_+^2$ in the form $\tilde{b} \bu_{c,+}^1$ and $\tilde{b} \bu_{c,+}^2$. Similarly, we can write the primitive vectors for $P_c$ corresponding to $e^\epsilon R_\theta \tilde{a} \bu_-^1$ and $e^\epsilon R_\theta \tilde{a} \bu_-^2$ in the form $e^{-\epsilon}R_\theta \tilde{b} \bu_{c,-}^1$ and $e^{-\epsilon} R_\theta \tilde{b} \bu_{c,-}^2$. A simple calculation shows
\begin{equation}
\begin{split}
\bu_{c,l}^1 &= \frac{d_l}{f_l \rho_l} \bhatx\\
\bu_{c,l}^2 &= \frac{2(h_l \bhatx + \bhaty)}{g_l \rho_l}.
\end{split}
\end{equation}
Additionally,
\begin{equation}\label{appeq:commensurate-BZ-area}
|\text{BZ}_c| = |\text{BZ}_l|/N_l
\end{equation}
for $l \in \{+, -\}$.

\subsection{Equivalence class of $\bK_l$ modulo $P_c$}\label{app:equivalence-class-K_l}
Note that the equivalence class of $\bK_l$ modulo $P_c$ is invariant under rotations by $2\pi/3$ about $\bhatz$. It follows that $\bK_l$ is either in $P_c$ or is equivalent to one of the two distinct corners of $\text{BZ}_c$. We first find the conditions under which $\bK_l \in P_c$.

Since $\bK_+ = \tilde{b}(2\bhatx + \bhaty)/3$ and $\bK_- = e^{-\epsilon} R_\theta \tilde{b}(2\bhatx + \bhaty)/3$, it follows that $\bK_l \in P_c$ if and only if
\begin{equation}
\frac{2}{3}\bhatx + \frac{1}{3}\bhaty = n_1 \bu_{c,l}^1 + n_2 \bu_{c,l}^2
\end{equation}
for some integers $n_1$ and $n_2$. This equation can be solved for rational $n_1$ and $n_2$ as
\begin{equation}
\begin{split}
n_1 &= \frac{f_l \rho_l}{3d_l} (2 - h_l)\\
n_2 &= \frac{g_l \rho_l}{6}.
\end{split}
\end{equation}
Since $g_l \rho_l$ is always even, $n_2 \in \Z$ is equivalent to $3 | \rho_l$. Now suppose $3 | \rho_l$ and we will show that $n_1 \in \Z$. If $3 \nmid d_l$ then $(3d_l) | \rho_l$ so that $n_1 \in \Z$. On the other hand, if $3 | d_l$ then we must have $3 | \mu_l$ and $3 \nmid \nu_l$. If we reduce \cref{appeq:h_l-congruence} modulo $3$, we find $2\nu_l h_l \equiv \nu_l \pmod{3}$ or $h_l \equiv 2 \pmod{3}$. In this case we again have $n_1 \in \Z$. We conclude $\bK_l \in P_c$ if and only if $3 | \rho_l$.

Next, we will show that at most one of $\rho_-$ and $\rho_+$ is divisible by $3$. We have
\begin{equation}\label{appeq:mu-nu-rho-plus-minus}
\begin{split}
\frac{\mu_l + \nu_l i\sqrt{3}}{\rho_l} &= \left(\frac{\mu_{-l} + \nu_{-l} i\sqrt{3}}{\rho_{-l}}\right)^{-1}\\
&= \frac{\rho_{-l} \mu_{-l} -\rho_{-l} \nu_{-l} i\sqrt{3}}{\mu_{-l}^2 + 3\nu_{-l}^2}
\end{split}
\end{equation}
and suppose that $3 | \rho_{-l}$. If $3 \nmid \mu_{-l}$ then $3 \nmid (\mu_{-l}^2 + 3\nu_{-l}^2)$ and so $3 \nmid \rho_l$. On the other hand if $3 | \mu_{-l}$ then $3 \nmid \nu_{-l}$ so that $9 \nmid (\mu_{-l}^2 + 3\nu_{-l}^2)$. The single factor of $3$ in $\mu_{-l}^2 + 3\nu_{-l}^2$ is then canceled by that in $\rho_{-l}$ so that again $3 \nmid \rho_l$.

Finally, we consider the case in which $3 \nmid \rho_- \rho_+$ so that $\bK_-, \bK_+ \not\in P_c$. Since $\bK_-$ and $\bK_+$ are both equivalent to corners of $\text{BZ}_c$, we must have $\bK_- -t \bK_+ \in P_c$ for some $t \in \{+, -\}$ which we will now find. Using \cref{appeq:scaled-b_inv-R_theta-b-with-l}, we can write
\begin{equation}
\bK_- = \tilde{b}(e^{-\epsilon} \tilde{b}^{-1} R_\theta \tilde{b})(2\bhatx + \bhaty)/3 = \tilde{b}(2\mu_+ \bhatx + (\mu_+ + 3\nu_+)\bhaty)/(3\rho_+).
\end{equation}
It follows that
\begin{equation}
\frac{2(\mu_+ - t\rho_+)}{3\rho_+}\bhatx + \frac{\mu_+ + 3\nu_+ - t\rho_+}{3\rho_+}\bhaty = n_1 \bu_{c,+}^1 + n_2 \bu_{c,+}^2
\end{equation}
for some integers $n_1$ and $n_2$. Solving for $n_2$ gives
\begin{equation}
n_2 = \frac{g_+}{6}(\mu_+ + 3\nu_+ - t \rho_+).
\end{equation}
In particular, this implies $3 | (\mu_+ - t\rho_+)$ or equivalently
\begin{equation}\label{appeq:determine-t-mod-3}
t \equiv \mu_+ \rho_+ \pmod{3}.
\end{equation}
Note that \cref{appeq:mu-nu-rho-plus-minus} implies that $3 \nmid \mu_+$ whenever $3 \nmid \rho_- \rho_+$, so \cref{appeq:determine-t-mod-3} indeed determines a valid value for $t$. Additionally, we can see from \cref{appeq:mu-nu-rho-plus-minus} that
\begin{equation}
\mu_- \rho_- \equiv \mu_+ \rho_+ \pmod{3}
\end{equation}
whenever $3 \nmid \rho_- \rho_+$.

We summarize the results of this section by enumerating the four possible types of commensurate configuration with regard to the equivalence classes of $\bK_-$ and $\bK_+$ modulo $P_c$.
\begin{enumerate}
\item [I$+$.] $3 \nmid \rho_-\rho_+$ and $\mu_- \rho_- \equiv \mu_+ \rho_+ \equiv 1 \pmod{3}$. In this case, $\bK_-, \bK_+ \not\in P_c$ but $\bK_- - \bK_+ \in P_c$.
\item [I$-$.] $3 \nmid \rho_-\rho_+$ and $\mu_- \rho_- \equiv \mu_+ \rho_+ \equiv -1 \pmod{3}$. In this case, $\bK_-, \bK_+ \not\in P_c$ but $\bK_- + \bK_+ \in P_c$.
\item [II$+$.] $3 \nmid \rho_-$ and $3 | \rho_+$. In this case, $\bK_- \not\in P_c$ and $\bK_+ \in P_c$.
\item [II$-$.] $3 | \rho_-$ and $3 \nmid \rho_+$. In this case, $\bK_- \in P_c$ and $\bK_+ \not\in P_c$.
\end{enumerate}

\section{Moir\'e models for each type of commensurate configuration}\label{app:general-moire-models}
In this section, we derive moir\'e models for each of the four types of commensurate configuration in \cref{app:equivalence-class-K_l}. The arguments given in this section are a simple generalization of those in Ref. \cite{Scheer2022} which considered the case of TBG twisted near an arbitrary commensurate angle. See also \cref{app:coupled-valley-spinless,app:coupled-valley-spinful,app:opposite-velocity-model} for detailed derivations of the moir\'e models near the configurations $\namesqthree$ and $\namezero$.

We consider a bilayer structure consisting of two 2D crystalline materials with triangular Bravais lattices. For simplicity, we assume that the top layer is graphene, although the following arguments apply also to other similar materials. The interatomic spacing of graphene is $a_0 \approx \SI{0.142}{\nano\meter}$ and the lattice constant of graphene is $\latconst = a_0 \sqrt{3} \approx \SI{0.246}{\nano\meter}$. For simplicity, we choose the chemical potential so that we can regard the graphene Fermi energy as $0$. As in \cref{app:notations}, the bottom layer has lattice constant $e^\epsilon \latconst$ and is rotated counterclockwise by angle $\theta$ relative to the top layer. We take $\epsilon$ and $\theta$ as in \cref{eq:moire-condition} where $\epsilon_0$ and $\theta_0$ are parameters for a commensurate configuration, as described in \cref{app:commensuration-conditions}. We write $P_-^0 = e^{\delta\epsilon}R_{-\delta\theta}P_-$ and $\text{BZ}_-^0 = e^{\delta\epsilon}R_{-\delta\theta}\text{BZ}_-$ for the bottom layer reciprocal lattice and Brillouin zone in the commensurate case. The commensurate reciprocal lattice is $P_c = P_-^0 + P_+$ and \cref{appeq:commensurate-BZ-area} implies
\begin{equation}\label{appeq:commensurate-BZ-area-repeated}
|\text{BZ}_c| = |\text{BZ}_+|/N_+ = |\text{BZ}_-^0|/N_-
\end{equation}
where $\text{BZ}_c$ is the Brillouin zone for $P_c$.

We are interested in the low energy sector of the bilayer Hamiltonian that originates from the graphene Dirac cones at the $\bK_+$ and $-\bK_+$ points. We assume there is some effective Slater-Koster model that describes this low energy physics \cite{Slater1954}. For simplicity, we neglect spin degrees of freedom for now and consider them in \cref{app:general-moire-model-spin}. We use Wannier functions $\ket{\br, l, \alpha}$ for $l \in \{+, -\}$, $\br \in L_l$, and $\alpha \in \mathcal{O}_l$ as a basis for our model, where $\mathcal{O}_l$ is the set of orbitals on layer $l$. The Wannier function $\ket{\br, l, \alpha}$ is localized at position $\br + \btau_\alpha^l$ in the $xy$ plane. The most important orbitals in $\mathcal{O}_+$ are the carbon $p_z$ orbitals located on the two graphene sublattices, since these orbitals give rise to the Dirac cones \cite{CastroNeto2009}. We denote these orbitals by $A$ and $B$ and take
\begin{equation}
\btau_A^+ = a_0 \bhaty,\quad \btau_B^+ = -a_0\bhaty
\end{equation}
following the convention in \cref{app:honeycomb-one-orbital}.

The Slater-Koster Hamiltonian takes the form
\begin{equation}
\braket{\br', l', \alpha' | H_{\text{SK}} | \br, l, \alpha} = t_{l', \alpha', l, \alpha}(\br' + \btau_{\alpha'}^{l'} - \br - \btau_\alpha^l)
\end{equation}
for some complex valued functions $t_{l',\alpha',l,\alpha}$. We transform to momentum space by defining Bloch states
\begin{equation}
\ket{\bk, l, \alpha} = \frac{1}{\sqrt{|\text{BZ}_l|}} \sum_{\br \in L_l} e^{i\bk \cdot (\br + \btau_\alpha^l)} \ket{\br, l, \alpha}.
\end{equation}
The interlayer coupling can then be written
\begin{equation}\label{appeq:momentum-slater-koster}
\braket{\bk', -l, \alpha' | H_{\text{SK}} | \bk, l, \alpha} = \sum_{\bG_- \in P_-} \sum_{\bG_+ \in P_+} \frac{\hat{t}_{-l,\alpha',l,\alpha}(\bk + \bG_l)}{\sqrt{|\Omega_-||\Omega_+|}} e^{i\btau_{\alpha'}^{-l} \cdot \bG_{-l}} e^{-i\btau_\alpha^l \cdot \bG_l}\delta^2(\bk + \bG_l - \bk' - \bG_{-l})
\end{equation}
where $\hat{t}_{l',\alpha',l,\alpha}$ is the Fourier transform of $t_{l',\alpha',l,\alpha}$. Note that \cref{appeq:momentum-slater-koster} generalizes Eq. (9) in Ref. \cite{Scheer2022}.

We assume that $\hat{t}_{l',\alpha',l,\alpha}(\bQ)$ depends only on $|\bQ|$ and decreases rapidly as $|\bQ|$ grows. This implies that the magnitude of a term in \cref{appeq:momentum-slater-koster} is large when $|\bk + \bG_l|$ is small. By a simple generalization of the arguments in Secs. C and D of Ref. \cite{Scheer2022}, $H_{\text{SK}}$ generates a significant coupling between states at momentum $\eta \bK_+ + \bp$ with small $|\bp|$ in layer $+$ and momentum $\bk$ in layer $l$ if one of the following two statements holds.
\begin{enumerate}
	\item $l = +$ and $\bk = \eta\bk_0 + \bp - (e^{-\delta\epsilon}R_{\delta\theta} - 1)\eta\bQ$ for some $\bk_0 \in \bK_+ + P_c$ and
	\begin{equation}\label{appeq:Q-hopping-plus}
	\bQ \in (\bK_+ - \bk_0 + P_+) \cap P_-^0
	\end{equation}
	with small $|\bQ|$.
	
	\item $l = -$ and $\bk = e^{-\delta\epsilon} R_{\delta\theta} \eta\bk_0 + \bp - (e^{-\delta\epsilon}R_{\delta\theta} - 1)\eta\bQ$ for some $\bk_0 \in \bK_+ + P_c$ and
	\begin{equation}\label{appeq:Q-hopping-minus}
	\bQ \in (\bK_+ + P_+) \cap (\bk_0 + P_-^0)
	\end{equation}
	with small $|\bQ|$.
\end{enumerate}
We define
\begin{equation}\label{appeq:define-S-sets}
\mathcal{S}_+ = (\bK_+ + P_c) \cap \text{BZ}_+,\quad \mathcal{S}_- = (\bK_+ + P_c) \cap \text{BZ}_-^0
\end{equation}
and note that $\mathcal{S}_l$ contains $N_l$ elements by \cref{appeq:commensurate-BZ-area-repeated}. All states in layer $l$ relevant to the physics originating with the graphene Dirac cones have momentum near an element of $\mathcal{S}_l$.

Next, we will apply these results to derive the form of the moir\'e model in three cases. In each case, we will show that the Hamiltonian can be written as a direct sum of Hamiltonians of a certain canonical form. We first present this form and then proceed to the three cases.

\subsection{Two Dirac cone moir\'e Hamiltonian}\label{app:two-Dirac-cone}
The canonical form for a moir\'e Hamiltonian with two Dirac cones is
\begin{equation}\label{appeq:two-Dirac-hamiltonian}
\begin{split}
H &= \int d^2\br \ket{\br} \mathcal{H}(\br) \bra{\br}\\
\mathcal{H}(\br) &= \begin{pmatrix}
S_+(\br) - i\hbar v_+ \bsigma \cdot \nabla & T(\br)\\
T^\dagger(\br) & S_-(\br) -i\hbar v_- \bsigma \cdot \nabla
\end{pmatrix}.
\end{split}
\end{equation}
Here, $\ket{\br}$ is a four-dimensional row vector of states located at position $\br$, $T(\br)$ and $S_\pm(\br)$ are $2\times 2$ spatially varying moir\'e potentials, and $v_\pm$ are Fermi velocities for the two Dirac cones. The potentials can be expanded as
\begin{equation}\label{appeq:two-Dirac-T-S-expansion}
T(\br) = \sum_{\bq \in P_M^+} T_\bq e^{i\br \cdot \bq},\quad S_\pm(\br) = \sum_{\bq \in P_M} S_{\pm,\bq} e^{i\br \cdot \bq}
\end{equation}
for $2\times 2$ complex matrices $T_\bq$ and $S_{\pm,\bq}$ with $S_{\pm,\bq}^\dagger = S_{\pm,-\bq}$. The moir\'e reciprocal lattice $P_M$ is given by
\begin{equation}\label{appeq:define-P_M}
P_M = \{n_1 \bq_1 + n_2 \bq_2 + n_3 \bq_3 | n_1, n_2, n_3 \in \Z, n_1 + n_2 + n_3 = 0\},
\end{equation}
and
\begin{equation}\label{appeq:define-P_M^+}
P_M^+ = \bq_1 + P_M = \bq_2 + P_M = \bq_3 + P_M.
\end{equation}
The $\bq_1$, $\bq_2$, and $\bq_3$ vectors are defined by
\begin{equation}\label{appeq:two-Dirac-q_j}
\bq_j = (e^{-\delta\epsilon}R_{\delta\theta}-1)R_{\xi_0 + \zeta_j}\sqrt{N_-}\bK_+
\end{equation}
where $\zeta_j = \frac{2\pi}{3}(j-1)$, $\xi_0$ is an angle, and $N_-$ is the number of bottom layer Bravais lattice sites in each commensurate unit cell for the configuration with parameters $\epsilon_0$ and $\theta_0$ (see \cref{app:primitive-vectors-L_c}). The magnitudes of the $T_\bq$ and $S_{\pm,\bq}$ matrices typically decay rapidly with $|\bq|$ which allows us to truncate the infinite sums in \cref{appeq:two-Dirac-T-S-expansion}. We use a straightforward generalization of the method described in Appendix M of Ref. \cite{Scheer2022} to compute band structures of $H$ in the moir\'e Brillouin zone $\text{BZ}_M$, which is defined as the Wigner-Seitz unit cell of $P_M$.

\subsection{Type I$\pm$ with bottom layer Dirac cones}\label{app:I-pm-Dirac}
We first suppose that the commensurate configuration with parameters $\epsilon_0$ and $\theta_0$ is of type I$x$ for $x \in \{+, -\}$. We define $\bK_-^0 = e^{\delta\epsilon}R_{-\delta\theta}\bK_-$ so that $x\bK_-^0 \in \mathcal{S}_-$. Note that the graphene layer has a Dirac cone at $\bK_+$ at $0$ energy and a large gap around $0$ at all other momenta in $\mathcal{S}_+$. We assume that the bottom layer has a Dirac cone at $x\bK_-$ near $0$ energy and a large gap around $0$ at all other momenta in $e^{-\delta\epsilon}R_{\delta\theta}\mathcal{S}_-$. The low energy physics associated with the graphene Dirac cones is then described by an effective continuum model involving only top layer momenta near $\bK_+$ and $-\bK_+$ and bottom layer momenta near $\bK_-$ and $-\bK_-$. Furthermore, the model has two decoupled valleys, one of which contains $\bK_+$ and $x\bK_-$ and the other of which contains $-\bK_+$ and $-x\bK_-$. For simplicity, we focus on the valley containing $\bK_+$ and $x\bK_-$.

We introduce continuum states $\ket{\bp, l, \alpha}$ for $\bp \in \R^2$, $l \in \{+, -\}$, and $\alpha \in \{A, B\}$ which satisfy the normalization condition
\begin{equation}
\braket{\bp', l', \alpha' | \bp, l, \alpha} = \delta^2(\bp' - \bp)\delta_{l',l}\delta_{\alpha',\alpha}.
\end{equation}
The continuum state $\ket{\bp, +, \alpha}$ represents the Bloch state $\ket{\bK_+ + \bp, +, \alpha}$ so that the Dirac cone at $\bK_+$ takes the form
\begin{equation}
\int d^2\bp \begin{pmatrix}
\ket{\bp, +, A} & \ket{\bp, +, B}
\end{pmatrix}
(\hbar v_+ \bsigma \cdot \bp)
\begin{pmatrix}
\bra{\bp, +, A} \\ \bra{\bp, +, B}
\end{pmatrix}
\end{equation}
where $v_+$ is the graphene Fermi velocity. The continuum states $\ket{\bp, -, A}$ and $\ket{\bp, -, B}$ represent appropriate linear combinations of bottom layer Bloch states with momenta $x\bK_- + \bp$ such that the Dirac cone at $x\bK_-$ takes the form
\begin{equation}
\int d^2\bp \begin{pmatrix}
\ket{\bp, -, A} & \ket{\bp, -, B}
\end{pmatrix}
(E_- + \hbar v_- \bsigma \cdot \bp)
\begin{pmatrix}
\bra{\bp, -, A} \\ \bra{\bp, -, B}
\end{pmatrix}
\end{equation}
where $v_-$ is the bottom layer Fermi velocity and $E_-$ is a small energy offset. Note that a Dirac cone in two dimensions with any rotation angle or helicity is related to one of the form $\hbar v \bsigma \cdot \bp$ by a unitary transformation, so one can always choose $\ket{\bp, -, A}$ and $\ket{\bp, -, B}$ to satisfy this requirement.

If we group the continuum states with momentum $\bp$ into a row vector
\begin{equation}
\ket{\bp} = \begin{pmatrix}
\ket{\bp, +, A} & \ket{\bp, +, B} & \ket{\bp, -, A} & \ket{\bp, -, B}
\end{pmatrix},
\end{equation}
the effective Hamiltonian can be written in the form
\begin{equation}
\begin{split}
H &= \int d^2\bp' d^2\bp \ket{\bp'} \mathcal{H}(\bp', \bp) \bra{\bp}\\
\mathcal{H}(\bp', \bp) &= \begin{pmatrix}
\hbar v_+ \bsigma \cdot \bp & 0\\
0 & \hbar v_- \bsigma \cdot \bp
\end{pmatrix}\delta^2(\bp'-\bp)
+ \sum_{\bq \in P_M^+} \begin{pmatrix}
0 & T_\bq\\
0 & 0
\end{pmatrix}\delta^2(\bp' - \bp - \bq)\\
&+ \sum_{\bq \in P_M^+} \begin{pmatrix}
0 & 0\\
T_\bq^\dagger & 0
\end{pmatrix}\delta^2(\bp' - \bp + \bq)
+ \sum_{\bq \in P_M} \begin{pmatrix}
S_{+,\bq} & 0\\
0 & S_{-,\bq}
\end{pmatrix}\delta^2(\bp' - \bp - \bq)
\end{split}
\end{equation}
where
\begin{align}
P_M &= (e^{-\delta\epsilon}R_{\delta\theta} - 1)\left(P_+ \cap P_-^0\right)\label{appeq:define-P_M-I-pm}\\
P_M^+ &= (e^{-\delta\epsilon}R_{\delta\theta} - 1)\left((\bK_+ + P_+) \cap (x\bK_-^0 + P_-^0)\right).\label{appeq:define-P_M^+-I-pm}
\end{align}
$T_\bq$ and $S^\pm_\bq$ denote complex $2\times 2$ matrices with norms that generally decrease as $|\bq|$ increases. The values of these matrices include contributions from states with momenta near values in $\mathcal{S}_+$ or $e^{-\delta\epsilon}R_{\delta\theta}\mathcal{S}_-$ which are not explicitly included in the effective Hamiltonian.

By the results of \cref{app:primitive-vectors-Pminus-Pplus}, $P_+ \cap P_-^0$ is a triangular Bravais lattice with a unit cell $N_-$ times larger than that of $P_+$. Furthermore, $(\bK_+ + P_+)\cap(x\bK_-^0 + P_-^0)$ can be written in the form $\bv + P_+ \cap P_-^0$ for some nonzero vector $\bv$ such that $R_{2\pi/3}\bv - \bv \in P_+ \cap P_-^0$. It follows that there is some angle $\xi_0$ such that
\begin{equation}
\begin{split}
P_+ \cap P_-^0 &= R_{\xi_0}\sqrt{N}_- P_+\\
(\bK_+ + P_+) \cap (x\bK_-^0 + P_-^0) &= R_{\xi_0} \sqrt{N}_- (\bK_+ + P_+).
\end{split}
\end{equation}
We then have
\begin{equation}
\begin{split}
P_M &= (e^{-\delta\epsilon}R_{\delta\theta} - 1)R_{\xi_0} \sqrt{N_-}P_+\\
P_M^+ &= (e^{-\delta\epsilon}R_{\delta\theta} - 1)R_{\xi_0} \sqrt{N}_-(\bK_+ + P_+)
\end{split}
\end{equation}
which is equivalent to \cref{appeq:define-P_M,appeq:define-P_M^+,appeq:two-Dirac-q_j}. As an example, when $\namezero$ we have $N_- = 1$ and one can take $\xi_0 = 0$. This particular case will be discussed further in \cref{app:opposite-velocity-model}.

Finally, we introduce real space continuum states
\begin{equation}
\ket{\br, l, \alpha} = \frac{1}{2\pi} \int d^2\bp e^{-i\bp \cdot \br}\ket{\bp, l, \alpha}
\end{equation}
which satisfy the normalization condition
\begin{equation}
\braket{\br', l', \alpha' | \br, l, \alpha} = \delta^2(\br'-\br)\delta_{l',l}\delta_{\alpha',\alpha}.
\end{equation}
If we group the continuum states with position $\br$ into a row vector
\begin{equation}\label{eq:define-r-states-I-pm-Dirac}
\ket{\br} = \begin{pmatrix}
\ket{\br, +, A} & \ket{\br, +, B} & \ket{\br, -, A} & \ket{\br, -, B}
\end{pmatrix},
\end{equation}
the Hamiltonian takes the form of \cref{appeq:two-Dirac-hamiltonian} with $T(\br)$ and $S_\pm(\br)$ potentials given by \cref{appeq:two-Dirac-T-S-expansion}.

We conclude that the moir\'e model for the valley containing $\bK_+$ and $x\bK_-$ is of the form in \cref{app:two-Dirac-cone} with the two Dirac cones coming from opposite layers. By a similar argument, the Hamiltonian for the valley containing $-\bK_+$ and $-x\bK_-$ also takes this form, though the $\xi_0$ angles for the two valleys differ by $\pi$.

\subsection{Type I$\pm$ or II$-$ with gapped bottom layer}\label{app:I-pm-II-minus-gapped}
Next, suppose that the commensurate configuration with parameters $\epsilon_0$ and $\theta_0$ is of type I$+$, I$-$, or II$-$. This is equivalent to the requirement that $\bK_+ \not\in P_c$. In contrast to \cref{app:I-pm-Dirac}, we assume that the bottom layer has a large gap around $0$ at all momenta in $e^{-\delta\epsilon}R_{\delta\theta}\mathcal{S}_-$. In this case, the low energy physics associated with the graphene Dirac cones is described by an effective continuum model involving only top layer momenta near $\bK_+$ and $-\bK_+$. Furthermore, the model has two decoupled valleys, one of which contains $\bK_+$ and the other of which contains $-\bK_+$. Following a similar argument to that in \cref{app:I-pm-Dirac}, we find that the moir\'e model for both valleys can be written in the form described in \cref{app:two-Dirac-cone} with $T(\br) = 0$. In this case, the two Dirac cones correspond to the two graphene valleys, $v_+ = v_-$, and the angle $\xi_0$ is constrained only by
\begin{equation}
P_+ \cap P_-^0 = R_{\xi_0}\sqrt{N}_- P_+.
\end{equation}

\subsection{Type II$+$ with gapped bottom layer}\label{app:II-plus-gapped}
Finally, suppose that the commensurate configuration with parameters $\epsilon_0$ and $\theta_0$ is of type II$+$ so that $-\bK_+ \in \mathcal{S}_+$. As in \cref{app:I-pm-II-minus-gapped}, we assume that the bottom layer has a large gap around $0$ at all momenta in $e^{-\delta\epsilon}R_{\delta\theta}\mathcal{S}_-$. In this case, the low energy physics associated with the graphene Dirac cones is described by an effective continuum model involving only top layer momenta near $\bK_+$ and $-\bK_+$. In contrast to \cref{app:I-pm-Dirac,app:I-pm-II-minus-gapped}, the model has only one valley so there is no degree of freedom associated with valley. Following a similar argument to that in \cref{app:I-pm-Dirac}, we find that the moir\'e model can be written in the form described in \cref{app:two-Dirac-cone}. In this case, the two Dirac cones correspond to the two graphene valleys so that $v_+ = v_-$. Additionally, although \cref{appeq:define-P_M-I-pm} still holds, \cref{appeq:define-P_M^+-I-pm} is modified to
\begin{equation}
\begin{split}
P_M^+ &= (e^{-\delta\epsilon}R_{\delta\theta} - 1)\left((\bK_+ - (-\bK_+) + P_+) \cap P_-^0\right)\\
&= (e^{-\delta\epsilon}R_{\delta\theta} - 1)\left((-\bK_+ + P_+) \cap P_-^0\right).
\end{split}
\end{equation}
As a result, the angle $\xi_0$ is now constrained by
\begin{equation}
\begin{split}
P_+ \cap P_-^0 &= R_{\xi_0}\sqrt{N}_- P_+\\
(-\bK_+ + P_+) \cap P_-^0 &= R_{\xi_0} \sqrt{N}_- (\bK_+ + P_+).
\end{split}
\end{equation}
As an example, when $\namesqthree$ we have $N_- = 1$ and one can take $\xi_0 = \pi$. This particular case will be discussed further in \cref{app:coupled-valley-spinless,app:coupled-valley-spinful}.

\subsection{Including spin}\label{app:general-moire-model-spin}
So far in \cref{app:general-moire-models} we have neglected spin degrees of freedom. In each case, if both layers have negligible spin-orbit coupling, the system can be described as a direct sum of two copies of the spinless model, one for spin $\uparrow$ and the other for spin $\downarrow$. If there is significant spin-orbit coupling (but the Dirac cones remain unchanged), the two spinless models can have different parameters and can be coupled together. In this paper, we will focus on cases in which the $z$ component of spin is preserved. In that case, the two spinless models can have different parameters but will not be coupled. If the system has time-reversal symmetry, the Hamiltonians for the two spins will be related by time-reversal. This is explained in more detail in \cref{app:coupled-valley-spinful} for the case in which $\namesqthree$.

\section{Moir\'e coordinate system}\label{app:moire-coordinate-system}
In order to make the Hamiltonian in \cref{app:two-Dirac-cone} more closely resemble the BM model for small angle TBG, we will define a coordinate system $x_M$, $y_M$ in which $\bq_1$ is on the positive $y_M$ axis. Let
\begin{equation}
\xi = \arg\left(e^{-\delta\epsilon + i\delta\theta}-1\right) + \xi_0 - \pi/2
\end{equation}
and
\begin{equation}
\bhatx_M = R_\xi \bhatx,\quad \bhaty_M = R_\xi \bhaty.
\end{equation}
Note that $\bq_1 = |\bq_1| \bhaty_M$ and $\bhatx_M \times \bhaty_M = \bhatz$. We refer to the coordinate system $x_M$, $y_M$ defined by $\bhatx_M$ and $\bhaty_M$ as the moir\'e coordinate system and illustrate it in \cref{fig:sqrt3-diagram}\textbf{(d)}, \textbf{(e)}.

Note that
\begin{equation}
\bsigma = e^{-i(\xi/2)\sigma_z} \bsigma_M e^{i(\xi/2)\sigma_z}
\end{equation}
where $\bsigma_M = \sigma_x \bhatx_M + \sigma_y \bhaty_M$. Applying the unitary change of basis
\begin{equation}
\ket{\br}_\xi = \ket{\br} \sigma_0 \otimes e^{-i(\xi/2)\sigma_z},
\end{equation}
\cref{appeq:two-Dirac-hamiltonian} can be rewritten
\begin{equation}
\begin{split}
H &= \int d^2\br \ket{\br}_\xi \mathcal{H}_\xi(\br) \bra{\br}_\xi\\
\mathcal{H}_\xi(\br) &= \left(\sigma_0 \otimes e^{i(\xi/2)\sigma_z}\right) \mathcal{H}(\br) \left(\sigma_0 \otimes e^{-i(\xi/2)\sigma_z}\right)\\
&= \begin{pmatrix}
S_{\xi,+}(\br) -i\hbar v_+ \bsigma_M \cdot \nabla & T_\xi(\br)\\
T^\dagger_\xi(\br) & S_{\xi,-}(\br) -i\hbar v_- \bsigma_M \cdot \nabla
\end{pmatrix}.
\end{split}
\end{equation}
The $T_\xi(\br)$ and $S^\pm_\xi(\br)$ potentials are given by
\begin{equation}
\begin{split}
T_\xi(\br) &= e^{i(\xi/2)\sigma_z} T(\br) e^{-i(\xi/2)\sigma_z}\\
&= \sum_{\bq \in P_M^+} T_{\xi,\bq} e^{i\br \cdot \bq}\\
S_{\xi,\pm}(\br) &= e^{i(\xi/2)\sigma_z} S_\pm(\br) e^{-i(\xi/2)\sigma_z}\\
&= \sum_{\bq \in P_M} S_{\xi,\pm,\bq} e^{i\br \cdot \bq}
\end{split}
\end{equation}
with
\begin{equation}
\begin{split}
T_{\xi,\bq} &= e^{i(\xi/2)\sigma_z} T_\bq e^{-i(\xi/2)\sigma_z}\\
S_{\xi,\pm,\bq} &= e^{i(\xi/2)\sigma_z} S_{\pm,\bq} e^{-i(\xi/2)\sigma_z}.
\end{split}
\end{equation}
We have now expressed the Hamiltonian in the moir\'e coordinate system. All models and figures in the main text use this transformation.

\section{Coupled-valley graphene model without spin}\label{app:coupled-valley-spinless}
We follow the construction of \cref{app:general-moire-models} for a bilayer structure in which the top layer is graphene and the bottom layer is some other 2D crystalline material with a triangular Bravais lattice. In this section, we focus on moir\'e patterns near the second commensurate configuration in \cref{apptbl:commensurate-configurations} which is type II$+$ and has $\namesqthree$. For this configuration, we have $N_+ = 3$ and $N_- = 1$, so that the sets $\mathcal{S}_\pm$ defined in \cref{appeq:define-S-sets} are given by
\begin{equation}\label{appeq:sqrt3-30-deg-S-sets}
\mathcal{S}_+ = \{\bK_+, -\bK_+, \bGamma_+\},\quad \mathcal{S}_- = \{\bGamma_-\}.
\end{equation}
As in \cref{app:II-plus-gapped}, we assume that the bottom layer has a gap around $0$ at $\bGamma_-$ so that the low energy physics is described by a moir\'e model of the form in \cref{app:two-Dirac-cone} where the two Dirac cones correspond to the two graphene valleys. In this section, we start with a moir\'e model involving degrees of freedom in both layers. We then use second order Schrieffer-Wolff perturbation theory \cite{Schrieffer1966,Bravyi2011} to find an explicit form for the graphene moir\'e model. Finally, we use symmetry and corepresentation theory to constrain the model parameters. We neglect spin degrees of freedom in this section and consider them in \cref{app:coupled-valley-spinful}.

\subsection{Continuum Hamiltonian}
When $\delta\epsilon =\delta\theta = 0$, the delta function in \cref{appeq:momentum-slater-koster} ensures that the only bottom layer momentum that is coupled to $\bk = \eta \bK_+ + \bp$ is $\bk' = \bGamma_- + \bp$, where $\eta \in \{+, -\}$ denotes the two graphene valleys. However, when $\delta\epsilon$ and $\delta\theta$ are not both zero, it is possible to couple a top layer state with momentum $\bk = \eta \bK_+ + \bp$ to a bottom layer state with momentum $\bk' = \bGamma_- + \bp'$ as long as
\begin{equation}
\bk + \bG_+ = \bk' + \bG_-
\end{equation}
for some $\bG_- \in P_-$ and $\bG_+ \in P_+$. As in \cref{app:general-moire-models}, we assume that $\hat{t}_{l',\alpha',l,\alpha}(\bQ)$ depends only on $|\bQ|$ and decays rapidly as $|\bQ|$ grows. It is then sufficient to consider only terms in \cref{appeq:momentum-slater-koster} for which $|\bk + \bG_+|$ is small. For small $|\bp|$, the dominant terms are those with
\begin{equation}
\begin{split}
\bG_+ &= \eta(R_{\zeta_j} - 1)\bK_+\\
e^{\delta\epsilon} R_{-\delta\theta} \bG_- &= \eta R_{\zeta_j}\bK_+
\end{split}
\end{equation}
where $\zeta_j = \frac{2\pi}{3}(j-1)$. For these terms, we have
\begin{equation}\label{appeq:define-q-vectors}
\begin{split}
\bp' &= \bp + \eta \bq_j\\
\bq_j &= R_{\zeta_j}(1 - e^{-\delta\epsilon}R_{\delta\theta}) \bK_+.
\end{split}
\end{equation}
Recall from \cref{app:II-plus-gapped} that for $\namesqthree$ we can take $\xi_0 = \pi$. Since additionally we have $N_- = 1$, the definition of $\bq_j$ in \cref{appeq:define-q-vectors} is consistent with \cref{appeq:two-Dirac-q_j}.

With this motivation, we now introduce a continuum model that describes the low energy physics of the bilayer system associated with the graphene Dirac cones. We define continuum states $\ket{\bp, \eta, \alpha}_c$ for $\bp \in \R^2$, $\eta \in \{+, 0, -\}$, and $\alpha \in \mathcal{O}_l$ where $l = +$ when $\eta \in \{+, -\}$ and $l = -$ when $\eta = 0$. The state $\ket{\bp, \eta, \alpha}_c$ represents the Bloch state $\ket{\eta \bK_+ + \bp, l, \alpha}$ for small $\bp$, and the continuum states satisfy the normalization condition
\begin{equation}
\braket{\bp', \eta', \alpha' |_c \bp, \eta, \alpha}_c = \delta^2(\bp'-\bp)\delta_{\eta',\eta}\delta_{\alpha',\alpha}.
\end{equation}
We group the continuum states with momentum $\bp$ into a row vector
\begin{equation}
\ket{\bp}_c = \begin{pmatrix}
\ket{\bp, +, A}_c & \ket{\bp, +, B}_c & \ket{\bp, -, A}_c & \ket{\bp, -, B}_c & \ket{\bp, 0, \alpha_1}_c & \cdots & \ket{\bp, 0, \alpha_n}_c
\end{pmatrix}
\end{equation}
where $\mathcal{O}_- = \{\alpha_1, \dots, \alpha_n\}$ is an index set for the orbitals on the bottom layer. The continuum Hamiltonian then takes the form
\begin{equation}\label{appeq:continuum-hamiltonian-momentum}
\begin{split}
H_c &= \int d^2\bp' d^2\bp \ket{\bp'}_c \mathcal{H}_c(\bp', \bp)\bra{\bp}_c\\
\mathcal{H}_c(\bp', \bp) &= \begin{pmatrix}
\hbar v_F \bsigma \cdot \bp & 0 & 0\\
0 & -\hbar v_F \bsigma^* \cdot \bp & 0\\
0 & 0 & \mathcal{H}_-
\end{pmatrix}\delta^2(\bp'-\bp)\\
&+ \sum_{j=1}^3 \begin{pmatrix}
0 & 0 & 0\\
0 & 0 & T_{-,\bq_j}\\
T_{+,\bq_j}^\dagger & 0 & 0
\end{pmatrix}\delta^2(\bp' - \bp - \bq_j) + \sum_{j=1}^3 \begin{pmatrix}
0 & 0 & T_{+,\bq_j}\\
0 & 0 & 0\\
0 & T_{-,\bq_j}^\dagger & 0
\end{pmatrix}\delta^2(\bp'-\bp + \bq_j).
\end{split}
\end{equation}
Here, $v_F$ is the graphene Fermi velocity and $\mathcal{H}_-$ is the $n \times n$ Hamiltonian for the bottom layer at the $\bGamma_-$ point, which we approximate as independent of momentum. The $T_{\eta,\bq_j}$ are $2\times n$ complex matrices describing the coupling between valley $\eta$ of graphene and $\mathcal{H}_-$.

We can write the continuum Hamiltonian in a simpler form by transforming to real space. We define real space continuum states
\begin{equation}
\ket{\br, \eta, \alpha}_c = \frac{1}{2\pi} \int d^2\bp e^{-i\bp \cdot \br}\ket{\bp, \eta, \alpha}_c
\end{equation}
which satisfy
\begin{equation}
\braket{\br', \eta', \alpha'|_c \br, \eta, \alpha} = \delta^2(\br'-\br)\delta_{\eta',\eta}\delta_{\alpha',\alpha}
\end{equation}
and group them into a row vector
\begin{equation}
\ket{\br}_c = \begin{pmatrix}
\ket{\br, +, A}_c & \ket{\br, +, B}_c & \ket{\br, -, A}_c & \ket{\br, -, B}_c & \ket{\br, 0, \alpha_1}_c & \cdots & \ket{\br, 0, \alpha_n}_c
\end{pmatrix}.
\end{equation}
The continuum Hamiltonian then becomes
\begin{equation}\label{appeq:continuum-hamiltonian-real}
\begin{split}
H_c &= \int d^2\br \ket{\br}_c \mathcal{H}_c(\br) \bra{\br}_c\\
\mathcal{H}_c(\br) &= \begin{pmatrix}
-i\hbar v_F \bsigma \cdot \nabla & 0 & T_+(\br)\\
0 & i\hbar v_F \bsigma^* \cdot \nabla & T_-(\br)\\
T_+^\dagger(\br) & T_-^\dagger(\br) & \mathcal{H}_-
\end{pmatrix}\\
T_\eta(\br) &= \sum_{j=1}^3 T_{\eta,\bq_j} e^{-i\eta \br \cdot \bq_j}.
\end{split}
\end{equation}

Assuming that the spectrum of $\mathcal{H}_-$ has a sufficiently large gap around $0$, we can treat the bottom layer states perturbatively. Applying Schrieffer-Wolff perturbation theory \cite{Schrieffer1966,Bravyi2011} to second order in the interlayer couplings and neglecting the graphene energies, we find the effective Hamiltonian for the graphene degrees of freedom
\begin{equation}\label{appeq:H_g-r_g}
\begin{split}
H_g &= \int d^2\br \ket{\br}_g \mathcal{H}_g(\br) \bra{\br}_g\\
\mathcal{H}_g(\br) &= \begin{pmatrix}
-i\hbar v_F \bsigma \cdot \nabla & 0\\
0 & i\hbar v_F \bsigma^* \cdot \nabla
\end{pmatrix} - \begin{pmatrix}
T_+(\br) \\ T_-(\br)
\end{pmatrix}
\mathcal{H}_-^{-1}
\begin{pmatrix}
T_+^\dagger(\br) & T_-^\dagger(\br)
\end{pmatrix}\\
\ket{\br}_g &= \begin{pmatrix}
\ket{\br, +, A}_c & \ket{\br, +, B}_c & \ket{\br, -, A}_c & \ket{\br, -, B}_c
\end{pmatrix}.
\end{split}
\end{equation}
Finally, we take a unitary change of basis to bring the model into a form with two Dirac cones of the same helicity. To do so, we define new states $\ket{\br}$ by applying a $\sigma_y$ transformation to the sublattice degree of freedom in the $\eta = -$ valley. Specifically, we take
\begin{equation}\label{appeq:transformed-r-states}
\ket{\br} = \ket{\br}_g U_0 \text{ where } U_0 = \begin{pmatrix}
\sigma_0 & 0\\
0 & \sigma_y
\end{pmatrix}.
\end{equation}
We then have
\begin{equation}\label{appeq:canonical-moire-model}
\begin{split}
H_g &= \int d^2\br \ket{\br} \mathcal{H}(\br) \bra{\br}\\
\mathcal{H}(\br) &= \begin{pmatrix}
S_+(\br) - i\hbar v_F \bsigma \cdot \nabla & T(\br)\\
T^\dagger(\br) & S_-(\br) -i\hbar v_F \bsigma \cdot \nabla
\end{pmatrix}
\end{split}
\end{equation}
which is of the form in \cref{appeq:two-Dirac-hamiltonian} with $v_+ = v_- = v_F$. The $T(\br)$ and $S_\eta(\br)$ potentials are given by
\begin{equation}\label{appeq:S-T-2nd-order}
\begin{split}
T(\br) &= -T_+(\br) \mathcal{H}_-^{-1} T_-^\dagger(\br)\sigma_y\\
S_+(\br) &= -T_+(\br) \mathcal{H}_-^{-1} T_+^\dagger(\br)\\
S_-(\br) &= -\sigma_y T_-(\br) \mathcal{H}_-^{-1} T_-^\dagger(\br)\sigma_y.
\end{split}
\end{equation}
Note that these potentials can be expanded in the form given by \cref{appeq:two-Dirac-T-S-expansion}. However, in this case all $T_\bq$ and $S_{\pm,\bq}$ matrices vanish except for those of the form
\begin{equation}\label{appeq:S-T-expansion-nonzero}
T_{\bq_j}, \quad T_{-2\bq_j}, \quad S_{\pm,\bq_j - \bq_k}.
\end{equation}

Finally, it is worth noting that bottom layer states near $\bGamma_-$ are coupled to top layer states near $\bGamma_+$ in addition to top layer states near $\bK_+$ and $-\bK_+$. This is clear from \cref{appeq:momentum-slater-koster} as well as from \cref{appeq:sqrt3-30-deg-S-sets}. As a result, in principle one should also include top layer states near $\bGamma_+$ in \cref{appeq:continuum-hamiltonian-momentum}. However, since these states all have large energies and are not directly coupled to the graphene Dirac cones, they make no contribution to a low energy model derived using second order Schrieffer-Wolff perturbation theory. For this reason, we only consider top layer states near $\bK_+$ and $-\bK_+$ and bottom layer states near $\bGamma_-$.

\subsection{Symmetry constraints}\label{app:symmetry-constraints}
We now consider the non-translational symmetries of the Hamiltonian in \cref{appeq:canonical-moire-model}. The possible crystalline symmetry generators are $C_{3z}$ (rotation by angle $2\pi/3$ about $\bhatz$), $C_{2z}$ (rotation by angle $\pi$ about $\bhatz$), $M_x$ (reflection through the $yz$ plane), and $M_y$ (reflection through the $xz$ plane). Additionally, we consider the antiunitary time-reversal symmetry $\mathcal{T}$ which satisfies $\mathcal{T}^2 = 1$. By considering the action of these symmetries on the graphene Wannier functions and Bloch states, we can deduce the appropriate definitions for the symmetry operators on the states $\ket{\br}_g$ in \cref{appeq:H_g-r_g}. We define
\begin{equation}\label{appeq:symmetry-generators-graphene}
\begin{split}
C_{3z} \ket{\br}_g &= \ket{R_{2\pi/3}\br}_g e^{i(2\pi/3)\sigma_z \otimes \sigma_z}\\
C_{2z} \ket{\br}_g &= \ket{-\br}_g \sigma_x \otimes \sigma_x\\
M_x \ket{\br}_g &= \ket{\mathcal{R}_\bhatx \br}_g \sigma_x \otimes \sigma_0\\
M_y \ket{\br}_g &= \ket{\mathcal{R}_\bhaty \br}_g \sigma_0 \otimes \sigma_x\\
\mathcal{T}\ket{\br}_g &= \ket{\br}_g \sigma_x \otimes \sigma_0.
\end{split}
\end{equation}
Here, the first (second) Pauli matrix in each tensor product acts on valley (sublattice). These operators can also be written in terms of the $\ket{\br}$ states in \cref{appeq:transformed-r-states} as
\begin{equation}\label{appeq:symmetry-generators-transformed}
\begin{split}
C_{3z} \ket{\br} &= \ket{R_{2\pi/3}\br} \sigma_0 \otimes e^{i(2\pi/3) \sigma_z}\\
C_{2z} \ket{\br} &= -\ket{-\br} \sigma_y \otimes \sigma_z\\
M_x \ket{\br} &= \ket{\mathcal{R}_\bhatx \br} \sigma_x \otimes \sigma_y\\
M_y \ket{\br} &= \ket{\mathcal{R}_\bhaty \br} \sigma_z \otimes \sigma_x\\
\mathcal{T}\ket{\br} &= -i\ket{\br} \sigma_y \otimes \sigma_y.
\end{split}
\end{equation}

Suppose that $U$ is a unitary or antiunitary symmetry which acts as $U\ket{\br} = \ket{\mathcal{O}\br} M$ for an orthogonal matrix $\mathcal{O}$ and a unitary matrix $M$. Then $[U, H_g] = 0$ is equivalent to $\mathcal{H}(\mathcal{O}\br) = M \mathcal{H}(\br) M^\dagger$ in the unitary case and $\mathcal{H}(\mathcal{O}\br) = M \mathcal{H}^*(\br) M^\dagger$ in the antiunitary case. Applying this to the symmetry generators gives the following constraints on the $T_\bq$ and $S^\eta_\bq$ matrices defined by \cref{appeq:two-Dirac-T-S-expansion}.
\begin{equation}\label{appeq:symmetry-constraints-coupled-valley}
\begin{split}
[C_{3z}, H_g] = 0 &\iff T_{R_{2\pi/3}\bq} = e^{i(2\pi/3)\sigma_z} T_\bq e^{-i(2\pi/3)\sigma_z} \text{ and } S_{\eta,R_{2\pi/3}\bq} = e^{i(2\pi/3)\sigma_z} S_{\eta,\bq} e^{-i(2\pi/3)\sigma_z}\\
[C_{2z}, H_g] = 0 &\iff T_\bq = -\sigma_z T^\dagger_\bq \sigma_z \text{ and } S_{+,-\bq} = \sigma_z S_{-,\bq} \sigma_z\\
[M_x, H_g] = 0&\iff T_{\mathcal{R}_\bhaty \bq} = \sigma_y T_\bq^\dagger \sigma_y \text{ and } S_{+,\mathcal{R}_\bhaty \bq} = \sigma_y S_{-,-\bq} \sigma_y\\
[M_y, H_g] = 0 &\iff T_{\mathcal{R}_\bhaty \bq} = -\sigma_x T_\bq \sigma_x \text{ and } S_{\eta,\mathcal{R}_\bhaty \bq} = \sigma_x S_{\eta,\bq} \sigma_x\\
[\mathcal{T}, H_g] = 0 &\iff T_\bq = -\sigma_y T^T_\bq \sigma_y \text{ and } S_{+,\bq} = \sigma_y S^*_{-,-\bq} \sigma_y.
\end{split}
\end{equation}

We will always make the assumption that $C_{3z}$ and $\mathcal{T}$ symmetries are preserved. By \cref{appeq:symmetry-constraints-coupled-valley}, $\mathcal{T}$ symmetry implies
\begin{equation}\label{appeq:S-minus-from-plus}
S_-(\br) = \sigma_y S^*_+(\br) \sigma_y
\end{equation}
so it suffices to characterize $T(\br)$ and $S_+(\br)$. Keeping only the terms in \cref{appeq:S-T-expansion-nonzero}, we can use $C_{3z}$, $\mathcal{T}$, and Hermiticity to write
\begin{equation}\label{appeq:S-T-explicit-full}
\begin{split}
T_{\bq_j} &= u_{1,x} \bsigma \cdot \bhatn_{\zeta_j} + u_{1,y} \bsigma \cdot \bhatn_{\zeta_j + \pi/2} + u_{1,z} \sigma_z\\
T_{-2\bq_j} &= u_{2,x} \bsigma \cdot \bhatn_{\zeta_j} + u_{2,y} \bsigma \cdot \bhatn_{\zeta_j + \pi/2} + u_{2,z} \sigma_z\\
S_{+,\bzero} &= u_{0,0} \sigma_0 + u_{0, z}\sigma_z\\
S_{+,\bq_{1+j}-\bq_{2+j}} &= u_{\sqrt{3}, 0} \sigma_0 + u_{\sqrt{3},x} \bsigma\cdot\bhatn_{\zeta_j} + u_{\sqrt{3},y}\bsigma \cdot \bhatn_{\zeta_j + \pi/2} + u_{\sqrt{3},z}\sigma_z\\
S_{+,\bq_{2+j}-\bq_{1+j}} &= S^\dagger_{+,\bq_{1+j}-\bq_{2+j}}
\end{split}
\end{equation}
where $u_{0,0}$ and $u_{0,z}$ are real parameters and all of the other $u$ parameters are in general complex. We now consider the symmetry constraints in several cases.
\begin{enumerate}
\item The symmetry group is generated by $C_{3z}$ and $\mathcal{T}$. In this case,
\begin{equation}\label{appeq:u-constraints-C3z-T}
u_{0,0}, u_{0,z} \in \R.
\end{equation}

\item The symmetry group is generated by $C_{3z}$, $\mathcal{T}$, and $C_{2z}$. In this case,
\begin{equation}\label{appeq:u-constraints-C3z-T-C2z}
\begin{split}
u_{0,0}, u_{1,x}, u_{1,y}, u_{2,x}, u_{2,y}, u_{\sqrt{3},0}, u_{\sqrt{3},x}, u_{\sqrt{3},y} &\in \R\\
u_{1,z}, u_{2,z}, u_{\sqrt{3}, z} &\in i\R\\
u_{0,z} &= 0.
\end{split}
\end{equation}

\item The symmetry group is generated by $C_{3z}$, $\mathcal{T}$, and $M_x$. In this case,
\begin{equation}\label{appeq:u-constraints-C3z-T-Mx}
\begin{split}
u_{0,0}, u_{0,z}, u_{1,y}, u_{2,y} &\in \R\\
u_{1,x}, u_{1,z}, u_{2,x}, u_{2,z} &\in i\R\\
u_{\sqrt{3},y} &= 0.
\end{split}
\end{equation}

\item The symmetry group is generated by $C_{3z}$, $\mathcal{T}$, and $M_y$. In this case,
\begin{equation}\label{appeq:u-constraints-C3z-T-My}
\begin{split}
u_{0,0}, u_{\sqrt{3},0}, u_{\sqrt{3},x} &\in \R\\
u_{\sqrt{3}, y}, u_{\sqrt{3},z} &\in i\R\\
u_{0, z} = u_{1, x} = u_{2,x} &= 0.
\end{split}
\end{equation}

\item The symmetry group is generated by $C_{3z}$, $\mathcal{T}$, $C_{2z}$, $M_x$, and $M_y$. In this case,
\begin{equation}\label{appeq:u-constraints-C3z-T-C2z-Mx-My}
\begin{split}
u_{0,0}, u_{1,y}, u_{2,y}, u_{\sqrt{3},0}, u_{\sqrt{3},x} &\in \R\\
u_{1,z}, u_{2,z}, u_{\sqrt{3},z} &\in i\R\\
u_{0, z} = u_{1, x} = u_{2,x} = u_{\sqrt{3}, y} &= 0.
\end{split}
\end{equation}
In this case, we summarize \cref{appeq:S-T-explicit-full} as
\begin{equation}\label{appeq:S-T-simplified}
\begin{split}
T_{m\bq_j} &= w_{|m|,y} \bsigma \cdot \bhatn_{\zeta_j + \pi/2} + w_{|m|,z} i \sigma_z\\
S_{+,\bzero} &= w_{0,0} \sigma_0\\
S_{+,\gamma(\bq_{1+j} - \bq_{2+j})} &= w_{\sqrt{3},0}\sigma_0 + w_{\sqrt{3},x}\bsigma\cdot \bhatn_{\zeta_j} + \gamma w_{\sqrt{3},z} i\sigma_z
\end{split}
\end{equation}
for $m \in \{1, -2\}$ and $\gamma \in \{+, -\}$, where the real $w$ parameters are defined by
\begin{equation}
\begin{split}
w_{1,y} = u_{1,y},\quad w_{1,z} = -i u_{1,z},\quad w_{2,y} &= u_{2,y},\quad w_{2,z} = -iu_{2,z},\quad w_{0,0} = u_{0,0}\\
w_{\sqrt{3},0} = u_{\sqrt{3},0},\quad w_{\sqrt{3},x} &= u_{\sqrt{3},x},\quad w_{\sqrt{3},z} = -i u_{\sqrt{3},z}.
\end{split}
\end{equation}
\end{enumerate}
Note that $C_{2z} = M_x M_y$ so that this is a full enumeration of the cases with $C_{3z}$ and $\mathcal{T}$ symmetries. See \cref{app:point-groups} for a discussion of these symmetry groups and \cref{apptbl:characters} for their character tables.

\subsection{Transformation to the moir\'e coordinate system}\label{app:moire-coordinate-translation-spinless}
We now translate these results into the moir\'e coordinate system defined in \cref{app:moire-coordinate-system}. \cref{appeq:S-minus-from-plus,appeq:S-T-explicit-full} become
\begin{equation}
S_{\xi,-}(\br) = \sigma_y S^*_{\xi,+}(\br) \sigma_y
\end{equation}
and
\begin{equation}\label{appeq:S-T-explicit-full-moire-coordinates}
\begin{split}
T_{\xi, \bq_j} &= u_{1,x} \bsigma \cdot \bhatn_{\zeta_j - \xi} + u_{1,y} \bsigma \cdot \bhatn_{\zeta_j + \pi/2 - \xi} + u_{1,z} \sigma_z\\
T_{\xi, -2\bq_j} &= u_{2,x} \bsigma \cdot \bhatn_{\zeta_j - \xi} + u_{2,y} \bsigma \cdot \bhatn_{\zeta_j + \pi/2 - \xi} + u_{2,z} \sigma_z\\
S_{\xi, +,\bzero} &= u_{0,0} \sigma_0 + u_{0, z}\sigma_z\\
S_{\xi, +,\bq_{1+j}-\bq_{2+j}} &= u_{\sqrt{3}, 0} \sigma_0 + u_{\sqrt{3},x} \bsigma\cdot\bhatn_{\zeta_j - \xi} + u_{\sqrt{3},y}\bsigma \cdot \bhatn_{\zeta_j + \pi/2 - \xi} + u_{\sqrt{3},z}\sigma_z\\
S_{\xi, +,\bq_{2+j}-\bq_{1+j}} &= S^\dagger_{\xi, +,\bq_{1+j}-\bq_{2+j}}.
\end{split}
\end{equation}
Additionally, \cref{appeq:S-T-simplified} becomes
\begin{equation}
\begin{split}
T_{\xi, m\bq_j} &= w_{|m|,y} \bsigma \cdot \bhatn_{\zeta_j + \phi_1} + w_{|m|,z} i \sigma_z\\
S_{\xi, +,\bzero} &= w_{0,0} \sigma_0\\
S_{\xi, +,\gamma(\bq_{1+j} - \bq_{2+j})} &= w_{\sqrt{3},0}\sigma_0 + w_{\sqrt{3},x}\bsigma\cdot \bhatn_{\zeta_j + \phi_1-\pi/2} + i \gamma w_{\sqrt{3},z} \sigma_z
\end{split}
\end{equation}
for $m \in \{1, -2\}$ and $\gamma \in \{+, -\}$, where
\begin{equation}\label{appeq:define-phi_1}
\phi_1 = \pi/2 - \xi = \arg\left(e^{-\delta\epsilon - i\delta\theta}-1\right)
\end{equation}
since $\xi_0 = \pi$.

\subsection{Coirrep decomposition for the bottom layer}\label{app:coirrep-decomposition}
In this section, we show that the $T(\br)$ and $S_+(\br)$ potentials can be written as a sum over the irreducible corepresentations (coirreps) of the symmetry group at the $\bGamma_-$ point in the bottom layer. To do so, we analyze the non-translational symmetries of the bilayer continuum Hamiltonian in \cref{appeq:continuum-hamiltonian-real}. Generalizing \cref{appeq:symmetry-generators-graphene}, the relevant symmetry generators take the form
\begin{equation}\label{appeq:symmetry-generators-bilayer}
\begin{split}
C_{3z} \ket{\br}_g &= \ket{R_{2\pi/3}\br}_g e^{i(2\pi/3)\sigma_z \otimes \sigma_z} \oplus C_{3z,-}\\
C_{2z} \ket{\br}_g &= \ket{-\br}_g (\sigma_x \otimes \sigma_x) \oplus C_{2z,-}\\
M_x \ket{\br}_g &= \ket{\mathcal{R}_\bhatx \br}_g (\sigma_x \otimes \sigma_0) \oplus M_{x,-}\\
M_y \ket{\br}_g &= \ket{\mathcal{R}_\bhaty \br}_g (\sigma_0 \otimes \sigma_x) \oplus M_{y,-}\\
\mathcal{T}\ket{\br}_g &= \ket{\br}_g (\sigma_x \otimes \sigma_0) \oplus \mathcal{T}_-
\end{split}
\end{equation}
for some unitary matrices $C_{3z,-}$, $C_{2z,-}$, $M_{x,-}$, $M_{y,-}$, and $\mathcal{T}_-$ which act on the bottom layer. In a manner similar to that in \cref{app:symmetry-constraints}, one can derive the following constraints on the $T_{\eta,\bq_j}$ matrices arising from the commutation of each symmetry with $H_c$.
\begin{equation}\label{appeq:symmetry-constraints-bilayer}
\begin{split}
[C_{3z}, H_c] = 0 \iff& T_{\eta,\bq_{j+1}} = e^{i\eta(2\pi/3)\sigma_z}T_{\eta, \bq_j}C_{3z,-}^\dagger\\
[C_{2z}, H_c] = 0 \iff& T_{\eta,\bq_j} = \sigma_x T_{-\eta,\bq_j}C_{2z,-}^\dagger\\
[M_x, H_c] = 0 \iff& T_{\eta,\bq_1} = T_{-\eta,\bq_1}M_{x,-}^\dagger\\
&T_{\eta,\bq_2} = T_{-\eta,\bq_3}M_{x,-}^\dagger\\
&T_{\eta,\bq_3} = T_{-\eta,\bq_2}M_{x,-}^\dagger\\
[M_y, H_c] = 0 \iff& T_{\eta,\bq_1} = \sigma_x T_{\eta,\bq_1}M_{y,-}^\dagger\\
&T_{\eta,\bq_2} = \sigma_x T_{\eta,\bq_3}M_{y,-}^\dagger\\
&T_{\eta,\bq_3} = \sigma_x T_{\eta,\bq_2}M_{y,-}^\dagger\\
[\mathcal{T}, H_c] = 0 \iff& T_{\eta,\bq_j} = T_{-\eta,\bq_j}^*\mathcal{T}_-^\dagger.
\end{split}
\end{equation}

In the presence of $C_{3z}$ and $\mathcal{T}$ symmetries, all of the $T_{\eta,\bq_j}$ matrices are determined by $T_{+,\bq_1}$. Specifically, we have
\begin{equation}
\begin{split}
T_{\eta,\bq_j} &= e^{i\eta \zeta_j \sigma_z} T_{\eta,\bq_1} C_{3z,-}^{1-j}\\
T_{-,\bq_1} &= T_{+,\bq_1}^* \mathcal{T}_-^\dagger.
\end{split}
\end{equation}
Any additional symmetries then imply constraints on $T_{+,\bq_1}$. Specifically,
\begin{align}
[C_{2z}, H_c] = 0 &\implies T_{+,\bq_1} = \sigma_x T_{+,\bq_1}^*(C_{2z,-}\mathcal{T}_-)^\dagger \label{appeq:C2zT-constraint}\\
[M_x, H_c] = 0 &\implies T_{+,\bq_1} = T_{+,\bq_1}^*(M_{x,-}\mathcal{T}_-)^\dagger \label{appeq:MxT-constraint}\\
[M_y, H_c] = 0 &\implies T_{+,\bq_1} = \sigma_x T_{+,\bq_1}M_{y,-}^\dagger.\label{appeq:My-constraint}
\end{align}

By \cref{appeq:continuum-hamiltonian-real,appeq:S-T-2nd-order}, we have
\begin{align}
T(\br) &= -\sum_{j,k=1}^3 T_{+,\bq_j} \mathcal{H}_-^{-1} T_{-,\bq_k}^\dagger \sigma_y e^{-i\br \cdot (\bq_k + \bq_j)}\\
&= -\sum_{j,k=1}^3 e^{i\zeta_j \sigma_z} T_{+,\bq_1} C_{3z,-}^{1-j}\mathcal{H}_-^{-1} C_{3z,-}^{k-1} \mathcal{T}_- T_{+,\bq_1}^T \sigma_y e^{-i\zeta_k \sigma_z} e^{-i\br \cdot (\bq_k + \bq_j)}\label{appeq:T-expansion}\\
S_+(\br) &= -\sum_{j,k=1}^3 T_{+,\bq_j} \mathcal{H}_-^{-1} T_{+,\bq_k}^\dagger e^{i\br \cdot (\bq_k - \bq_j)}\\
&= -\sum_{j,k=1}^3 e^{i\zeta_j \sigma_z} T_{+,\bq_1} C_{3z,-}^{1-j}\mathcal{H}_-^{-1} C_{3z,-}^{k-1} T_{+,\bq_1}^\dagger e^{-i\zeta_k \sigma_z} e^{i\br \cdot (\bq_k - \bq_j)}.\label{appeq:S-expansion}
\end{align}
It follows that we can compute the $T(\br)$ and $S_+(\br)$ potentials from $T_{+,\bq_1}$ along with $\mathcal{H}_-$ and the symmetry corepresentation on the bottom layer.

By Schur's lemma for corepresentation theory, there is a unitary change of basis for the bottom layer subspace that diagonalizes $\mathcal{H}_-$ while also block diagonalizing the symmetry corepresentation on the bottom layer into coirreps \cite{Newmarch1982}. The contributions to $T(\br)$ and $S_+(\br)$ in \cref{appeq:T-expansion,appeq:S-expansion} from each coirrep simply add. We can therefore consider the form for $T(\br)$ and $S_+(\br)$ independently for each coirrep. In the following sections, we explicitly find the form for the $u$ parameters in \cref{appeq:S-T-explicit-full} for each coirrep of each possible symmetry group that contains $C_{3z}$ and $\mathcal{T}$.

\begin{table}
	\centering
	\begin{tabular}{c|cccc}
	$31'$ & $1$ & $C_{3z}$ & $C_{3z}^{-1}$ & $-1$\\
	\hline $A_1$ & $1$ & $1$ & $1$ & $1$\\
	${^2E}{^1E}$ & $2$ & $-1$ & $-1$ & $2$\\
	\hline $\overline{EE}$ & $2$ & $-2$ & $-2$ & $-2$\\
	${^1\overline{E}}{^2\overline{E}}$ & $2$ & $1$ & $1$ & $-2$\\
	\end{tabular}
	\hspace{1cm}
	\begin{tabular}{c|ccccccc}
	$61'$ & $1$ & $C_{3z}$ & $C_{3z}^{-1}$ & $C_{2z}$ & $C_{6z}^{-1}$ & $C_{6z}$ & $-1$\\
	\hline $A$ & $1$ & $1$ & $1$ & $1$ & $1$ & $1$ & $1$\\
	$B$ & $1$ & $1$ & $1$ & $-1$ & $-1$ & $-1$ & $1$\\
	${^2E_1}{^1E_1}$ & $2$ & $-1$ & $-1$ & $2$ & $-1$ & $-1$ & $2$\\
	${^2E_2}{^1E_2}$ & $2$ & $-1$ & $-1$ & $-2$ & $1$ & $1$ & $2$\\
	\hline ${^2\overline{E}_1}{^1\overline{E}_1}$ & $2$ & $-2$ & $-2$ & $0$ & $0$ & $0$ & $-2$\\
	${^1\overline{E}_3}{^2\overline{E}_3}$ & $2$ & $1$ & $1$ & $0$ & $\sqrt{3}$ & $\sqrt{3}$ & $-2$\\
	${^1\overline{E}_2}{^2\overline{E}_2}$ & $2$ & $1$ & $1$ & $0$ & $-\sqrt{3}$ & $-\sqrt{3}$ & $-2$\\
	\end{tabular}
	\hspace{1cm}
	\begin{tabular}{c|ccccc}
	$mm21'$ & $1$ & $C_{2z}$ & $M_x$ & $M_y$ & $-1$\\
	\hline $A_1$ & $1$ & $1$ & $1$ & $1$ & $1$\\
	$A_2$ & $1$ & $1$ & $-1$ & $-1$ & $1$\\
	$B_1$ & $1$ & $-1$ & $-1$ & $1$ & $1$\\
	$B_2$ & $1$ & $-1$ & $1$ & $-1$ & $1$\\
	\hline $\overline{E}$ & $2$ & $0$ & $0$ & $0$ & $-2$
	\end{tabular}\\
	\vspace{0.5cm}
	\begin{tabular}{c|cccc}
	$3m1'$ & $1$ & $C_{3z}$ & $M$ & $-1$\\
	\hline $A_1$ & $1$ & $1$ & $1$ & $1$\\
	$A_2$ & $1$ & $1$ & $-1$ & $1$\\
	$E$ & $2$ & $-1$ & $0$ & $2$\\
	\hline ${^2\overline{E}}{^1\overline{E}}$ & $2$ & $-2$ & $0$ & $-2$\\
	$\overline{E}_1$ & $2$ & $1$ & $0$ & $-2$\\
	\end{tabular}
	\hspace{1cm}
	\begin{tabular}{c|ccccccc}
	$6mm1'$ & $1$ & $C_{3z}$ & $C_{2z}$ & $C_{6z}$ & $M_x$ & $M_y$ & $-1$\\
	\hline $A_1$ & $1$ & $1$ & $1$ & $1$ & $1$ & $1$ & $1$\\
	$A_2$ & $1$ & $1$ & $1$ & $1$ & $-1$ & $-1$ & $1$\\
	$B_2$ & $1$ & $1$ & $-1$ & $-1$ & $-1$ & $1$ & $1$\\
	$B_1$ & $1$ & $1$ & $-1$ & $-1$ & $1$ & $-1$ & $1$\\
	$E_2$ & $2$ & $-1$ & $2$ & $-1$ & $0$ & $0$ & $2$\\
	$E_1$ & $2$ & $-1$ & $-2$ & $1$ & $0$ & $0$ & $2$\\
	\hline $\overline{E}_3$ & $2$ & $-2$ & $0$ & $0$ & $0$ & $0$ & $-2$\\
	$\overline{E}_2$ & $2$ & $1$ & $0$ & $-\sqrt{3}$ & $0$ & $0$ & $-2$\\
	$\overline{E}_1$ & $2$ & $1$ & $0$ & $\sqrt{3}$ & $0$ & $0$ & $-2$
	\end{tabular}
	\caption{Character tables enumerating the coirreps of the spinful magnetic point groups $31'$, $61'$, $mm21'$, $3m1'$, and $6mm1'$ described in \cref{app:point-groups}. Each column represents a corepresentation conjugacy class \cite{Newmarch1982}, and for brevity we do not include columns for conjugacy classes which are $-1$ times a shown conjugacy class. As explained in \cref{app:point-groups}, the rows in each table for which the entry corresponding to $-1$ is positive (i.e., the spinless coirreps) also form the character tables for the spinless magnetic point groups $31'_0$, $61'_0$, $mm21'_0$, $3m1'_0$, and $6mm1'_0$. In each table, a horizontal line is included to visually separate the spinless coirreps (above the line) from the spinful coirreps (below the line). These tables can be found on the Bilbao Crystallographic Server \cite{Aroyo2006,Aroyo2006a,Gallego2012,Elcoro2021,Xu2020}.}
	\label{apptbl:characters}
\end{table}

\subsubsection{The symmetry group is generated by $C_{3z}$ and $\mathcal{T}$}\label{app:symmetry-group-C3z-T}
In this case, the symmetry group is isomorphic to the spinless magnetic point group $31'_0$ defined in \cref{app:point-groups}. We see from \cref{apptbl:characters} that there are two coirreps, namely $A_1$ and ${^2E}{^1E}$.
\begin{enumerate}
\item For $A_1$, we take $C_{3z,-} = 1$, $\mathcal{T}_- = 1$, and
\begin{equation}\label{appeq:T1-dag-1D}
T_{+,\bq_1} = \begin{pmatrix}
v_1\\ v_2
\end{pmatrix}
\end{equation}
for complex parameters $v_1$ and $v_2$. Taking $\mathcal{H}_- = E_0$, \cref{appeq:T-expansion,appeq:S-expansion} imply that $T(\br)$ and $S_\eta(\br)$ take the form of \cref{appeq:S-T-explicit-full} with
\begin{equation}\label{appeq:u-from-v-1D}
\begin{split}
u_{1,x} &= \frac{i}{E_0}(v_1^2 - v_2^2),\quad u_{1,y} = -\frac{1}{E_0}(v_1^2 + v_2^2),\quad u_{1,z} = \frac{i}{E_0}v_1v_2\\
u_{2,x} &= \frac{i}{2E_0}(v_1^2 - v_2^2),\quad u_{2,y} = -\frac{1}{2E_0}(v_1^2 + v_2^2),\quad u_{2,z} = -\frac{i}{E_0}v_1v_2\\
u_{0,0} &= -\frac{3}{2E_0}(|v_1|^2 + |v_2|^2),\quad u_{0,z} = -\frac{3}{2E_0}(|v_1|^2 - |v_2|^2)\\
u_{\sqrt{3}, 0} &= \frac{1 - i\sqrt{3}}{4E_0}|v_1|^2 + \frac{1+i\sqrt{3}}{4E_0}|v_2|^2,\quad u_{\sqrt{3},x} = -\frac{1}{2E_0}(v_1 v_2^* + v_1^* v_2)\\
u_{\sqrt{3},y} &= -\frac{i}{2E_0}(v_1 v_2^* - v_1^* v_2),\quad u_{\sqrt{3}, z} = \frac{1 - i\sqrt{3}}{4E_0}|v_1|^2 - \frac{1+i\sqrt{3}}{4E_0}|v_2|^2.
\end{split}
\end{equation}

\item For ${^2E}{^1E}$, we take $C_{3z,-} = e^{-i(2\pi/3)\sigma_z}$, $\mathcal{T}_- = \sigma_x$, and
\begin{equation}\label{appeq:T1-dag-2D}
T_{+,\bq_1} = v_0 \sigma_0 + v_x \sigma_x + v_y \sigma_y + v_z\sigma_z
\end{equation}
for complex parameters $v_0$, $v_x$, $v_y$, and $v_z$. Taking $\mathcal{H}_- = E_0\sigma_0$, \cref{appeq:T-expansion,appeq:S-expansion} imply that $T(\br)$ and $S_\eta(\br)$ take the form of \cref{appeq:S-T-explicit-full} with
\begin{equation}\label{appeq:u-from-v-2D}
\begin{split}
u_{1,x} &= -\frac{2}{E_0}(v_0 v_y + i v_x v_z),\quad u_{1,y} = \frac{2}{E_0}(v_0 v_x - i v_y v_z),\quad u_{1,z} = \frac{i}{E_0}(v_0^2 - 2v_x^2 -2v_y^2 - v_z^2)\\
u_{2,x} &= \frac{2}{E_0}(v_0 v_y + i v_x v_z),\quad u_{2,y} = -\frac{2}{E_0}(v_0 v_x - i v_y v_z),\quad u_{2,z} = -\frac{i}{E_0}(v_0^2 + v_x^2 + v_y^2 - v_z^2)\\
u_{0,0} &= -\frac{3}{E_0}(|v_0|^2 + |v_x|^2 + |v_y|^2 + |v_z|^2),\quad u_{0,z} = -\frac{3}{E_0}(v_0 v_z^* + v_0^* v_z + i(v_x v_y^* - v_x^* v_y))\\
u_{\sqrt{3},0} &= \frac{1}{2E_0}(|v_0|^2 - 2|v_x|^2 - 2|v_y|^2 + |v_z|^2 + i\sqrt{3}(v_0 v_z^* + v_0^* v_z))\\
u_{\sqrt{3},x} &= \frac{1}{2E_0}(v_0 v_x^* + v_0^* v_x + i(v_y v_z^* - v_y^* v_z) -\sqrt{3}(v_0 v_y^* - v_0^* v_y) - i\sqrt{3}(v_x v_z^* + v_x^* v_z))\\
u_{\sqrt{3},y} &= \frac{1}{2E_0}(v_0 v_y^* + v_0^* v_y -i(v_x v_z^* - v_x^* v_z) + \sqrt{3}(v_0 v_x^* - v_0^* v_x) - i\sqrt{3}(v_y v_z^* + v_y^* v_z))\\
u_{\sqrt{3},z} &= \frac{1}{2E_0}(v_0 v_z^* + v_0^* v_z - 2i(v_x v_y^* - v_x^* v_y) + i\sqrt{3}(|v_0|^2 + |v_z|^2))
\end{split}
\end{equation}
\end{enumerate}

\subsubsection{The symmetry group is generated by $C_{3z}$, $\mathcal{T}$, and $C_{2z}$}\label{app:symmetry-group-C3z-T-C2z}
In this case, the symmetry group is isomorphic to the spinless magnetic point group $61'_0$ defined in \cref{app:point-groups}. We see from \cref{apptbl:characters} that there are four coirreps, namely $A$, $B$, ${^2E_1}{^1E_1}$, and ${^2E_2}{^1E_2}$.
\begin{enumerate}
\item For $A$, we take $C_{3z,-} = 1$, $\mathcal{T}_- = 1$, and $C_{2z,-} = 1$. \cref{appeq:T1-dag-1D,appeq:u-from-v-1D} apply, and additionally \cref{appeq:C2zT-constraint} implies
\begin{equation}\label{appeq:C2z-A-constraint}
v_2 = v_1^*.
\end{equation}

\item For $B$, we take $C_{3z,-} = 1$, $\mathcal{T}_- = 1$, and $C_{2z,-} = -1$. \cref{appeq:T1-dag-1D,appeq:u-from-v-1D} apply, and additionally \cref{appeq:C2zT-constraint} implies
\begin{equation}\label{appeq:C2z-B-constraint}
v_2 = -v_1^*.
\end{equation}

\item For ${^2E_1}{^1E_1}$, we take $C_{3z,-} = e^{-i(2\pi/3)\sigma_z}$, $\mathcal{T}_- = \sigma_x$, and $C_{2z,-} = \sigma_0$. \cref{appeq:T1-dag-2D,appeq:u-from-v-2D} apply, and additionally \cref{appeq:C2zT-constraint} implies
\begin{equation}\label{appeq:C2z-2E1-1E1-constraint}
\begin{split}
v_0, v_x, v_y &\in \R\\
v_z &\in i\R.
\end{split}
\end{equation}

\item For ${^2E_2}{^1E_2}$, we take $C_{3z,-} = e^{-i(2\pi/3)\sigma_z}$, $\mathcal{T}_- = \sigma_x$, and $C_{2z,-} = -\sigma_0$. \cref{appeq:T1-dag-2D,appeq:u-from-v-2D} apply, and additionally \cref{appeq:C2zT-constraint} implies
\begin{equation}\label{appeq:C2z-2E2-1E2-constraint}
\begin{split}
v_z &\in \R\\
v_0, v_x, v_y &\in i\R.
\end{split}
\end{equation}
\end{enumerate}

\subsubsection{The symmetry group is generated by $C_{3z}$, $\mathcal{T}$, and $M_x$}\label{app:symmetry-group-C3z-T-Mx}
In this case, the symmetry group is isomorphic to the spinless magnetic point group $3m1'_0$ defined in \cref{app:point-groups}. We see from \cref{apptbl:characters} that there are three coirreps, namely $A_1$, $A_2$, and $E$.
\begin{enumerate}
\item For $A_1$, we take $C_{3z,-} = 1$, $\mathcal{T}_- = 1$, and $M_{x,-} = 1$. \cref{appeq:T1-dag-1D,appeq:u-from-v-1D} apply, and additionally \cref{appeq:MxT-constraint} implies
\begin{equation}\label{appeq:Mx-A1-constraint}
v_1, v_2 \in \R.
\end{equation}

\item For $A_2$, we take $C_{3z,-} = 1$, $\mathcal{T}_- = 1$, and $M_{x,-} = -1$. \cref{appeq:T1-dag-1D,appeq:u-from-v-1D} apply, and additionally \cref{appeq:MxT-constraint} implies
\begin{equation}\label{appeq:Mx-A2-constraint}
v_1, v_2 \in i\R.
\end{equation}

\item For $E$, we take $C_{3z,-} = e^{-i(2\pi/3)\sigma_z}$, $\mathcal{T}_- = \sigma_x$, and $M_{x,-} = -\sigma_x$. \cref{appeq:T1-dag-2D,appeq:u-from-v-2D} apply, and additionally \cref{appeq:MxT-constraint} implies
\begin{equation}\label{appeq:Mx-E-constraint}
\begin{split}
v_y & \in \R\\
v_0, v_x, v_z & \in i\R.
\end{split}
\end{equation}
\end{enumerate}

\subsubsection{The symmetry group is generated by $C_{3z}$, $\mathcal{T}$, and $M_y$}\label{app:symmetry-group-C3z-T-My}
In this case, the symmetry group is isomorphic to the spinless magnetic point group $3m1'_0$ defined in \cref{app:point-groups}. We see from \cref{apptbl:characters} that there are three coirreps, namely $A_1$, $A_2$, and $E$.
\begin{enumerate}
\item For $A_1$, we take $C_{3z,-} = 1$, $\mathcal{T}_- = 1$, and $M_{y,-} = 1$. \cref{appeq:T1-dag-1D,appeq:u-from-v-1D} apply, and additionally \cref{appeq:My-constraint} implies
\begin{equation}\label{appeq:My-A1-constraint}
v_2 = v_1.
\end{equation}

\item For $A_2$, we take $C_{3z,-} = 1$, $\mathcal{T}_- = 1$, and $M_{y,-} = -1$. \cref{appeq:T1-dag-1D,appeq:u-from-v-1D} apply, and additionally \cref{appeq:My-constraint} implies
\begin{equation}\label{appeq:My-A2-constraint}
v_2 = -v_1.
\end{equation}

\item For $E$, we take $C_{3z,-} = e^{-i(2\pi/3)\sigma_z}$, $\mathcal{T}_- = \sigma_x$, and $M_{y,-} = \sigma_x$. \cref{appeq:T1-dag-2D,appeq:u-from-v-2D} apply, and additionally \cref{appeq:My-constraint} implies
\begin{equation}\label{appeq:My-E-constraint}
v_y = v_z = 0.
\end{equation}
\end{enumerate}

\subsubsection{The symmetry group is generated by $C_{3z}$, $\mathcal{T}$, $C_{2z}$, $M_x$, and $M_y$}\label{app:symmetry-group-C3z-T-C2z-Mx-My}
In this case, the symmetry group is isomorphic to the spinless magnetic point group $6mm1'_0$ defined in \cref{app:point-groups}. We see from \cref{apptbl:characters} that there are six coirreps, namely $A_1$, $A_2$, $B_2$, $B_1$, $E_2$, and $E_1$.
\begin{enumerate}
\item For $A_1$, we take $C_{3z,-} =1$, $\mathcal{T}_- = 1$, $C_{2z,-} = 1$, $M_{x,-} = 1$, and $M_{y,-} = 1$. \cref{appeq:T1-dag-1D,appeq:u-from-v-1D} apply, and additionally \cref{appeq:C2zT-constraint,appeq:MxT-constraint,appeq:My-constraint} imply \cref{appeq:C2z-A-constraint,appeq:Mx-A1-constraint,appeq:My-A1-constraint}.

\item For $A_2$, we take $C_{3z,-} =1$, $\mathcal{T}_- = 1$, $C_{2z,-} = 1$, $M_{x,-} = -1$, and $M_{y,-} = -1$. \cref{appeq:T1-dag-1D,appeq:u-from-v-1D} apply, and additionally \cref{appeq:C2zT-constraint,appeq:MxT-constraint,appeq:My-constraint} imply \cref{appeq:C2z-A-constraint,appeq:Mx-A2-constraint,appeq:My-A2-constraint}.

\item For $B_2$, we take $C_{3z,-} =1$, $\mathcal{T}_- = 1$, $C_{2z,-} = -1$, $M_{x,-} = -1$, and $M_{y,-} = 1$. \cref{appeq:T1-dag-1D,appeq:u-from-v-1D} apply, and additionally \cref{appeq:C2zT-constraint,appeq:MxT-constraint,appeq:My-constraint} imply \cref{appeq:C2z-B-constraint,appeq:Mx-A2-constraint,appeq:My-A1-constraint}.

\item For $B_1$, we take $C_{3z,-} =1$, $\mathcal{T}_- = 1$, $C_{2z,-} = -1$, $M_{x,-} = 1$, and $M_{y,-} = -1$. \cref{appeq:T1-dag-1D,appeq:u-from-v-1D} apply, and additionally \cref{appeq:C2zT-constraint,appeq:MxT-constraint,appeq:My-constraint} imply \cref{appeq:C2z-B-constraint,appeq:Mx-A1-constraint,appeq:My-A2-constraint}.

\item For $E_2$, we take $C_{3z,-} =e^{-i(2\pi/3)\sigma_z}$, $\mathcal{T}_- = \sigma_x$, $C_{2z,-} = \sigma_0$, $M_{x,-} = \sigma_x$, and $M_{y,-} = \sigma_x$. \cref{appeq:T1-dag-2D,appeq:u-from-v-2D} apply, and additionally \cref{appeq:C2zT-constraint,appeq:My-constraint} imply \cref{appeq:C2z-2E1-1E1-constraint,appeq:My-E-constraint} while \cref{appeq:MxT-constraint} implies
\begin{equation}
\begin{split}
v_0, v_x, v_z & \in \R\\
v_y & \in i\R.
\end{split}
\end{equation}

\item For $E_1$, we take $C_{3z,-} =e^{-i(2\pi/3)\sigma_z}$, $\mathcal{T}_- = \sigma_x$, $C_{2z,-} = -\sigma_0$, $M_{x,-} = -\sigma_x$, and $M_{y,-} = \sigma_x$. \cref{appeq:T1-dag-2D,appeq:u-from-v-2D} apply, and additionally \cref{appeq:C2zT-constraint,appeq:MxT-constraint,appeq:My-constraint} imply \cref{appeq:C2z-2E2-1E2-constraint,appeq:Mx-E-constraint,appeq:My-E-constraint}.
\end{enumerate}

We can summarize the results for this case more succinctly using the $w$ parameters in \cref{appeq:S-T-simplified}. For coirreps $A_1$, $A_2$, $B_2$, and $B_1$, we have
\begin{equation}
\begin{split}
w_{1,y} &= -\frac{2v_1^2}{E_0},\quad w_{1,z} = \frac{M_{y,-} v_1^2}{E_0},\quad w_{2,y} = -\frac{v_1^2}{E_0},\quad w_{2,z} = -\frac{M_{y,-} v_1^2}{E_0},\quad w_{0,0} = -\frac{3 |v_1|^2}{E_0}\\
w_{\sqrt{3},0} &= \frac{|v_1|^2}{2E_0},\quad w_{\sqrt{3},x} = -\frac{M_{y,-}|v_1|^2}{E_0},\quad w_{\sqrt{3},z} = -\frac{|v_1|^2\sqrt{3}}{2E_0}
\end{split}
\end{equation}
where $v_1 \in \R$ for $A_1$ and $B_1$, $v_1 \in i\R$ for $A_2$ and $B_2$, $M_{y,-} = 1$ for $A_1$ and $B_2$, and $M_{y,-} = -1$ for $A_2$ and $B_1$. Likewise, for coirreps $E_2$ and $E_1$, we have
\begin{equation}\label{appeq:w-parameters-6mm1'}
\begin{split}
w_{1,y} &= \frac{2v_0 v_x}{E_0},\quad w_{1,z} = \frac{v_0^2 - 2v_x^2}{E_0},\quad w_{2,y} = -\frac{2v_0 v_x}{E_0},\quad w_{2,z} = -\frac{v_0^2 + v_x^2}{E_0},\quad w_{0,0} = -\frac{3 (|v_0|^2 + |v_x|^2)}{E_0}\\
w_{\sqrt{3},0} &= \frac{|v_0|^2 - 2|v_x|^2}{2E_0},\quad w_{\sqrt{3},x} = \frac{v_0 v_x^* + v_0^* v_x}{2E_0},\quad w_{\sqrt{3},z} = \frac{|v_0|^2\sqrt{3}}{2E_0}
\end{split}
\end{equation}
where $v_0, v_x \in \R$ for $E_2$ and $v_0, v_x \in i\R$ for $E_1$.

\section{Coupled-valley graphene model with spin}\label{app:coupled-valley-spinful}
In \cref{app:coupled-valley-spinless} we ignored the spin degrees of freedom in order to simplify the discussion. In this section, we present the changes necessary to the derivation in \cref{app:coupled-valley-spinless} to include spin and spin-orbit coupling.

\subsection{Continuum model}
We use $\ket{\uparrow}$ and $\ket{\downarrow}$ to denote orthonormal spin states in the $z$ direction and define the row vector of states
\begin{equation}
\ket{s_z} = \begin{pmatrix}
\ket{\uparrow} & \ket{\downarrow}
\end{pmatrix}.
\end{equation}
Generalizing \cref{appeq:continuum-hamiltonian-real}, the spinful continuum Hamiltonian is
\begin{equation}\label{appeq:continuum-hamiltonian-real-spin}
\begin{split}
\tilde{H}_c &= \int d^2\br (\ket{\br}_c \otimes \ket{s_z}) \tilde{\mathcal{H}}_c(\br) (\bra{\br}_c \otimes \bra{s_z})\\
\tilde{\mathcal{H}}_c(\br) &= \begin{pmatrix}
-i\hbar v_F \bsigma \cdot \nabla \otimes \sigma_0 & 0 & \tilde{T}_+(\br)\\
0 & i\hbar v_F \bsigma^* \cdot \nabla  \otimes \sigma_0 & \tilde{T}_-(\br)\\
\tilde{T}_+^\dagger(\br) & \tilde{T}_-^\dagger(\br) & \tilde{\mathcal{H}}_-
\end{pmatrix}\\
\tilde{T}_\eta(\br) &= \sum_{j=1}^3 \tilde{T}_{\eta,\bq_j} e^{-i\eta \br \cdot \bq_j}.
\end{split}
\end{equation}
Here, the $\tilde{T}_{\eta,\bq_j}$ are $4\times (2n)$ complex matrices and $\tilde{\mathcal{H}}_-$ is the spinful Hamiltonian for the bottom layer at $\bGamma_-$. After applying perturbation theory and a change of basis, we find a spinful continuum Hamiltonian for the graphene degrees of freedom of the form
\begin{equation}\label{appeq:canonical-moire-model-spin}
\begin{split}
\tilde{H}_g &= \int d^2\br (\ket{\br} \otimes \ket{s_z}) \tilde{\mathcal{H}}(\br)(\bra{\br} \otimes \bra{s_z})\\
\tilde{\mathcal{H}}(\br) &= \begin{pmatrix}
\tilde{S}_+(\br) - i\hbar v_F \bsigma \cdot \nabla \otimes \sigma_0 & \tilde{T}(\br)\\
\tilde{T}^\dagger(\br) & \tilde{S}_-(\br) -i\hbar v_F \bsigma \cdot \nabla \otimes \sigma_0
\end{pmatrix}
\end{split}
\end{equation}
where the $\tilde{T}$ and $\tilde{S}_\eta$ potentials are given by
\begin{equation}\label{appeq:S-T-2nd-order-spin}
\begin{split}
\tilde{T}(\br) &= -\tilde{T}_+(\br) \tilde{\mathcal{H}}_-^{-1} \tilde{T}_-^\dagger(\br)(\sigma_y \otimes \sigma_0)\\
\tilde{S}_+(\br) &= -\tilde{T}_+(\br) \tilde{\mathcal{H}}_-^{-1} \tilde{T}_+^\dagger(\br)\\
\tilde{S}_-(\br) &= -(\sigma_y \otimes \sigma_0) \tilde{T}_-(\br) \tilde{\mathcal{H}}_-^{-1} \tilde{T}_-^\dagger(\br)(\sigma_y \otimes \sigma_0).
\end{split}
\end{equation}
We expand the potentials as
\begin{equation}\label{appeq:canonical-S-T-spin}
\tilde{T}(\br) = \sum_{\bq \in P_M^+} \tilde{T}_\bq e^{i\br \cdot \bq},\quad \tilde{S}_\eta(\br) = \sum_{\bq \in P_M} \tilde{S}_{\eta,\bq} e^{i\br \cdot \bq}
\end{equation}
for $4\times 4$ complex matrices $\tilde{T}_\bq$ and $\tilde{S}_{\eta,\bq}$ with $\tilde{S}^\dagger_{\eta,\bq} = \tilde{S}_{\eta,-\bq}$.

\subsection{Symmetry constraints}
In the presence of spin, we supplement \cref{appeq:symmetry-generators-graphene,appeq:symmetry-generators-transformed} with
\begin{equation}\label{appeq:symmetry-generators-spin}
\begin{split}
C_{3z} \ket{s_z} &= \ket{s_z} e^{-i(\pi/3)\sigma_z}\\
C_{2z} \ket{s_z} &= \ket{s_z} e^{-i(\pi/2)\sigma_z}\\
M_x \ket{s_z} &= \ket{s_z} e^{-i(\pi/2)\sigma_x}\\
M_y \ket{s_z} &= \ket{s_z} e^{-i(\pi/2)\sigma_y}\\
\mathcal{T} \ket{s_z} &= \ket{s_z} \sigma_y.
\end{split}
\end{equation}
\cref{appeq:symmetry-constraints-coupled-valley} then generalizes to
\begin{equation}\label{appeq:symmetry-constraints-spin}
\begin{split}
[C_{3z}, \tilde{H}_g] = 0 \iff& \tilde{T}_{R_{2\pi/3}\bq} = \left(e^{i(2\pi/3)\sigma_z} \otimes e^{i(2\pi/3)\sigma_z}\right) \tilde{T}_\bq \left(e^{-i(2\pi/3)\sigma_z} \otimes e^{-i(2\pi/3)\sigma_z}\right)\\
&\text{and } \tilde{S}_{\eta,R_{2\pi/3}\bq} = \left(e^{i(2\pi/3)\sigma_z} \otimes e^{i(2\pi/3)\sigma_z}\right) \tilde{S}_{\eta,\bq} \left(e^{-i(2\pi/3)\sigma_z} \otimes e^{-i(2\pi/3)\sigma_z}\right)\\
[C_{2z}, \tilde{H}_g] = 0 \iff& \tilde{T}_\bq = -(\sigma_z \otimes \sigma_z) \tilde{T}^\dagger_\bq (\sigma_z \otimes \sigma_z) \text{ and } \tilde{S}_{+,-\bq} = (\sigma_z \otimes \sigma_z) \tilde{S}_{-,\bq} (\sigma_z \otimes \sigma_z)\\
[M_x, \tilde{H}_g] = 0 \iff& \tilde{T}_{\mathcal{R}_\bhaty \bq} = (\sigma_y \otimes \sigma_x) \tilde{T}_\bq^\dagger (\sigma_y \otimes \sigma_x) \text{ and } \tilde{S}_{+,\mathcal{R}_\bhaty \bq} = (\sigma_y \otimes \sigma_x) \tilde{S}_{-,-\bq} (\sigma_y \otimes \sigma_x)\\
[M_y, \tilde{H}_g] = 0 \iff& \tilde{T}_{\mathcal{R}_\bhaty \bq} = -(\sigma_x \otimes \sigma_y) \tilde{T}_\bq (\sigma_x \otimes \sigma_y) \text{ and } \tilde{S}_{\eta,\mathcal{R}_\bhaty \bq} = (\sigma_x \otimes \sigma_y) \tilde{S}_{\eta,\bq} (\sigma_x \otimes \sigma_y)\\
[\mathcal{T}, \tilde{H}_g] = 0 \iff& \tilde{T}_\bq = -(\sigma_y \otimes \sigma_y) \tilde{T}^T_\bq (\sigma_y \otimes \sigma_y) \text{ and } \tilde{S}_{+,\bq} = (\sigma_y \otimes \sigma_y) \tilde{S}^*_{-,-\bq} (\sigma_y \otimes \sigma_y),
\end{split}
\end{equation}
where the newly added second Pauli matrix in each tensor product acts on spin.

As in \cref{app:symmetry-constraints}, we will always make the assumption that $C_{3z}$ and $\mathcal{T}$ symmetries are preserved. By \cref{appeq:symmetry-constraints-spin}, $\mathcal{T}$ symmetry implies
\begin{equation}\label{appeq:S-minus-from-plus-spin}
\tilde{S}_-(\br) = (\sigma_y \otimes \sigma_y) \tilde{S}^*_+(\br) (\sigma_y \otimes \sigma_y)
\end{equation}
so it suffices to characterize $\tilde{T}(\br)$ and $\tilde{S}_+(\br)$. If we expand these potentials to second order as in \cref{appeq:S-T-2nd-order-spin,appeq:canonical-S-T-spin}, we can use $C_{3z}$, $\mathcal{T}$, and Hermiticity to write

\begin{equation}\label{appeq:S-T-explicit-full-spin}
\begin{split}
\tilde{T}_{\bq_j} &= T_{\bq_j} \otimes \sigma_0 + \sigma_0 \otimes (\tilde{u}_{1,0x}\bsigma\cdot \bhatn_{\zeta_j} + \tilde{u}_{1,0y}\bsigma\cdot\bhatn_{\zeta_j+\pi/2} + \tilde{u}_{1,0z}\sigma_z)\\
\tilde{T}_{-2\bq_j} &= T_{-2\bq_j} \otimes \sigma_0 + \sigma_0 \otimes (\tilde{u}_{2,0x}\bsigma\cdot \bhatn_{\zeta_j} + \tilde{u}_{2,0y}\bsigma\cdot\bhatn_{\zeta_j+\pi/2} + \tilde{u}_{2,0z}\sigma_z)\\
\tilde{S}_{+,\bzero} &= S_{+,\bzero} \otimes \sigma_0 + \tilde{u}_{0,0z}\sigma_0 \otimes \sigma_z + \tilde{u}_{0,zz}\sigma_z \otimes \sigma_z + \tilde{u}_{0,xx}(\sigma_x \otimes \sigma_x + \sigma_y \otimes \sigma_y) + \tilde{u}_{0,xy}(\sigma_x \otimes \sigma_y - \sigma_y \otimes \sigma_x)\\
\tilde{S}_{+,\bq_{1+j}-\bq_{2+j}} &= S_{+,\bq_{1+j}-\bq_{2+j}} \otimes \sigma_0 + \sigma_0 \otimes (\tilde{u}_{\sqrt{3},0x}\bsigma\cdot \bhatn_{\zeta_j} + \tilde{u}_{\sqrt{3},0y}\bsigma\cdot\bhatn_{\zeta_j+\pi/2} + \tilde{u}_{\sqrt{3},0z}\sigma_z)\\
&+ \bsigma \cdot \bhatn_{\zeta_j} \otimes (\tilde{u}_{\sqrt{3},xx}\bsigma\cdot \bhatn_{\zeta_j} + \tilde{u}_{\sqrt{3},xy}\bsigma\cdot\bhatn_{\zeta_j+\pi/2} + \tilde{u}_{\sqrt{3},xz}\sigma_z)\\
&+ \bsigma \cdot \bhatn_{\zeta_j + \pi/2} \otimes (\tilde{u}_{\sqrt{3},yx}\bsigma\cdot \bhatn_{\zeta_j} + \tilde{u}_{\sqrt{3},yy}\bsigma\cdot\bhatn_{\zeta_j+\pi/2} + \tilde{u}_{\sqrt{3},yz}\sigma_z)\\
&+ \sigma_z \otimes (\tilde{u}_{\sqrt{3},zx}\bsigma\cdot \bhatn_{\zeta_j} + \tilde{u}_{\sqrt{3},zy}\bsigma\cdot\bhatn_{\zeta_j+\pi/2} + \tilde{u}_{\sqrt{3},zz}\sigma_z)\\
\tilde{S}_{+,\bq_{2+j}-\bq_{1+j}} &= \tilde{S}^\dagger_{+,\bq_{1+j}-\bq_{2+j}}
\end{split}
\end{equation}
where $\tilde{u}_{0,0z}$, $\tilde{u}_{0,zz}$, $\tilde{u}_{0,xx}$, and $\tilde{u}_{0,xy}$ are real parameters, all of the other $\tilde{u}$ parameters are in general complex, and where the forms for the $T_\bq$ and $S_{+,\bq}$ matrices are given in \cref{appeq:S-T-explicit-full}. We now consider the symmetry constraints in the same five cases as in \cref{app:symmetry-constraints}.

\begin{enumerate}
\item The symmetry group is generated by $C_{3z}$ and $\mathcal{T}$. In this case, we have \cref{appeq:u-constraints-C3z-T} and
\begin{equation}
\tilde{u}_{0,0z}, \tilde{u}_{0,zz}, \tilde{u}_{0,xx}, \tilde{u}_{0,xy} \in \R.
\end{equation}

\item The symmetry group is generated by $C_{3z}$, $\mathcal{T}$, and $C_{2z}$. In this case, we have \cref{appeq:u-constraints-C3z-T-C2z} and
\begin{equation}
\begin{split}
\tilde{u}_{0,zz}, \tilde{u}_{0,xx}, \tilde{u}_{0,xy}, \tilde{u}_{1,0x}, \tilde{u}_{1,0y}, \tilde{u}_{2,0x}, \tilde{u}_{2,0y}, \tilde{u}_{\sqrt{3},0x}, \tilde{u}_{\sqrt{3},0y}, \tilde{u}_{\sqrt{3},xx}, \tilde{u}_{\sqrt{3},xy}, \tilde{u}_{\sqrt{3},yx}, \tilde{u}_{\sqrt{3},yy}, \tilde{u}_{\sqrt{3},zz} &\in \R\\
\tilde{u}_{1,0z}, \tilde{u}_{2,0z}, \tilde{u}_{\sqrt{3},0z}, \tilde{u}_{\sqrt{3},xz}, \tilde{u}_{\sqrt{3},yz}, \tilde{u}_{\sqrt{3},zx}, \tilde{u}_{\sqrt{3},zy} &\in i\R\\
\tilde{u}_{0,0z} &= 0.
\end{split}
\end{equation}

\item The symmetry group is generated by $C_{3z}$, $\mathcal{T}$, and $M_x$. In this case, we have \cref{appeq:u-constraints-C3z-T-Mx} and
\begin{equation}
\begin{split}
\tilde{u}_{0,0z}, \tilde{u}_{0,zz}, \tilde{u}_{0,xy}, \tilde{u}_{1,0x}, \tilde{u}_{2,0x} &\in \R\\
\tilde{u}_{1,0y}, \tilde{u}_{1,0z}, \tilde{u}_{2,0y}, \tilde{u}_{2,0z} & \in i\R\\
\tilde{u}_{0,xx} = \tilde{u}_{\sqrt{3},0x} = \tilde{u}_{\sqrt{3},xx} = \tilde{u}_{\sqrt{3},yy} = \tilde{u}_{\sqrt{3},yz} = \tilde{u}_{\sqrt{3},zx} &= 0.
\end{split}
\end{equation}

\item The symmetry group is generated by $C_{3z}$, $\mathcal{T}$, and $M_y$. In this case, we have \cref{appeq:u-constraints-C3z-T-My} and
\begin{equation}
\begin{split}
\tilde{u}_{0,zz}, \tilde{u}_{0,xy}, \tilde{u}_{\sqrt{3},0y}, \tilde{u}_{\sqrt{3},xy}, \tilde{u}_{\sqrt{3},yx}, \tilde{u}_{\sqrt{3},yz}, \tilde{u}_{\sqrt{3},zx}, \tilde{u}_{\sqrt{3},zz} &\in \R\\
\tilde{u}_{\sqrt{3},0x}, \tilde{u}_{\sqrt{3},0z}, \tilde{u}_{\sqrt{3},xx}, \tilde{u}_{\sqrt{3},xz}, \tilde{u}_{\sqrt{3},yy}, \tilde{u}_{\sqrt{3},zy} &\in i\R\\
\tilde{u}_{0,0z} = \tilde{u}_{0,xx} = \tilde{u}_{1,0y} = \tilde{u}_{2,0y} &= 0.
\end{split}
\end{equation}

\item The symmetry group is generated by $C_{3z}$, $\mathcal{T}$, $C_{2z}$, $M_x$, and $M_y$. In this case, we have \cref{appeq:u-constraints-C3z-T-C2z-Mx-My} and
\begin{equation}\label{appeq:u-constraints-C3z-T-C2z-Mx-My-spin}
\begin{split}
\tilde{u}_{0,zz}, \tilde{u}_{0,xy}, \tilde{u}_{1,0x}, \tilde{u}_{2,0x}, \tilde{u}_{\sqrt{3},0y}, \tilde{u}_{\sqrt{3},xy}, \tilde{u}_{\sqrt{3},yx}, \tilde{u}_{\sqrt{3},zz} &\in \R\\
\tilde{u}_{1,0z}, \tilde{u}_{2,0z}, \tilde{u}_{\sqrt{3},0z}, \tilde{u}_{\sqrt{3},xz}, \tilde{u}_{\sqrt{3},zy} &\in i\R\\
\tilde{u}_{0,0z} = \tilde{u}_{0,xx} = \tilde{u}_{1,0y} = \tilde{u}_{2,0y} = \tilde{u}_{\sqrt{3},0x} = \tilde{u}_{\sqrt{3},xx} = \tilde{u}_{\sqrt{3},yy} = \tilde{u}_{\sqrt{3},yz} = \tilde{u}_{\sqrt{3},zx} &= 0.
\end{split}
\end{equation}
\end{enumerate}

\subsection{Conservation of $z$ component of spin}\label{app:spin-conservation}
In this section we consider the case in which the Hamiltonian commutes with $S_z$, the generator of spin rotations about the $z$ axis. This operator is defined by
\begin{equation}
S_z \ket{\br}_c = \ket{\br}_c,\quad S_z \ket{s_z} = \ket{s_z} \hbar \sigma_z/2
\end{equation}
and we have
\begin{equation}
[S_z, \tilde{H}_g] = 0 \iff [\tilde{T}_\bq, \sigma_0 \otimes \sigma_z] = 0 \text{ and } [\tilde{S}_{\eta,\bq}, \sigma_0 \otimes \sigma_z] = 0.
\end{equation}
When $\tilde{H}_g$ commutes with $S_z$, we can write
\begin{equation}
\tilde{H}_g = H_{g,\uparrow} \otimes \ket{\uparrow}\bra{\uparrow} + H_{g,\downarrow} \otimes \ket{\downarrow}\bra{\downarrow}
\end{equation}
where $H_{g,\uparrow}$ and $H_{g,\downarrow}$ are spinless Hamiltonians of the form in \cref{appeq:canonical-moire-model}. That is to say, we can write
\begin{equation}
\begin{split}
H_{g,s} &= \int d^2\br \ket{\br} \mathcal{H}_s(\br) \bra{\br}\\
\mathcal{H}_s(\br) &= \begin{pmatrix}
S_{s,+}(\br) - i\hbar v_F \bsigma \cdot \nabla & T_s(\br)\\
T^\dagger_s(\br) & S_{s,-}(\br) -i\hbar v_F \bsigma \cdot \nabla
\end{pmatrix}
\end{split}
\end{equation}
for $s \in \{\uparrow,\downarrow\}$, where
\begin{equation}\label{appeq:S-T-spin-decomposition}
\begin{split}
\tilde{T}(\br) &= T_\uparrow(\br) \otimes \frac{\sigma_0 + \sigma_z}{2} + T_\downarrow(\br) \otimes \frac{\sigma_0 - \sigma_z}{2}\\
\tilde{S}_\eta(\br) &= S_{\uparrow,\eta}(\br) \otimes \frac{\sigma_0 + \sigma_z}{2} + S_{\downarrow,\eta}(\br) \otimes \frac{\sigma_0 - \sigma_z}{2}.
\end{split}
\end{equation}
We expand the $T_s$ and $S_{s,\eta}$ potentials as
\begin{equation}\label{appeq:canonical-S-T-spin-decomposition}
T_s(\br) = \sum_{\bq \in P_M^+} T_{s,\bq} e^{i\br \cdot \bq},\quad S_{s,\eta}(\br) = \sum_{\bq \in P_M} S_{s,\eta,\bq} e^{i\br \cdot \bq}
\end{equation}
for $2\times 2$ complex matrices $T_{s,\bq}$ and $S_{s,\eta,\bq}$ with $S^\dagger_{s,\eta,\bq} = S_{s,\eta,-\bq}$.

Now suppose that $\tilde{H}_g$ commutes with $S_z$, $C_{3z}$, and $\mathcal{T}$. By \cref{appeq:S-minus-from-plus-spin}, $\mathcal{T}$ symmetry implies
\begin{equation}\label{appeq:S-minus-from-plus-spin-conservation}
S_{s,-}(\br) = \sigma_y S^*_{-s,+}(\br) \sigma_y
\end{equation}
so it suffices to characterize the $T_s$ and $S_{s,+}$ potentials. Additionally, \cref{appeq:S-T-explicit-full-spin} holds with
\begin{equation}
\tilde{u}_{1,0x} = \tilde{u}_{1,0y} = \tilde{u}_{2,0x} = \tilde{u}_{2,0y} = \tilde{u}_{0,xx} = \tilde{u}_{0,xy} = \tilde{u}_{\sqrt{3},0x} = \tilde{u}_{\sqrt{3},0y} = \tilde{u}_{\sqrt{3},xx} = \tilde{u}_{\sqrt{3},xy} = \tilde{u}_{\sqrt{3},zx} = \tilde{u}_{\sqrt{3},zy} = 0.
\end{equation}
We can then write
\begin{equation}\label{appeq:S-T-spin-conservation}
\begin{split}
T_{s,\bq_j} &= T_{\bq_j} + s\tilde{u}_{1,0z}\sigma_0\\
T_{s,-2\bq_j} &= T_{-2\bq_j} + s\tilde{u}_{2,0z}\sigma_0\\
S_{s,+,\bzero} &= S_{+,\bzero} + s\tilde{u}_{0,0z}\sigma_0 + s\tilde{u}_{0,zz}\sigma_z\\
S_{s,+,\bq_{1+j}-\bq_{2+j}} &= S_{+,\bq_{1+j}-\bq_{2+j}} + s\tilde{u}_{\sqrt{3},0z}\sigma_0 + s\tilde{u}_{\sqrt{3},xz}\bsigma \cdot \bhatn_{\zeta_j} + s\tilde{u}_{\sqrt{3},yz}\bsigma \cdot \bhatn_{\zeta_j + \pi/2} + s\tilde{u}_{\sqrt{3},zz}\sigma_z\\
S_{s,+,\bq_{2+j}-\bq_{1+j}} &= S^\dagger_{s,+,\bq_{1+j}-\bq_{2+j}}
\end{split}
\end{equation}
for $s \in \{\uparrow, \downarrow\}$, where we have made the identification $\uparrow = +$, $\downarrow = -$, and where the $T_\bq$ and $S_{+,\bq}$ matrices are given by \cref{appeq:S-T-explicit-full}. In the special case that $\tilde{H}_g$ commutes with $S_z$, $C_{3z}$, $\mathcal{T}$, $C_{2z}$, $M_x$, and $M_y$, \cref{appeq:u-constraints-C3z-T-C2z-Mx-My-spin,appeq:S-T-simplified} allows us to summarize this more succinctly as
\begin{equation}\label{appeq:S-T-simplified-spin}
\begin{split}
T_{s,m\bq_j} &= s\tilde{w}_{|m|,0} i\sigma_0 + w_{|m|,y} \bsigma \cdot \bhatn_{\zeta_j + \pi/2} + w_{|m|,z} i \sigma_z\\
S_{s,+,\bzero} &= w_{0,0}\sigma_0 + s\tilde{w}_{0,z} \sigma_z\\
S_{s,+,\gamma(\bq_{1+j} - \bq_{2+j})} &= (w_{\sqrt{3},0} + s\gamma\tilde{w}_{\sqrt{3},0} i)\sigma_0 + (w_{\sqrt{3},x} + s\gamma\tilde{w}_{\sqrt{3},x} i)\bsigma\cdot \bhatn_{\zeta_j} + (s\tilde{w}_{\sqrt{3},z} + \gamma w_{\sqrt{3},z} i)\sigma_z
\end{split}
\end{equation}
for $m \in \{1, -2\}$, $\gamma \in \{+, -\}$, and $s \in \{\uparrow,\downarrow\}$, where the real $\tilde{w}$ parameters are defined by
\begin{equation}
\tilde{w}_{1,0} = -i\tilde{u}_{1,z},\quad \tilde{w}_{2,0} = -i\tilde{u}_{2,z},\quad \tilde{w}_{0,z} = \tilde{u}_{0,zz},\quad \tilde{w}_{\sqrt{3},0} = -i\tilde{u}_{\sqrt{3},0z},\quad \tilde{w}_{\sqrt{3},x} = -i\tilde{u}_{\sqrt{3},xz},\quad \tilde{w}_{\sqrt{3},z} = \tilde{u}_{\sqrt{3},zz}.
\end{equation}

\subsection{Transformation to the moir\'e coordinate system}
Since the model for spin $s \in \{\uparrow,\downarrow\}$ described in \cref{app:spin-conservation} is of the form described in \cref{app:two-Dirac-cone}, we can transform it into the moir\'e coordinate system defined in \cref{app:moire-coordinate-system} just as we did in \cref{app:moire-coordinate-translation-spinless}. \cref{appeq:S-minus-from-plus-spin-conservation,appeq:S-T-spin-conservation} imply
\begin{equation}
S_{\xi,s,-}(\br) = \sigma_y S^*_{\xi,-s,+}(\br) \sigma_y
\end{equation}
and
\begin{equation}
\begin{split}
T_{\xi, s,\bq_j} &= T_{\xi,\bq_j} + s\tilde{u}_{1,0z}\sigma_0\\
T_{\xi, s,-2\bq_j} &= T_{-2\bq_j} + s\tilde{u}_{2,0z}\sigma_0\\
S_{\xi, s,+,\bzero} &= S_{\xi,+,\bzero} + s\tilde{u}_{0,0z}\sigma_0 + s\tilde{u}_{0,zz}\sigma_z\\
S_{\xi,s,+,\bq_{1+j}-\bq_{2+j}} &= S_{\xi,+,\bq_{1+j}-\bq_{2+j}} + s\tilde{u}_{\sqrt{3},0z}\sigma_0 + s\tilde{u}_{\sqrt{3},xz}\bsigma \cdot \bhatn_{\zeta_j-\xi} + s\tilde{u}_{\sqrt{3},yz}\bsigma \cdot \bhatn_{\zeta_j + \pi/2-\xi} + s\tilde{u}_{\sqrt{3},zz}\sigma_z\\
S_{\xi,s,+,\bq_{2+j}-\bq_{1+j}} &= S^\dagger_{\xi,s,+,\bq_{1+j}-\bq_{2+j}}
\end{split}
\end{equation}
where the $T_\bq$ and $S_{+,\bq}$ matrices are given by \cref{appeq:S-T-explicit-full-moire-coordinates}. Additionally, \cref{appeq:S-T-simplified-spin} implies
\begin{equation}
\begin{split}
T_{\xi,s,m\bq_j} &= s\tilde{w}_{|m|,0} i\sigma_0 + w_{|m|,y} \bsigma \cdot \bhatn_{\zeta_j + \phi_1} + w_{|m|,z} i \sigma_z\\
S_{\xi,s,+,\bzero} &= w_{0,0}\sigma_0 + s\tilde{w}_{0,z} \sigma_z\\
S_{\xi,s,+,\gamma(\bq_{1+j} - \bq_{2+j})} &= (w_{\sqrt{3},0} + s\gamma\tilde{w}_{\sqrt{3},0} i)\sigma_0 + (w_{\sqrt{3},x} + s\gamma\tilde{w}_{\sqrt{3},x} i)\bsigma\cdot \bhatn_{\zeta_j + \phi_1 - \pi/2} + (s\tilde{w}_{\sqrt{3},z} + \gamma w_{\sqrt{3},z} i)\sigma_z.
\end{split}
\end{equation}
for $m \in \{1, -2\}$, $\gamma \in \{+, -\}$, and $s \in \{\uparrow,\downarrow\}$, where $\phi_1$ is defined by \cref{appeq:define-phi_1}.

\subsection{Coirrep decomposition for the bottom layer}
In this section, we consider the effect of spin on the corepresentation theory analysis in \cref{app:coirrep-decomposition}. We begin by generalizing \cref{appeq:symmetry-generators-bilayer} to
\begin{equation}
\begin{split}
C_{3z} (\ket{\br}_g \otimes \ket{s_z}) &= \ket{R_{2\pi/3}\br}_g (e^{i(2\pi/3)\sigma_z \otimes \sigma_z} \otimes e^{-i(\pi/3)\sigma_z}) \oplus \tilde{C}_{3z,-}\\
C_{2z} (\ket{\br}_g \otimes \ket{s_z}) &= \ket{-\br}_g (\sigma_x \otimes \sigma_x \otimes e^{-i(\pi/2)\sigma_z}) \oplus \tilde{C}_{2z,-}\\
M_x (\ket{\br}_g \otimes \ket{s_z}) &= \ket{\mathcal{R}_\bhatx \br}_g (\sigma_x \otimes \sigma_0 \otimes e^{-i(\pi/2)\sigma_x}) \oplus \tilde{M}_{x,-}\\
M_y (\ket{\br}_g \otimes \ket{s_z}) &= \ket{\mathcal{R}_\bhaty \br}_g (\sigma_0 \otimes \sigma_x \otimes e^{-i(\pi/2)\sigma_y}) \oplus \tilde{M}_{y,-}\\
\mathcal{T} (\ket{\br}_g \otimes \ket{s_z}) &= \ket{\br}_g (\sigma_x \otimes \sigma_0 \otimes \sigma_y) \oplus \tilde{\mathcal{T}}_-
\end{split}
\end{equation}
for some unitary matrices $\tilde{C}_{3z,-}$, $\tilde{C}_{2z,-}$, $\tilde{M}_{x,-}$, $\tilde{M}_{y,-}$, and $\tilde{\mathcal{T}}_-$ which act on the bottom layer including spin. \cref{appeq:symmetry-constraints-bilayer} then generalizes to
\begin{equation}\label{appeq:symmetry-constraints-bilayer-spin}
\begin{split}
[C_{3z}, \tilde{H}_c] = 0 \iff& \tilde{T}_{\eta,\bq_{j+1}} = (e^{i\eta(2\pi/3)\sigma_z} \otimes e^{-i(\pi/3)\sigma_z}) \tilde{T}_{\eta,\bq_j} \tilde{C}_{3z,-}^\dagger\\
[C_{2z}, \tilde{H}_c] = 0 \iff& \tilde{T}_{\eta,\bq_j} = (\sigma_x \otimes e^{-i(\pi/2)\sigma_z}) \tilde{T}_{-\eta,\bq_j} \tilde{C}_{2z,-}^\dagger\\
[M_x, \tilde{H}_c] = 0 \iff& \tilde{T}_{\eta,\bq_1} = (\sigma_0 \otimes e^{-i(\pi/2)\sigma_x}) \tilde{T}_{-\eta,\bq_1} \tilde{M}_{x,-}^\dagger\\
&\tilde{T}_{\eta,\bq_2} = (\sigma_0 \otimes e^{-i(\pi/2)\sigma_x}) \tilde{T}_{-\eta,\bq_3} \tilde{M}_{x,-}\\
&\tilde{T}_{\eta,\bq_3} = (\sigma_0 \otimes e^{-i(\pi/2)\sigma_x}) \tilde{T}_{-\eta,\bq_2} \tilde{M}_{x,-}^\dagger\\
[M_y, \tilde{H}_c] = 0 \iff& \tilde{T}_{\eta,\bq_1} = (\sigma_x \otimes e^{-i(\pi/2)\sigma_y}) \tilde{T}_{\eta,\bq_1} \tilde{M}_{y,-}^\dagger\\
&\tilde{T}_{\eta,\bq_2} = (\sigma_x \otimes e^{-i(\pi/2)\sigma_y}) \tilde{T}_{\eta,\bq_3} \tilde{M}_{y,-}^\dagger\\
&\tilde{T}_{\eta,\bq_3} = (\sigma_x \otimes e^{-i(\pi/2)\sigma_y}) \tilde{T}_{\eta,\bq_2} \tilde{M}_{y,-}^\dagger\\
[\mathcal{T}, \tilde{H}_c] = 0 \iff& \tilde{T}_{\eta,\bq_j} =  (\sigma_0 \otimes \sigma_y) \tilde{T}_{-\eta,\bq_j}^* \tilde{\mathcal{T}}_-^\dagger.
\end{split}
\end{equation}
As in \cref{app:coirrep-decomposition}, in the presence of $C_{3z}$ and $\mathcal{T}$ symmetries, all of the $\tilde{T}_{\eta,\bq_j}$ matrices are determined by $\tilde{T}_{+,\bq_1}$ through the equations
\begin{equation}
\begin{split}
\tilde{T}_{\eta,\bq_j} &= (e^{i\eta \zeta_j \sigma_z} \otimes e^{-i(\zeta_j/2)\sigma_z}) \tilde{T}_{\eta,\bq_1} \tilde{C}_{3z,-}^{1-j}\\
\tilde{T}_{-,\bq_1} &= (\sigma_0 \otimes \sigma_y) \tilde{T}^*_{+,\bq_1} \tilde{\mathcal{T}}_-^\dagger.
\end{split}
\end{equation}
$C_{2z}$, $M_x$, and $M_y$ symmetries imply the constraints
\begin{align}
[C_{2z}, \tilde{H}_c] = 0 &\implies \tilde{T}_{+,\bq_1} = -(\sigma_x \otimes \sigma_x)\tilde{T}^*_{+,\bq_1} (\tilde{C}_{2z,-}\tilde{\mathcal{T}}_-)^\dagger\\
[M_x, \tilde{H}_c] = 0 &\implies \tilde{T}_{+,\bq_1} = (\sigma_0 \otimes \sigma_z) \tilde{T}^*_{+,\bq_1} (\tilde{M}_{x,-}\tilde{\mathcal{T}}_-)^\dagger\\
[M_y, \tilde{H}_c] = 0 &\implies \tilde{T}_{+,\bq_1} = -i(\sigma_x \otimes \sigma_y)\tilde{T}_{+,\bq_1} \tilde{M}_{y,-}^\dagger.
\end{align}
By \cref{appeq:continuum-hamiltonian-real-spin,appeq:S-T-2nd-order-spin}, we have
\begin{align}
\tilde{T}(\br) &= -\sum_{j,k=1}^3 \tilde{T}_{+,\bq_j} \tilde{\mathcal{H}}_-^{-1} \tilde{T}_{-,\bq_k}^\dagger (\sigma_y \otimes \sigma_0) e^{-i\br \cdot (\bq_k + \bq_j)}\\
&= -\sum_{j,k=1}^3 \left(e^{i\zeta_j \sigma_z} \otimes e^{-i(\zeta_j/2)\sigma_z}\right) \tilde{T}_{+,\bq_1} \tilde{C}_{3z,-}^{1-j}\tilde{\mathcal{H}}_-^{-1} \tilde{C}_{3z,-}^{k-1} \tilde{\mathcal{T}}_- \tilde{T}_{+,\bq_1}^T (\sigma_y \otimes \sigma_y) \left(e^{-i\zeta_k \sigma_z} \otimes e^{i(\zeta_k/2)\sigma_z}\right) e^{-i\br \cdot (\bq_k + \bq_j)}\label{appeq:T-expansion-spin}\\
\tilde{S}_+(\br) &= -\sum_{j,k=1}^3 \tilde{T}_{+,\bq_j} \tilde{\mathcal{H}}_-^{-1} \tilde{T}_{+,\bq_k}^\dagger e^{i\br \cdot (\bq_k - \bq_j)}\\
&= -\sum_{j,k=1}^3 \left(e^{i\zeta_j \sigma_z} \otimes e^{-i(\zeta_j/2)\sigma_z}\right) \tilde{T}_{+,\bq_1} \tilde{C}_{3z,-}^{1-j}\tilde{\mathcal{H}}_-^{-1} \tilde{C}_{3z,-}^{k-1} \tilde{T}_{+,\bq_1}^\dagger \left(e^{-i\zeta_k \sigma_z} \otimes e^{i(\zeta_k/2)\sigma_z}\right) e^{i\br \cdot (\bq_k - \bq_j)}.\label{appeq:S-expansion-spin}
\end{align}

By the same argument as in \cref{app:coirrep-decomposition}, it is possible to use \cref{appeq:T-expansion-spin,appeq:S-expansion-spin} to write $\tilde{T}(\br)$ and $\tilde{S}^+(\br)$ as a sum over the coirreps present on the bottom layer at $\bGamma_-$. Note that if we use $\mathcal{T}$ symmetry twice, we find
\begin{equation}
\tilde{T}_{+,\bq_1} = (\sigma_0 \otimes \sigma_y)((\sigma_0 \otimes \sigma_y)\tilde{T}_{+,\bq_1}^* \tilde{\mathcal{T}}_-^\dagger)^* \tilde{\mathcal{T}}_-^\dagger = \tilde{T}_{+,\bq_1}(-\tilde{\mathcal{T}}_- \tilde{\mathcal{T}}_-^*)^\dagger
\end{equation}
so that either $\tilde{\mathcal{T}}_- \tilde{\mathcal{T}}_-^* = -1$ or $\tilde{T}_{+,\bq_1} = 0$. As a result, it suffices to consider only the spinful coirreps in \cref{apptbl:characters}. We could write a straightforward adaptation of \cref{app:symmetry-group-C3z-T,app:symmetry-group-C3z-T-C2z,app:symmetry-group-C3z-T-Mx,app:symmetry-group-C3z-T-My,app:symmetry-group-C3z-T-C2z-Mx-My} to the spinful case, considering coirreps $\overline{EE}$ and ${^1\overline{E}}{^2\overline{E}}$ for $31'$, coirreps ${^2\overline{E}_1}{^1\overline{E}_1}$, ${^1\overline{E}_3}{^2\overline{E}_3}$, and ${^1\overline{E}_2}{^2\overline{E}_2}$ for $61'$, coirreps ${^2\overline{E}}{^1\overline{E}}$ and $\overline{E}_1$ for $3m1'$, and coirreps $\overline{E}_3$, $\overline{E}_2$, and $\overline{E}_1$ for $6mm1'$. Such an analysis would cover all possibilities including spin-orbit coupling in both the bottom layer and in the interlayer couplings. However, the formulas would contain sixteen complex parameters in the case of the spinful point group $31'$ and would not be particularly illuminating.

For these reasons, we focus instead on the more physically realistic case of weak spin-orbit coupling. Specifically, we make the approximation that spin-orbit coupling affects only the bottom layer Hamiltonian and not the interlayer coupling. We take
\begin{equation}\label{appeq:weak-soc}
\begin{split}
\tilde{T}_\eta(\br) &= T_\eta(\br) \otimes \sigma_0\\
\tilde{\mathcal{H}}_- &= \mathcal{H}_- \otimes \sigma_0 + \mathcal{H}_{SO,-}\\
\tilde{C}_{3z,-} &= C_{3z,-} \otimes e^{-i(\pi/3)\sigma_z}\\
\tilde{C}_{2z,-} &= C_{2z,-} \otimes e^{-i(\pi/2)\sigma_z}\\
\tilde{M}_{x,-} &= M_{x,-} \otimes e^{-i(\pi/2)\sigma_x}\\
\tilde{M}_{y,-} &= M_{y,-} \otimes e^{-i(\pi/2)\sigma_y}\\
\tilde{\mathcal{T}}_- &= \mathcal{T}_- \otimes \sigma_y
\end{split}
\end{equation}
where $\mathcal{H}_{SO,-}$ is the spin-orbit coupling Hamiltonian at $\bGamma_-$, which we assume is independent of momentum. As in \cref{app:coirrep-decomposition}, we block diagonalize the spinless part of the corepresentation (i.e., $C_{3z,-}$, $C_{2z,-}$, $M_{x,-}$, $M_{y,-}$, $\mathcal{T}_-$) along with $\mathcal{H}_-$. Since $\tilde{T}(\br)$ and $\tilde{S}_+(\br)$ in \cref{appeq:T-expansion-spin,appeq:S-expansion-spin} are additive over these spinless coirreps, we now consider each spinless coirrep separately.

We first consider the case of a one-dimensional spinless coirrep. In this case, the full corepresention in \cref{appeq:weak-soc} is two-dimensional. Kramers' theorem implies that spinful corepresentations cannot be one-dimensional, so the corepresentation must be irreducible. It follows that $\tilde{\mathcal{H}}_0 = E_0 \sigma_0$ and $\mathcal{H}_{SO,-} = 0$. \cref{appeq:T-expansion-spin,appeq:S-expansion-spin} imply
\begin{equation}
\begin{split}
\tilde{T}(\br) &= T(\br) \otimes \sigma_0\\
\tilde{S}_+(\br) &= S_+(\br) \otimes \sigma_0
\end{split}
\end{equation}
where $T(\br)$ and $S_+(\br)$ are given by \cref{appeq:T-expansion,appeq:S-expansion} with $\mathcal{H}_- = E_0$.

Next, suppose that the spinless coirrep is ${^2E}{^1E}$ for spinless point group $31'_0$. The full corepresentation in \cref{appeq:weak-soc} is then ${^2E}{^1E} \otimes {^1\overline{E}}{^2\overline{E}}$, which decomposes as
\begin{equation}
{^2E}{^1E} \otimes {^1\overline{E}}{^2\overline{E}} \cong \overline{EE} \oplus {^1\overline{E}}{^2\overline{E}}.
\end{equation}
Furthermore, if we take $C_{3z,-} = e^{-i(2\pi/3)\sigma_z}$ as in \cref{app:symmetry-group-C3z-T}, then
\begin{equation}\label{appeq:C3z-minus-spinful}
\tilde{C}_{3z,-} = e^{-i(2\pi/3)\sigma_z} \otimes e^{-i(\pi/3)\sigma_z} = \begin{pmatrix}
-1 & 0 & 0 & 0\\
0 & e^{-i(\pi/3)} & 0 & 0\\
0 & 0 & e^{i(\pi/3)} & 0\\
0 & 0 & 0 & -1
\end{pmatrix},
\end{equation}
which makes the coirrep decomposition explicit. It follows that we can write
\begin{equation}
\tilde{\mathcal{H}}_-^{-1} = \frac{1}{E_0}\sigma_0 \otimes \sigma_0 + \frac{\lambda}{E_0} \sigma_z \otimes \sigma_z
\end{equation}
where $E_0$ is the harmonic mean of the energies at $\bGamma_-$ and $\lambda$ is a real dimensionless parameter that controls the spin-orbit coupling strength. In the case of weak spin-orbit coupling, we will have $|\lambda| \ll 1$ and note that $\lambda = 0$ corresponds to vanishing spin-orbit coupling.

\cref{appeq:T-expansion-spin,appeq:S-expansion-spin} imply
\begin{equation}\label{appeq:S-T-weak-soc}
\begin{split}
\tilde{T}(\br) &= T(\br) \otimes \sigma_0 + \lambda T_{SO}(\br) \otimes \sigma_z\\
\tilde{S}_+(\br) &= S_+(\br) \otimes \sigma_0 + \lambda S_{SO,+}(\br) \otimes \sigma_z
\end{split}
\end{equation}
where $T(\br)$ and $S_+(\br)$ are given by \cref{appeq:T-expansion,appeq:S-expansion} with $\mathcal{H}_- = E_0 \sigma_0$ while $T_{SO}(\br)$ and $S_{SO,+}(\br)$ are given by \cref{appeq:T-expansion,appeq:S-expansion} with $\mathcal{H}_- = E_0 \sigma_z$. It is clear from \cref{appeq:S-T-weak-soc} that $[S_z, \tilde{H}_g] = 0$, which is the case that was discussed in \cref{app:spin-conservation}. If we parameterize $T_{+,\bq_1}$ as in \cref{appeq:T1-dag-2D} then $\tilde{T}(\br)$ and $\tilde{S}_+(\br)$ are given by \cref{appeq:S-T-spin-decomposition,appeq:canonical-S-T-spin-decomposition,appeq:S-T-spin-conservation} where the $T_\bq$ and $S_{+,\bq}$ matrices are given in terms of the $v_0$, $v_x$, $v_y$, and $v_z$ parameters by \cref{appeq:S-T-explicit-full,appeq:u-from-v-2D}, and additionally
\begin{equation}
\begin{split}
\tilde{u}_{1,0z} &= \frac{i\lambda}{E_0}(v_0^2 + 2v_x^2 + 2v_y^2 - v_z^2),\quad \tilde{u}_{2,0z} = -\frac{i\lambda}{E_0}(v_0^2 - v_x^2 - v_y^2 - v_z^2)\\
\tilde{u}_{0,0z} &= -\frac{3\lambda}{E_0}(v_0 v_z^* + v_0^* v_z - i(v_x v_y^* - v_x^* v_y)),\quad \tilde{u}_{0,zz} = -\frac{3\lambda}{E_0}(|v_0|^2 - |v_x|^2 - |v_y|^2 + |v_z|^2)\\
\tilde{u}_{\sqrt{3},0z} &= \frac{\lambda}{2E_0}(v_0 v_z^* + v_0^* v_z + 2i(v_x v_y^* - v_x^* v_y) + i\sqrt{3}(|v_0|^2 + |v_z|^2))\\ \tilde{u}_{\sqrt{3},xz} &= \frac{\lambda}{2E_0}(v_x v_z^* + v_x^* v_z -i(v_0 v_y^* - v_0^* v_y) +\sqrt{3}(v_y v_z^* - v_y^* v_z) -i\sqrt{3}(v_0 v_x^* + v_0^* v_x))\\
\tilde{u}_{\sqrt{3},yz} &= \frac{\lambda}{2E_0}(v_y v_z^* + v_y^* v_z + i(v_0 v_x^* - v_0^* v_x) -\sqrt{3}(v_x v_z^* - v_x^* v_z) -i\sqrt{3}(v_0 v_y^* + v_0^* v_y))\\
\tilde{u}_{\sqrt{3},zz} &= \frac{\lambda}{2E_0}(|v_0|^2 + 2|v_x|^2 + 2|v_y|^2 + |v_z|^2 + i\sqrt{3}(v_0 v_z^* + v_0^* v_z)).
\end{split}
\end{equation}

Finally, we consider the other possible spinless coirreps. The spinless coirrep could be ${^2E_1}{^1E_1}$ or ${^2E_2}{^1E_2}$ for $61'_0$, $E$ for $3m1'$, or $E_2$ or $E_1$ for $6mm1'$. However, the conventions we chose in \cref{app:symmetry-group-C3z-T-C2z,app:symmetry-group-C3z-T-Mx,app:symmetry-group-C3z-T-My,app:symmetry-group-C3z-T-C2z-Mx-My} imply that \cref{appeq:C3z-minus-spinful} is satisfied in each case, and in fact the above calculation applies for each coirrep. The distinctions are only in the constraints placed on the $v_0$, $v_x$, $v_y$, and $v_z$ parameters, which are listed for each coirrep in \cref{app:symmetry-group-C3z-T-C2z,app:symmetry-group-C3z-T-Mx,app:symmetry-group-C3z-T-My,app:symmetry-group-C3z-T-C2z-Mx-My}.

For the case of spinless point group $6mm1'$, \cref{appeq:S-T-spin-conservation} simplifies to \cref{appeq:S-T-simplified-spin} where the $w$ parameters are given by \cref{appeq:w-parameters-6mm1'} and
\begin{equation}
\begin{split}
\tilde{w}_{1,0} &= \frac{\lambda}{E_0}(v_0^2 + 2v_x^2),\quad \tilde{w}_{2,0} = -\frac{\lambda}{E_0}(v_0^2 - v_x^2),\quad \tilde{w}_{0,z} = -\frac{3\lambda}{E_0}(|v_0|^2 - |v_x|^2)\\
\tilde{w}_{\sqrt{3},0} &= \frac{\lambda\sqrt{3}}{2E_0}|v_0|^2,\quad \tilde{w}_{\sqrt{3},x} = -\frac{\lambda\sqrt{3}}{2E_0}(v_0 v_x^* + v_0^* v_x),\quad \tilde{w}_{\sqrt{3},z} = \frac{\lambda}{2E_0}(|v_0|^2 + 2|v_x|^2).
\end{split}
\end{equation}
Additionally, $v_0, v_x \in \R$ for coirrep $E_2$ and $v_0, v_x \in i\R$ for coirrep $E_1$.

\section{Opposite-velocity model}\label{app:opposite-velocity-model}
We now consider the construction of \cref{app:I-pm-Dirac} specifically in the case of the type I$+$ configuration with $\namezero$. In this case, we have $N_- = 1$ and we can take $\xi_0 = 0$ so that \cref{appeq:two-Dirac-q_j} implies
\begin{equation}
\bq_j = (e^{-\epsilon}R_\theta-1)R_{\zeta_j}\bK_+.
\end{equation}
For simplicity, we ignore spin in this section. We do not need to assume that either layer is graphene, though we do require both layers to have triangular Bravais lattices and Dirac cones at their $\bK$ and $-\bK$ points centered at their Fermi energies. Furthermore, we require that the symmetries $C_{3z}$ (rotation by angle $2\pi/3$ about $\bhatz$), $C_{2z}$ (rotation by angle $\pi$ about $\bhatz$), $M_x$ (reflection through the $yz$ plane), $M_y$ (reflection through the $xz$ plane), and $\mathcal{T}$ (antiunitary time-reversal with $\mathcal{T}^2 = 1$) take the same form on the Dirac cones as they do in graphene. As in \cref{app:I-pm-Dirac}, we focus on the valley containing $\bK_+$ and $\bK_-$, which is preserved by $C_{3z}$, $C_{2z}\mathcal{T}$, and $M_y$. Using the states $\ket{\br}$ in \cref{eq:define-r-states-I-pm-Dirac}, we then have
\begin{equation}\label{eq:symmetry-generators-one-valley}
\begin{split}
C_{3z}\ket{\br} &= \ket{R_{2\pi/3}\br} \sigma_0 \otimes e^{i(2\pi/3)\sigma_z}\\
C_{2z}\mathcal{T}\ket{\br} &= \ket{-\br} \sigma_0 \otimes \sigma_x\\
M_y\ket{\br} &= \ket{\mathcal{R}_\bhaty \br} \sigma_0 \otimes \sigma_x.
\end{split}
\end{equation}
Here, the first (second) Pauli matrix in each tensor product acts on layer (sublattice).

We now take a unitary change of basis
\begin{equation}
\ket{\br}_o = \ket{\br} \begin{pmatrix}
\sigma_0 & 0\\
0 & i\sigma_z
\end{pmatrix}
\end{equation}
under which the Hamiltonian in \cref{appeq:two-Dirac-hamiltonian} becomes
\begin{equation}
\begin{split}
H &= \int d^2\br \ket{\br}_o \tilde{\mathcal{H}}(\br) \bra{\br}_o\\
\tilde{\mathcal{H}}(\br) &= \begin{pmatrix}
\tilde{S}_+(\br) - i\hbar v_+ \bsigma \cdot \nabla & \tilde{T}(\br)\\
\tilde{T}^\dagger(\br) & \tilde{S}_-(\br) + i\hbar v_- \bsigma \cdot \nabla
\end{pmatrix}
\end{split}
\end{equation}
where
\begin{equation}
\tilde{S}_+(\br) = S_+(\br),\quad \tilde{S}_-(\br) = \sigma_z S_-(\br) \sigma_z,\quad \tilde{T}(\br) = iT(\br)\sigma_z.
\end{equation}
Additionally, \cref{eq:symmetry-generators-one-valley} implies
\begin{equation}
\begin{split}
C_{3z}\ket{\br}_o &= \ket{R_{2\pi/3}\br}_o \sigma_0 \otimes e^{i(2\pi/3)\sigma_z}\\
C_{2z}\mathcal{T}\ket{\br}_o &= \ket{-\br}_o \sigma_0 \otimes \sigma_x\\
M_y\ket{\br}_o &= \ket{\mathcal{R}_\bhaty \br}_o \sigma_z \otimes \sigma_x.
\end{split}
\end{equation}
Interestingly, this form for the $C_{3z}$, $C_{2z}\mathcal{T}$, and $M_y$ symmetries is equivalent to that implied by \cref{appeq:symmetry-generators-transformed} for the coupled-valley graphene model. With this motivation, we define the emergent $C_{2z}$ operator
\begin{equation}
\tilde{C}_{2z} = -\ket{-\br}\sigma_y \otimes \sigma_z.
\end{equation}

We expand the potentials as
\begin{equation}
\tilde{T}(\br) = \sum_{\bq \in P_M^+} \tilde{T}_{\bq} e^{i\br \cdot \bq},\quad \tilde{S}_l(\br) = \sum_{\bq \in P_M} \tilde{S}_{l,\bq} e^{i\br \cdot \bq}
\end{equation}
for $2\times 2$ complex matrices $T_{\bq}$ and $\tilde{S}_{l,\bq}$ with $\tilde{S}^\dagger_{l,\bq} = \tilde{S}_{l,-\bq}$. Applying the arguments of \cref{app:symmetry-constraints}, we find
\begin{equation}\label{appeq:symmetry-constraints-opposite-velocity}
\begin{split}
[C_{3z}, H] = 0 &\iff \tilde{T}_{R_{2\pi/3}\bq} = e^{i(2\pi/3)\sigma_z} \tilde{T}_\bq e^{-i(2\pi/3)\sigma_z} \text{ and } \tilde{S}_{l,R_{2\pi/3}\bq} = e^{i(2\pi/3)\sigma_z} \tilde{S}_{l,\bq} e^{-i(2\pi/3)\sigma_z}\\
[C_{2z}\mathcal{T}, H] = 0 &\iff \tilde{T}_\bq = \sigma_x \tilde{T}^*_\bq \sigma_x \text{ and } \tilde{S}_{l,\bq} = \sigma_x \tilde{S}^*_{l,\bq} \sigma_x\\
[M_y, H] = 0 &\iff \tilde{T}_{\mathcal{R}_\bhaty \bq} = -\sigma_x \tilde{T}_\bq \sigma_x \text{ and } \tilde{S}_{l,\mathcal{R}_\bhaty \bq} = \sigma_x \tilde{S}_{l,\bq} \sigma_x\\
[\tilde{C}_{2z}, H] = 0 &\iff \tilde{T}_\bq = -\sigma_z \tilde{T}^\dagger_\bq \sigma_z \text{, } \tilde{S}_{+,-\bq} = \sigma_z \tilde{S}_{-,\bq} \sigma_z \text{, and } v_- = -v_+.
\end{split}
\end{equation}
To lowest order, we only need to keep terms of the form
\begin{equation}
\tilde{T}_{\bq_j},\quad \tilde{S}_{l,\bzero}.
\end{equation}
Using $C_{3z}$, $C_{2z}\mathcal{T}$, and Hermiticity, we can then write
\begin{equation}
\begin{split}
\tilde{T}_{\bq_j} &= \tilde{w}_0\sigma_0 + \tilde{w}_1\bsigma \cdot \bhatn_{\zeta_j} + w_1 \bsigma \cdot \bhatn_{\zeta_j + \pi/2} + i w_0\sigma_z\\
\tilde{S}_{l,\bzero} &= \left(E_F + \frac{lE_\Delta}{2} \right) \sigma_0
\end{split}
\end{equation}
for real parameters $w_0$, $w_1$, $\tilde{w}_0$, $\tilde{w}_1$, $E_F$ and $E_\Delta$. If $M_y$ is also preserved then additionally $\tilde{w}_0 = \tilde{w}_1 = 0$. If we additionally have $E_\Delta = 0$ and $v_- = -v_+$ then $[\tilde{C}_{2z}, H] = 0$.

Finally, we transform to the moir\'e coordinate system as described in \cref{app:moire-coordinate-system}. The $\tilde{T}_{\bq_j}$ and $\tilde{S}_{l,\bzero}$ matrices are then replaced by
\begin{equation}
\begin{split}
\tilde{T}_{\xi,\bq_j} &= \tilde{w}_0\sigma_0 + \tilde{w}_1\bsigma \cdot \bhatn_{\zeta_j + \phi_1 - \pi/2} + w_1 \bsigma \cdot \bhatn_{\zeta_j + \phi_1} + i w_0\sigma_z\\
\tilde{S}_{\xi,l,\bzero} &= \left(E_F + \frac{lE_\Delta}{2} \right) \sigma_0
\end{split}
\end{equation}
where
\begin{equation}
\phi_1 = \pi/2 - \xi = \arg\left(1-e^{-\delta\epsilon - i\delta\theta}\right)
\end{equation}
since $\xi_0 = 0$.

\section{Magnetic point groups}\label{app:point-groups}
The spinful magnetic point group $6mm1'$ is generated by three elements $C_{6z}$, $M_y$, and $\mathcal{T}$ which satisfy the relations
\begin{equation}
\mathcal{T}^4 = 1,\quad C_{6z}^6 = M_y^2 = (C_{6z} M_y)^2 = \mathcal{T}^2,\quad C_{6z} \mathcal{T} = \mathcal{T} C_{6z},\quad M_y \mathcal{T} = \mathcal{T} M_y.
\end{equation}
We further define group elements
\begin{equation}
-1 = \mathcal{T}^2,\quad C_{3z} = C_{6z}^2,\quad C_{2z} = C_{6z}^3,\quad M_x = C_{2z}M_y^{-1}
\end{equation}
and subgroups
\begin{align}
31' &= \braket{C_{3z}, \mathcal{T}}\\
61' &= \braket{C_{6z}, \mathcal{T}}\\
mm21' &= \braket{M_x, M_y, \mathcal{T}}\\
3m1' &= \braket{C_{3z}, M_x, \mathcal{T}} \text{ or } \braket{C_{3z}, M_y, \mathcal{T}}\label{appeq:define-3m1'}
\end{align}
where $\braket{x_1, \dots , x_k}$ indicates the subgroup generated by elements $x_1, \dots, x_k$. Note that \cref{appeq:define-3m1'} gives two expressions for $3m1'$ which are different subgroups of $6mm1'$ but are isomorphic. Information about these groups is available on the Bilbao Crystallographic Server \cite{Aroyo2006,Aroyo2006a,Gallego2012,Elcoro2021,Xu2020}.

For each of these spinful magnetic point groups, we define a corresponding spinless magnetic point group produced by taking the quotient with the normal subgroup $\{1, -1\}$. Equivalently, we replace the condition $\mathcal{T}^4 = 1$ by $\mathcal{T}^2 = 1$. We use a subscript $0$ to differentiate the spinless groups from the spinful groups, so that we have defined $31'_0$, $61'_0$, $mm21'_0$, $3m1'_0$, and $6mm1'_0$.

The character tables for the spinful magnetic point groups, enumerating the coirreps in which $\mathcal{T}$ is represented by an antilinear operator, are given in \cref{apptbl:characters}. For the case of $3m1'$, we use $M$ in place of $M_x$ or $M_y$ to emphasize that both definitions yield the same character table. In each coirrep, the element $-1$ is either represented by $1$ or by $-1$. Coirreps in which $-1$ is represented by $1$ are also coirreps of the corresponding spinless magnetic point group. As a result, \cref{apptbl:characters} also effectively contains the character tables for the spinless magnetic point groups. Additionally, we say that a corepresentation is spinless (spinful) if it represents $-1$ by $1$ ($-1$).

\section{Parameter choices for the phase diagram of low energy bands}\label{app:phase-diagram-parameters}
In this section, we justify the parameter choices that were made to simplify the phase diagram of low energy bands in \cref{fig:sqrt3-phases}\textbf{(b)}. We consider the Hamiltonian in \cref{eq:coupled-valley-hamiltonian} in the case that the moir\'e potentials arise completely from a single 2D spinless coirrep. We first note that the Hamiltonian is invariant under $\varphi \mapsto \varphi + \pi$ so that for the purpose of computing a phase diagram we can take $\varphi \in [0, \pi)$. Similarly, the transformation $v_0 \mapsto iv_0$, $v_x \mapsto iv_x$ is equivalent to conjugation by $\sigma_z \otimes \sigma_0$ so we can take $v_0$ and $v_x$ to be real. Additionally, $\phi_1 \mapsto \phi_1 + \pi$ is equivalent to $\varphi \mapsto -\varphi$. Since we assume $\delta\theta = 0$ for the phase diagram, we can take $\delta\epsilon < 0$ and $\phi_1 = 0$. Finally, $E_0 \mapsto -E_0$ is equivalent to $\varphi \mapsto -\varphi$, conjugation by $\sigma_0 \otimes \sigma_z$, and an overall sign change. As a result, we are free to choose $E_0 < 0$.

\section{Kagome and honeycomb tight-binding models}\label{app:kagome-honeycomb-models}
\begin{figure}[h]
	\centering
	\includegraphics{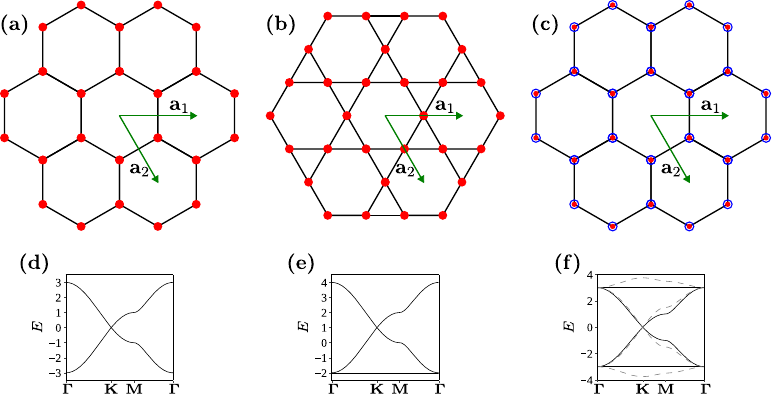}
	\caption{\textbf{(a)}-\textbf{(c)} Illustrations of the tight-binding models in \cref{app:kagome-honeycomb-models}. \textbf{(d)}-\textbf{(f)} Corresponding example band structures. \textbf{(a)} The honeycomb lattice one orbital model described in \cref{app:honeycomb-one-orbital}. \textbf{(b)} The kagome lattice one orbital model described in \cref{app:kagome-one-orbital}. \textbf{(c)} The honeycomb lattice two orbital model described in \cref{app:honeycomb-two-orbital}. \textbf{(d)} Band structure for the Hamiltonian in \cref{appeq:honeycomb-one-orbital} with $s = 0$, $t = 1$. \textbf{(e)} Band structure for the Hamiltonian in \cref{appeq:kagome-one-orbital} with $s = 0$, $t = 1$. \textbf{(f)} Band structures for the Hamiltonian in \cref{appeq:honeycomb-two-orbital} with $s = 0$, $t_0 = 1$. In solid black lines we have $t_x = 1$ and in dashed gray lines we have $t_x = 1.25$.}
	\label{appfig:tight-binding}
\end{figure}
In the following subsections we review three tight-binding models on kagome and honeycomb lattices. We denote the Bravais lattice, reciprocal lattice, and Brillouin zone for each model by $L$, $P$, and $\text{BZ}$, respectively. The primitive vectors of $L$ are $\ba_1$ and $\ba_2$ as given in \cref{eq:define-a1-a2} and illustrated in \cref{appfig:tight-binding}\textbf{(a)}-\textbf{(c)}. The primitive vectors for $P$ are $\bb_1$ and $\bb_2$ as given in \cref{eq:define-b1-b2} and the high symmetry momenta are
\begin{equation}
\bGamma = \bzero,\quad \bK = \frac{2}{3}\bb_1 + \frac{1}{3}\bb_2,\quad \bM = \frac{1}{2}\bb_1 + \frac{1}{2}\bb_2.
\end{equation}

\subsection{Honeycomb lattice one orbital model}\label{app:honeycomb-one-orbital}
We first consider a tight-binding model with a single spinless Wannier function on each site of a honeycomb lattice. The Wannier centers and nearest neighbor bonds are shown in \cref{appfig:tight-binding}\textbf{(a)} with red circles and black lines, respectively. The Wannier functions are all related by isometries to a single exponentially localized orbital which carries some 1D spinless coirrep $\rho \in \{A_1, A_2\}$ of $3m1'$ (see \cref{app:point-groups} and \cref{apptbl:characters}). The Wannier functions together transform under the induced corepresentation $(\rho)_{2b}$ of the magnetic space group $P6mm1'$ (No. 183.186 in the BNS setting \cite{Gallego2012}), which we now describe.

The lattice sites have positions $\br + \alpha\btau$ where $\br \in L$, $\alpha \in \{+, -\}$, and $\btau =\frac{1}{\sqrt{3}} R_{\pi/2}\ba_1$. We denote the Wannier function at position $\br + \alpha\btau$ by $\ket{\br, \alpha}$ and define Bloch states
\begin{equation}
\ket{\bk, \alpha} = \frac{1}{\sqrt{|\text{BZ}|}} \sum_{\br \in L} e^{i\bk \cdot (\br + \alpha\btau)}\ket{\br, \alpha}.
\end{equation}
Using coset representatives $T_\br$ and $T_\br C_{2z}$ where $T_\br$ denotes translation by $\br \in L$, the induced corepresentation $(\rho)_{2b}$ is determined by
\begin{equation}\label{appeq:3m1'-2b-1D}
\begin{split}
T_{\br} \ket{\bk, \alpha} &= e^{-i\bk \cdot \br}\ket{\bk, \alpha}\\
C_{3z} \ket{\bk, \alpha} &= \ket{R_{2\pi/3} \bk, \alpha}\\
C_{2z} \ket{\bk, \alpha} &= \ket{-\bk, -\alpha}\\
M_x \ket{\bk, \alpha} &= \text{tr}(\rho(M_x)) \ket{\mathcal{R}_\bhatx \bk, \alpha}\\
M_y \ket{\bk, \alpha} &= \text{tr}(\rho(M_x)) \ket{\mathcal{R}_\bhaty \bk, -\alpha}\\
\mathcal{T} \ket{\bk, \alpha} &= \ket{-\bk, \alpha}.
\end{split}
\end{equation}
According to the Bilbao Crystallographic Server \cite{Elcoro2021,Xu2020}, $(\rho)_{2b}$ is always an elementary band representation \cite{Bradlyn2017}.

If the Hamiltonian $H$ for this system commutes with $(\rho)_{2b}$ and has at most nearest neighbor hopping, then $H$ takes the form
\begin{equation}\label{appeq:honeycomb-one-orbital}
H\ket{\bk, \alpha} = s\ket{\bk, \alpha} + t\sum_{j=1}^3 e^{-i\alpha \bk \cdot R_{\zeta_j}\btau} \ket{\bk, -\alpha}
\end{equation}
for real parameters $s$ and $t$, where $\zeta_j = \frac{2\pi}{3}(j-1)$. \cref{appfig:tight-binding}\textbf{(d)} shows the band structure for the Hamiltonian in \cref{appeq:honeycomb-one-orbital} with $s = 0$ and $t = 1$. The band structure has two dispersive bands with at Dirac cone at $\bK$, van Hove singularities at $\bM$, and extrema at $\bGamma$. The spectrum is symmetric about $s$ because $H - s$ anticommutes with the chiral symmetry operator
\begin{equation}
C\ket{\bk, \alpha} = \alpha\ket{\bk, \alpha}.
\end{equation}

\subsection{Kagome lattice one orbital model}\label{app:kagome-one-orbital}
We next consider a tight-binding model with a single spinless Wannier function on each site of a kagome lattice. The Wannier centers and nearest neighbor bonds are shown in \cref{appfig:tight-binding}\textbf{(b)} with red circles and black lines, respectively. The Wannier functions are all related by isometries to a single exponentially localized orbital which carries some spinless coirrep $\rho \in \{A_1, A_2, B_1, B_2\}$ of $mm21'$ (see \cref{app:point-groups} and \cref{apptbl:characters}). The Wannier functions together transform under the induced corepresentation $(\rho)_{3c}$ of $P6mm1'$, which we now describe.

The lattice sites have positions $\br + \btau_j$ where $\br \in L$, $j \in \{1, 2, 3\}$, $\btau_j = \frac{1}{2}R_{\zeta_j}\ba_1$, and $\zeta_j = \frac{2\pi}{3}(j-1)$. We denote the Wannier function at position $\br + \btau_j$ by $\ket{\br, j}$ and define Bloch states
\begin{equation}
\ket{\bk, j} = \frac{1}{\sqrt{|\text{BZ}|}} \sum_{\br \in L} e^{i\bk \cdot (\br + \btau_j)}\ket{\br, j}.
\end{equation}
Using coset representatives $T_\br$, $T_\br C_{3z}$, and $T_\br C_{3z}^{-1}$ where $T_\br$ denotes translation by $\br \in L$, the induced corepresentation $(\rho)_{3c}$ is determined by
\begin{equation}\label{appeq:mm21'-3c}
\begin{split}
T_{\br} \ket{\bk, j} &= e^{-i\bk \cdot \br}\ket{\bk, j}\\
C_{3z} \ket{\bk, j} &= \ket{R_{2\pi/3} \bk, j+1}\\
C_{2z} \ket{\bk, j} &= \text{tr}(\rho(C_{2z})) \ket{-\bk, j}\\
M_x \ket{\bk, j} &= \text{tr}(\rho(M_x)) \ket{\mathcal{R}_\bhatx \bk, 2 - j}\\
M_y \ket{\bk, j} &= \text{tr}(\rho(M_y)) \ket{\mathcal{R}_\bhaty \bk, 2 - j}\\
\mathcal{T} \ket{\bk, j} &= \ket{-\bk, j}
\end{split}
\end{equation}
where the $j$ indices are cyclic modulo $3$. According to the Bilbao Crystallographic Server \cite{Elcoro2021,Xu2020}, $(\rho)_{3c}$ is always an elementary band representation \cite{Bradlyn2017}.

If the Hamiltonian $H$ for this system commutes with $(\rho)_{3c}$ and has at most nearest neighbor hopping, then $H$ takes the form
\begin{equation}\label{appeq:kagome-one-orbital}
H\ket{\bk, j} = s\ket{\bk, j} + 2t\cos(\bk \cdot \btau_{j-1})\ket{\bk, j+1} + 2t \cos(\bk \cdot \btau_{j+1})\ket{\bk, j-1}
\end{equation}
for real parameters $s$ and $t$. \cref{appfig:tight-binding}\textbf{(e)} shows the band structure for the Hamiltonian in \cref{appeq:kagome-one-orbital} with $s = 0$ and $t = 1$. The band structure has two dispersive bands with at Dirac cone at $\bK$, van Hove singularities at $\bM$, and extrema at $\bGamma$. Additionally, there is a flat band that has a quadratic touching with a dispersive band at $\bGamma$. We prove in \cref{app:kagome-one-orbital-flat} that the Hamiltonian in \cref{appeq:kagome-one-orbital} has a perfectly flat band with energy $s - 2t$.

\subsection{Honeycomb lattice two orbital model}\label{app:honeycomb-two-orbital}
Finally, we consider a tight-binding model with two spinless Wannier functions on each site of a honeycomb lattice. The Wannier centers and nearest neighbor bonds are shown in \cref{appfig:tight-binding}\textbf{(c)} with red and blue circles and black lines, respectively. The Wannier functions are all related by isometries to a pair of exponentially localized orbitals which transform under coirrep $E$ of $3m1'$ (see \cref{app:point-groups} and \cref{apptbl:characters}). The Wannier functions together transform under the induced corepresentation $(E)_{2b}$ of $P6mm1'$, which we now describe.

The lattice sites have positions $\br + \alpha\btau$ where $\br \in L$, $\alpha \in \{+, -\}$, and $\btau =\frac{1}{\sqrt{3}} R_{\pi/2}\ba_1$. We denote the two Wannier functions at position $\br + \alpha\btau$ by $\ket{\br, \alpha, \ell}$ where $\ell \in \{+, -\}$ labels the two orbitals, and we define Bloch states
\begin{equation}
\ket{\bk, \alpha, \ell} = \frac{1}{\sqrt{|\text{BZ}|}} \sum_{\br \in L} e^{i\bk \cdot (\br + \alpha\btau)}\ket{\br, \alpha, \ell}.
\end{equation}
For convenience, we choose the two orbitals so that $C_{3z}$, $M_x$, and $\mathcal{T}$ are represented by $e^{-i(2\pi/3)\sigma_z}$, $-\sigma_x$, and $\sigma_x \mathcal{K}$, respectively, where $\mathcal{K}$ denotes complex conjugation. This symmetry condition is satisfied, for example, by atomic orbitals $p_x + i \ell p_y$. Using coset representatives $T_\br$ and $T_\br C_{2z}$ where $T_\br$ denotes translation by $\br \in L$, the induced corepresentation $(E)_{2b}$ is determined by
\begin{equation}\label{appeq:3m1'-2b-2D}
\begin{split}
T_{\br} \ket{\bk, \alpha, \ell} &= e^{-i\bk \cdot \br}\ket{\bk, \alpha, \ell}\\
C_{3z} \ket{\bk, \alpha, \ell} &= e^{-i(2\pi/3)\ell}\ket{R_{2\pi/3} \bk, \alpha, \ell}\\
C_{2z} \ket{\bk, \alpha, \ell} &= \ket{-\bk, -\alpha, \ell}\\
M_x \ket{\bk, \alpha, \ell} &= -\ket{\mathcal{R}_\bhatx \bk, \alpha, -\ell}\\
M_y \ket{\bk, \alpha, \ell} &= -\ket{\mathcal{R}_\bhaty \bk, -\alpha, -\ell}\\
\mathcal{T} \ket{\bk, \alpha, \ell} &= \ket{-\bk, \alpha, -\ell}.
\end{split}
\end{equation}
According to the Bilbao Crystallographic Server \cite{Elcoro2021,Xu2020}, $(E)_{2b}$ is an elementary band representation \cite{Bradlyn2017}.

If the Hamiltonian $H$ for this system commutes with $(E)_{2b}$ and has at most nearest neighbor hopping, then $H$ takes the form
\begin{equation}\label{appeq:honeycomb-two-orbital}
H \ket{\bk, \alpha, \ell} = s\ket{\bk, \alpha, \ell} + \sum_{j=1}^3 e^{-i\alpha \bk \cdot R_{\zeta_j}\btau} (t_0 \ket{\bk, -\alpha, \ell} + t_x e^{-i\ell\zeta_j} \ket{\bk, -\alpha, -\ell})
\end{equation}
for real parameters $s$, $t_0$, and $t_x$, where $\zeta_j = \frac{2\pi}{3}(j-1)$. \cref{appfig:tight-binding}\textbf{(f)} shows band structures for the Hamiltonian in \cref{appeq:honeycomb-two-orbital} with $s = 0$, $t_0 = t_x = 1$ in solid black lines and $s = 0$, $t_0 = 1$, $t_x = 1.25$ in dashed gray lines. The middle two bands in both cases are dispersive and have a Dirac cone at $\bK$, van Hove singularities at $\bM$, and extrema at $\bGamma$. Additionally, the top and bottom bands have quadratic touchings with the middle two bands at $\bGamma$. The spectrum is symmetric about $s$ because $H - s$ anticommutes with the chiral symmetry operator
\begin{equation}
C\ket{\bk, \alpha, \ell} = \alpha\ket{\bk, \alpha, \ell}.
\end{equation}
When $t_0 = t_x = 1$, the top and bottom bands are flat, but when $t_0 = 1$, $t_x = 1.25$ they are dispersive. We prove in \cref{app:honeycomb-two-orbital-flat} that the Hamiltonian in \cref{appeq:honeycomb-two-orbital} has two perfectly flat bands with energies $s \pm 3 |t_0|$ when $|t_0| = |t_x|$.

\section{Kagome and honeycomb flat bands from bipartite crystalline lattices}\label{app:kagome-honeycomb-bcl}
In the following subsections we review the bipartite crystalline lattice construction of Ref. \cite{Calugaru2022} and then apply it to prove the existence of flat bands in the models described in \cref{app:kagome-one-orbital,app:honeycomb-two-orbital}.

\subsection{Bipartite crystalline lattice construction}\label{app:bcl-construction}
As in \cref{app:honeycomb-one-orbital,app:kagome-one-orbital,app:honeycomb-two-orbital}, we consider a tight-binding model supported on Wannier functions that are formed from some set of exponentially localized orbitals. We partition the orbitals into two subsets $\mathcal{O}_1$ and $\mathcal{O}_2$ and denote the Wannier functions arising from subset $\mathcal{O}_n$ by $\ket{\br, n, x}$ for $\br \in L$, $n \in \{1, 2\}$, and $1 \leq x \leq N_n$, where $L$ is the Bravais lattice and $N_n$ is the number of orbitals from $\mathcal{O}_n$ in each unit cell. The orbital $\ket{\br, n, x}$ has position $\br + \btau_{n,x}$ and we define Bloch states
\begin{equation}
\ket{\bk, n, x} = \frac{1}{\sqrt{|\text{BZ}|}}\sum_{\br\in L} e^{i\bk \cdot (\br + \btau_{n,x})} \ket{\br, n, x}.
\end{equation}

We assume the Hamiltonian $H$ has no hoppings between Wannier functions arising from the same subset $\mathcal{O}_n$, except for a constant chemical potential $\mu$ on Wannier functions arising from $\mathcal{O}_2$. A model of this form is called a bipartite crystalline lattice \cite{Calugaru2022}. We assume without loss of generality that $N_1 \geq N_2$. For a given crystal momentum $\bk$, $H$ is represented on Bloch states $\ket{\bk, n, x}$ by the matrix
\begin{equation}\label{eq:flat-band-matrix}
\mathcal{H}(\bk) = \begin{pmatrix}
0 & \mathcal{S}(\bk)\\
\mathcal{S}^\dagger(\bk) & -\mu
\end{pmatrix}
\end{equation}
for some $N_1 \times N_2$ complex matrix $\mathcal{S}(\bk)$. The first $N_1$ (last $N_2$) rows and columns of $\mathcal{H}(\bk)$ correspond to $\mathcal{O}_1$ ($\mathcal{O}_2$). By the rank-nullity theorem, the null space of $\mathcal{S}^\dagger(\bk)$ has dimension at least $N_1 - N_2$. Additionally, if $v$ is in the null space of $\mathcal{S}^\dagger(\bk)$ then
\begin{equation}
\mathcal{H}(\bk)\begin{pmatrix}
v \\ 0
\end{pmatrix} = 0.
\end{equation}
As a result, $H$ must have at least $N_1 - N_2$ perfectly flat bands with energy $0$ which are completely supported on the Wannier functions arising from $\mathcal{O}_1$.

We next apply Schrieffer-Wolff perturbation theory \cite{Schrieffer1966,Bravyi2011} to second order in $\mathcal{S}(\bk)$ to find the low energy Hamiltonian $\tilde{H}$ for the Wannier functions arising from $\mathcal{O}_1$. For a given cystal momentum $\bk$, $\tilde{H}$ is represented on Bloch states $\ket{\bk, 1, x}$ by the matrix
\begin{equation}
\tilde{\mathcal{H}}(\bk) = \frac{\mathcal{S}(\bk)\mathcal{S}^\dagger(\bk)}{\mu}.
\end{equation}
We now send $|\mu|, |\mathcal{S}(\bk)| \to \infty$ while fixing $\tilde{\mathcal{H}}(\bk)$. In this limit, the Schrieffer-Wolff perturbation theory at second order becomes exact. The resulting tight-binding model $\tilde{H}$ then has at least $N_1 - N_2$ perfectly flat bands with energy $0$.

\begin{figure}[h]
	\centering
	\includegraphics{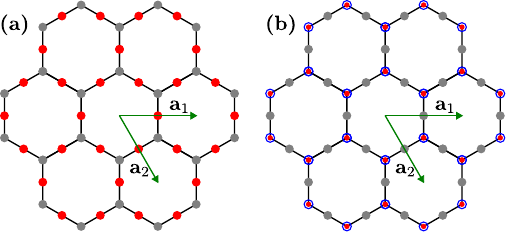}
	\caption{\textbf{(a)} Bipartite crystalline lattice for \cref{app:kagome-one-orbital-flat}. The $n=1$ ($n=2$) Wannier centers are shown with red (gray) circles. \textbf{(b)} Bipartite crystalline lattice for \cref{app:honeycomb-two-orbital-flat}. The $n=1$ ($n=2$) Wannier centers are shown with red and blue (gray) circles. In both diagrams, the black lines indicate nearest neighbor hoppings.}
	\label{appfig:bcl}
\end{figure}

\subsection{Kagome lattice one orbital model}\label{app:kagome-one-orbital-flat}
We take $\ket{\br, 1, x}$ to be the kagome lattice Wannier functions $\ket{\br, j}$ in \cref{app:kagome-one-orbital} so that $N_1 = 3$. Additionally, we take $\ket{\br, 2, x}$ to be the honeycomb lattice Wannier functions $\ket{\br, \alpha}$ in \cref{app:honeycomb-one-orbital} so that $N_2 = 2$. The Wannier functions and nearest neighbor hoppings are illustrated in \cref{appfig:bcl}\textbf{(a)} in red ($n = 1$) and gray ($n = 2$) circles, and black lines, respectively. Since the matrix elements of the Hamiltonian in \cref{appeq:kagome-one-orbital} do not depend on the choice of corepresentation, we are free to choose corepresentations $A_1$ of $mm21'$ and $A_1$ of $3m1'$ to induce the corepresentations of $P6mm1'$, without loss of generality. The full set of Wannier functions transforms under the direct sum of the two induced corepresentations.

If the Hamiltonian $H$ for this system commutes with the direct sum corepresentation, satisfies the requirements in \cref{app:bcl-construction}, and has at most first order hoppings, then we must have
\begin{equation}
H \ket{\bk, j} = \tilde{t} \sum_{\alpha = \pm} e^{i\bk \cdot \btau_{j,\alpha}} \ket{\bk, \alpha}, \quad H\ket{\bk, \alpha} = -\mu \ket{\bk, \alpha} + \tilde{t} \sum_{j=1}^3 e^{-i\bk \cdot \btau_{j,\alpha}} \ket{\bk, j}
\end{equation}
for a real parameter $\tilde{t}$, where
\begin{equation}\label{appeq:tau-alpha-j}
\btau_{j,\alpha} = \frac{\alpha}{2\sqrt{3}} R_{\zeta_j + \pi/2} \ba_1
\end{equation}
with $\zeta_j = \frac{2\pi}{3}(j-1)$. If we take $\mu = \tilde{t}^2/t$ for some fixed real value $t$ and then send $\tilde{t} \to \infty$, the low energy Hamiltonian $\tilde{H}$ described in \cref{app:bcl-construction} takes the form of \cref{appeq:kagome-one-orbital} with $s = 2t$. Since $\tilde{H}$ has $N_1 - N_2 = 1$ perfectly flat band with energy $0$ for any $t \in \R$, we conclude that the Hamiltonian in \cref{appeq:kagome-one-orbital} always has a perfectly flat band with energy $s - 2t$.

\subsection{Honeycomb lattice two orbital model}\label{app:honeycomb-two-orbital-flat}
Next, we take $\ket{\br, 1, x}$ to be the honeycomb lattice Wannier functions $\ket{\br, \alpha, \ell}$ in \cref{app:honeycomb-two-orbital} so that $N_1 = 4$. Additionally, we take $\ket{\br, 2, x}$ to be the kagome lattice Wannier functions $\ket{\br, j}$ in \cref{app:kagome-one-orbital} so that $N_2 = 3$. The Wannier functions and nearest neighbor hoppings are illustrated in \cref{appfig:bcl}\textbf{(b)} in red and blue ($n = 1$) and gray ($n = 2$) circles, and black lines, respectively. We choose corepresentations $E$ of $3m1'$ and $\rho \in \{A_1, A_2, B_1, B_2\}$ of $mm21'$ to induce the corepresentations of $P6mm1'$. The full set of Wannier functions transforms under the direct sum of the two induced corepresentations.

If the Hamiltonian $H$ for this system commutes with the direct sum corepresentation, satisfies the requirements in \cref{app:bcl-construction}, and has at most first order hoppings, then we must have
\begin{equation}
H \ket{\bk, \alpha, \ell} = \tilde{t}^* \sum_{j=1}^3 e^{-i\bk \cdot \btau_{j,\alpha}} e^{i\ell \zeta_j} \alpha^{\nu_1} \ell^{\nu_2} \ket{\bk, j}, \quad H\ket{\bk, j} = -\mu \ket{\bk, j} + \tilde{t} \sum_{\alpha = \pm}\sum_{\ell = \pm} e^{i\bk \cdot \btau_{j,\alpha}} e^{-i\ell\zeta_j}\alpha^{\nu_1} \ell^{\nu_2} \ket{\bk, \alpha, \ell}
\end{equation}
for a real or imaginary parameter $\tilde{t}$, where
\begin{equation}
\begin{cases}
\tilde{t} \in \R, \nu_1 = 0, \nu_2 = 0 & \text{when } \rho = A_2\\
\tilde{t} \in i\R, \nu_1 = 0, \nu_2 = 1 & \text{when } \rho = A_1\\
\tilde{t} \in \R, \nu_1 = 1, \nu_2 = 0 & \text{when } \rho = B_1\\
\tilde{t} \in i\R, \nu_1 = 1, \nu_2 = 1 & \text{when } \rho = B_2
\end{cases}
\end{equation}
(see \cref{apptbl:characters}), $\btau_{\alpha,j}$ is given by \cref{appeq:tau-alpha-j}, and $\zeta_j = \frac{2\pi}{3}(j-1)$. If we take $\mu = |\tilde{t}|^2/t$ for some fixed real value $t$ and then send $|\tilde{t}| \to \infty$, the low energy Hamiltonian $\tilde{H}$ described in \cref{app:bcl-construction} takes the form of \cref{appeq:honeycomb-two-orbital} with $s = 3t$, $t_0 = (-1)^{\nu_1}t$, $t_x = (-1)^{\nu_1 + \nu_2}t$. Since $\tilde{H}$ has a perfectly flat band with energy $0$ for any $t \in \R$ and $\nu_1, \nu_2 \in \{0, 1\}$, we conclude that the Hamiltonian in \cref{appeq:honeycomb-two-orbital} has two perfectly flat bands with energies $s \pm 3|t_0|$ when $|t_0| = |t_x|$.
\end{document}